\documentclass[12pt, oneside, a4paper]{report}
\usepackage[thref, framed, amsthm, thmmarks]{ntheorem}
\usepackage{amsmath, amsfonts, amssymb, cancel}
\usepackage{framed}
\usepackage{graphicx}
\usepackage{subfig}
\usepackage{makeidx}
\usepackage{fullpage}
\usepackage{tikz}
\usepackage{siunitx}
\usepackage[titletoc]{appendix}
\usepackage{booktabs}
\usepackage[nottoc]{tocbibind}
\usepackage{algorithm,algpseudocode}
\usepackage{longtable, multirow}
\usepackage[font=small]{caption}
\usepackage{fancyhdr}
\usepackage{cite}
\usepackage{CJKutf8}
\usepackage[overlap, CJK]{ruby}

\usepackage{afterpage}
\usepackage[hidelinks, bookmarksnumbered=true, pdftitle=Quantum Control based on Deep Reinforcement Learning
, pdfauthor=Zhikang Wang]{hyperref}
\makeindex
\DeclareMathOperator*{\argmax}{argmax}
\DeclareMathOperator*{\argmin}{argmin}
\theoremstyle{definition}
\newframedtheorem{theorem}{Theorem}[section]

\begin{document}

	\begin{titlepage}
		\begin{center}
	\vspace*{0.2cm}
	\huge
	Master Thesis\\
	{\begin{CJK}{UTF8}{ipxm}
			修士論文
	\end{CJK}}\\

	\vspace*{2.5cm}
	{Quantum Control based on Deep Reinforcement Learning}\\[0.8cm]
	
	{\begin{CJK}{UTF8}{ipxm}
			(深層強化学習に基づく量子制御)
		\end{CJK}}\\
	
	\vspace{4cm}
	
	\LARGE
	July, 2019\\[0.3cm]
	\LARGE
	{\begin{CJK}{UTF8}{ipxm}
			令和元年７月
	\end{CJK}}\\
	\vspace{1.5cm}
	\LARGE
	Zhikang Wang\\[0.3cm]
	{\begin{CJK}{UTF8}{ipxm}
			\ruby{王}{ワン}　\ruby{智康}{ヅーカン}
	\end{CJK}}\\

	\vfill
	\large
    \textit{Department of Physics, the University of Tokyo}\\[0.3cm]
    {\begin{CJK}{UTF8}{ipxm}
    		東京大学大学院　理学系研究科　物理学専攻
    \end{CJK}}

\end{center}

	\end{titlepage}

	%\begin{titlepage}
	%	\input{titlepage/titlepage}
	%\end{titlepage}
	
	\chapter*{Abstract}\label{abstract}
	\addcontentsline{toc}{chapter}{\numberline{}Abstract}
	With the development of quantum experimental techniques in recent years, artificial quantum systems have become more controllable, and the increased controllability has opened up a new regime of quantum physics that is both complex and realizable. Especially, quantum technology has been brought close to real-world applications, including quantum metrology, quantum-limited sensors \cite{QuantumSensing} and quantum computers \cite{Superconducting}, which are of great importance for future technology. However, quantum systems in the real world are susceptible to imperfections. They are vulnerable to noise and decoherence, and they are often constructed based on approximations, such as ignoring anharmonic factors or long-range interactions. These deficiencies are closely related to the fact that, when we take all these tricky factors into consideration, the quantum systems typically become too complicated to analyse and thus it is hard to find the best setting for them. To overcome this difficulty, deep reinforcement learning has been proposed as a universal solver to such problems. An artificial intelligence (AI) technology has the advantage in that it does not require to explicitly analyse the problem, and that it automatically searches for a good solution through trial and error \cite{RLReview}. On the other hand, deep learning as a new tool achieved its success only less than 10 years ago \cite{DeepLearningReview} and has been applied to physics over the last 3 years. Probably due to a knowledge gap between physicists and AI technology, deep learning has yet to be applied to many fundamental physical problems and yet to prove its efficacy or superior performance to conventional strategies. \\

In this thesis, we consider two simple but typical control problems and apply deep reinforcement learning to them, i.e., to cool and control a particle which is subject to continuous position measurement in a one-dimensional quadratic potential or in a quartic potential. We compare the performance of reinforcement learning control and conventional control strategies on the two problems, and show that the reinforcement learning achieves a performance comparable to the optimal control for the quadratic case, and outperforms conventional control strategies for the quartic case for which the optimal control strategy is unknown. To our knowledge, this is the first time deep reinforcement learning is applied to quantum control problems in continuous real space. Our research demonstrates that deep reinforcement learning can be used to control a stochastic quantum system in real space effectively as a measurement-feedback closed-loop controller, and our research also shows the ability of AI to discover new control strategies and properties of the quantum systems that are not well understood, and we can gain insights into these problems by learning from the AI, which opens up a new regime for scientific research.

	\chapter*{Acknowledgements}\label{acknowledgements}
	\addcontentsline{toc}{chapter}{\numberline{}Acknowledgements}
	I thank Yuto Ashida and professor Masahito Ueda for discussion on the research. Particularly, I would like to thank Y. Ashida for bringing my attention to the importance of the research topic presented here and to thank for professor M. Ueda's detailed comments on both the research and the writing of this thesis; without them the research would not be carried out and the thesis would not be completed. I also thank for financial support by the Global Science Graduate Course (GSGC) program and the support of computational resources by the Institute for Physics of Intelligence at the University of Tokyo, which have made the research possible. I am grateful to the Stack Exchange community and the Pytorch community for their valuable comments and technical helps on programming and implementation, which have solved myriads of the technical problems that I encountered and taught me how to build up the program. I would also like to thank Ziyin Liu for discussion on deep learning and programming and thank Ryusuke Hamazaki and Zongping Gong for discussion on the specific models investigated in the research, and I thank professor Mio Murao for discussion and her valuable advice which have helped me out to complete the research for all input cases that are discussed in Chapter 4. \\

My parents and friends have supported and encouraged me along the way in my life and in my research, and I wish to express my deepest gratitude to them.
	
	\tableofcontents
	
	\chapter{Introduction}\label{Introduction}
	\pagenumbering{arabic}
	In this chapter, we give a background review on the current developments of quantum control and deep learning in Section \ref{introBackground}, and then introduce the motivation and content of our research in Section \ref{introDeepLearningControl}, and finally present the outline of this thesis in Section \ref{outline}.\\ 

\section{Background}\label{introBackground}
\subsection{Quantum Control}\label{quantum control}
In general, a control problem is to find a control protocol that maximizes a score or minimizes a cost which encodes a prescribed target of control. The control as an output from the controller is a time sequence which influences evolution of the controlled system. When the evolution of the system is deterministic, the problem can be formulated as \cite{LinearQuadraticControl}\\
\begin{equation}\label{control}
u^*=\argmin_u \left[\int_{0}^{T} L(u,x,t)\, dt + L_T(u,x)\right],\quad dx=f(x,u)dt,
\end{equation}\\
where $u$ is the control variable, $L$ stands for the loss, $T$ is the total time considered in this control, $L_{T}$ is a specific loss for the final state, $f$ represents the rule of time evolution of the system, and the goal of a control problem is to find this $u^*$, which is called the optimal control, and $x$ and $u$ depend on time $t$. If $u$ does not use information obtained from the controlled system $x$ during its control, it is called \textit{open-loop} control \cite{ControlSystems}; if it depends on information obtained from the controlled system, it is called \textit{closed-loop} control, or \textit{feedback} control. The feedback control necessarily involves measurement, and therefore it is, in fact, the measurement-feedback control in the quantum case. For classical systems, the control problem can often be solved rather straightforwardly, while for quantum systems this is not the case. The main reason lies in the complexity of description of the controlled system. To describe a classical system, a few numbers as relevant physical quantities suffice, while for a quantum system, we need many more parameters to describe the superposition among different components, and if it is a many-body state we even need exponentially many parameters. This difficulty inhibits straightforward analysis to solve quantum control problems, except for a few cases where the quantum system can be simplified. \\

This quantum control problem did not attract as much attention as it deserves until the development of controllable artificial quantum systems in the last two decades \cite{ColdAtomQuantumControl}. Examples of the controllable systems are quantum dots in semiconductors and photonic crystals, trapped ion systems, superconducting qubits, optical lattices, nitrogen-vacancy (NV) centers, coupling optical systems, cavity optomechanical systems and so on \cite{QuantumDots,QuantumSpinDots,QuantumSimulation,TrappedIon,NVCenter,Superconducting,OpticalQuantumComputation,CavityOptomechanics}. They are used as platforms for simulation of quantum dynamics or considered as potential candidates for quantum computation devices. However, none of them is perfect. They always come with sources of noise and contain small error terms that cannot be controlled in a straightforward manner. For example, superconducting qubits use the lowest two energy levels of its small superconducting circuit as the logical $|0\rangle$ and $|1\rangle$ states for quantum computation, but there is always a non-zero probability for the state to jump to energy levels higher than $|0\rangle$ and $|1\rangle$ and to go out of control. To find a control strategy to suppress such problems, typically approximations and assumptions are made to simplify the situations, often involving use of perturbative expansions or reduced subspace effective Hamiltonians, and then a control is calculated. For the superconducting qubit example, the control pulses are optimized so that one or two nearest energy levels above the operating qubit levels are suppressed. Concerning decoherence, the dynamical decoupling control is a good example, which considers the fast limit of control and series expansion \cite{DynamicDecoupling}. Another example could be the spin echo control technique, which specifically deals with time-independent inhomogeneous imperfections \cite{SpinEcho}. All these methods come along with important assumptions and approximations. Moreover, it should be mentioned that usually the analysis does not straightforwardly give the optimal control as defined in Eq. (\ref{control}), but only gives some hints so that reasonable control protocols can be designed by hand.\\

When the above analysis-based method does not work, numerical search algorithms provide an alternative. They assume that the control is a series of pulses, or a superposition of trigonometric waves, which can be parametrized with a sequence of parameters, and they consider small variations of the control parameters to get closer to the control target, such as fidelity, and then they do it iteratively to gradually modify the control parameters. These algorithms include QOCT \cite{QuantumOptimalControlTheory}, CRAB \cite{CRAB} and GRAPE \cite{GRAPE}. However, since these methods are either effectively or directly based on gradients, if the situation is complicated, they can easily be trapped in local optima and do not give satisfactory solutions, as exemplified in Ref. \cite{EvolutionaryQuantumControl}. Nevertheless, they have been shown to be useful in many simple practical scenarios where no other means are available to find a control, even including simple chaotic systems \cite{ChaosControl}. One weakness of these methods is that they can only be used as open-loop controls, for which the starting point and the endpoint of control are prescribed beforehand. Actually it is almost impossible to optimize the control and at the same time make it conditioned on all types of measurement outcomes.\\

On the whole, it can be recognized that there is no universal approach to quantum control, and most of the current methods are ad hoc for specific situations. Therefore, if the controlled system becomes more complicated and involved, it would be much more difficult to find a satisfactory control using the above strategies, and it is desirable if we can find some general and better strategy to overcome the difficulty of analysing a complex system in order to obtain a control. \\

\subsection{Deep Learning}
Deep learning, namely machine learning with deep neural networks, has become popular
and used extensively in recent years, especially for tasks that were previously considered difficult for AI to do. For instance, deep learning has established new records on many problem-solving contests \cite{ImageNetClassification,ZeroResourceSpeech} and defeated human champions in games \cite{chess-like}, and it is still being researched and developing rapidly. Deep learning will be introduced more formally and in detail in Chapter 3, and therefore we only give a brief introduction here to present the general idea.\\ 

\begin{figure}[tb]
	\centering
	\subfloat{
		\begin{minipage}[c]{0.11\linewidth}
			\centering
			\includegraphics[width=\linewidth]{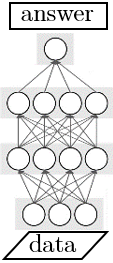}
	\end{minipage}}\qquad {\LARGE $\Rightarrow$}\qquad
	\subfloat{
		\begin{minipage}[c]{0.144\linewidth}
			\centering
			\includegraphics[width=\linewidth]{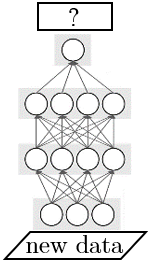}
	\end{minipage}}
	\caption{Working of a deep learning system. It learns by fitting its internal connection parameters (weights and biases) into the given data-answer pairs, such that the computation of the network precisely relates every data to its corresponding answer for the whole training dataset that it learns. After training, it is used to predict something on new data.}
	\label{fig:deeplearning}
\end{figure}

Generally speaking, deep learning uses a deep neural network as a powerful function approximator that can learn patterns of previous data to give predictions on new data, as illustrated in Fig. \ref{fig:deeplearning}. It learns by modifying its internal parameters to fit given data-answer pairs as training data. Due to the complexity and universality of deep neural networks, it has turned out that this simple learning procedure can make the neural network correctly learn various complex relations between the data and the answer. For example, based on this simple method, it can be used to recognize objects, modify images and do translation \cite{Translation, ImagenetSota, DeepPS}, and evaluate the advantages and disadvantages in chess \cite{chess-like}. It is generally believed that neural networks can almost learn any functions, as long as functions are sufficiently smooth and do not appear to be pathological. However, as a drawback, deep learning always gives approximate solutions and does not explain its reason. Deep learning is typically not precise, and is generally a blackbox technique which we cannot explain well so far \cite{DeepLearningBook}.\\

\section{Combination of Deep Learning and Quantum Control} \label{introDeepLearningControl}
To use deep learning for a control problem, the reinforcement learning scheme needs to be implemented, and it will be explained in detail in Chapter 3. Reinforcement learning with deep learning uses a neural network to evaluate which control option is good and which control option is bad during the control, and it learns by exploring its environment, which stands for the controlled system behaviour. It explores its environment to accumulate experience, and it learns the accumulated experience, and its goal is set to maximize the control target when making control decisions. Overall, it learns and explores different possibilities of its environment automatically, and can learn underlying rules of the environment and often avoid local optima. Therefore, it can be seen as an alternative to the gradient-based quantum control algorithms in Section \ref{quantum control}. One advantage of reinforcement learning based control is that, it can deal with both open-loop control and closed-loop control in the same way. Since the neural network needs to take information about the controlled system to give a control output, it does not matter if the controlled system is changed suddenly due to measurement-backaction: if the system changed, the control output from the neural network is also changed and that is all. The versatility of AI makes all kinds of control scenarios possible without the need for human design, which is difficult with only conventional methods. \\
\begin{figure}[thb]
	\centering
	\includegraphics[width=0.3\linewidth]{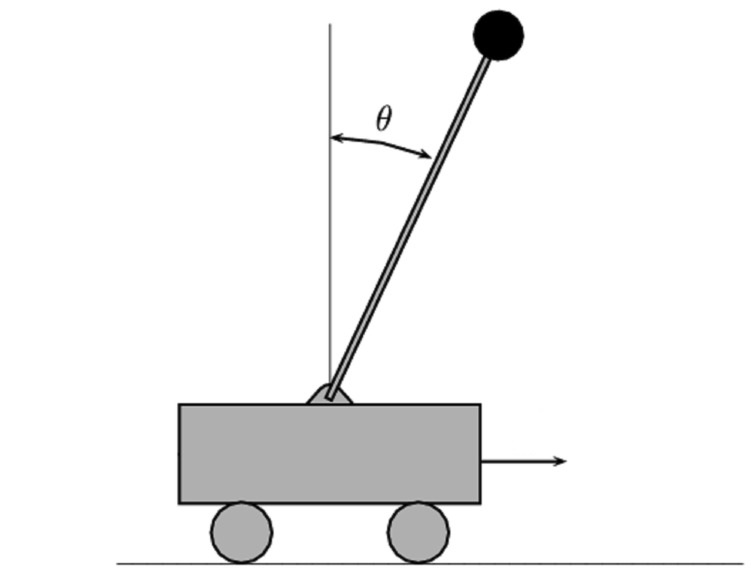}\qquad\qquad
	\includegraphics[width=0.3\linewidth]{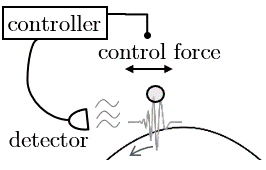}
	\caption{The classical cart-pole system and the controlled inverted potential quantum system with measurement. In either case the system is unstable, and the particle tends to fall off from the center. The target of control is to move the cart, or apply an external force, so that the particle stays at the center. For the quantum case, position measurement is necessary to prevent the wavefunction from expanding, and at the same time serves as a source of random noise.}
	\label{fig:cart-pole-system}
\end{figure}

Existing researches on deep-reinforcement-based quantum control are not many, and almost all of them only consider discrete systems composed of spins and qubits, and mostly focus on error correction or noise-resistant manipulation under some noise models \cite{GateControlDeepLearning, ErrorCorrectionDeepLearning, SpinControlDeepLearning}, which are clearly for practical purposes. Also, most of them only involve deterministic evolution of the states. In our research, we consider a system in continuous position space subject to measurement, which is yet to be investigated, and we use deep reinforcement learning to control the system and compare its control strategy with existing conventional controls to gain insight into what is learned by the AI and how it may outperform existing methods. \\

Specifically, we consider a particle in a 1D quadratic potential under continuous position measurement. The measurement introduces stochasticity into the system, and makes the system more realistic. When the potential of the system is upright, i.e. its minimum lies at the center, the system is just a usual harmonic oscillator; when the potential is inverted, the system is essentially an inverted pendulum and becomes analogous to the standard cart-pole system \cite{CartPole} which is a benchmark for reinforcement learning control (Fig. \ref{fig:cart-pole-system}). In the former case, the target of control is set to cool down the harmonic oscillator, which is ground-state cooling and is important as a real problem in experiments \cite{MeasurementFeedbackControlOptomechanic}. In the latter case, the target of control is to keep the particle at the center of the potential, which amounts to stabilizing the unstable system, and in both cases, the controller uses an external force exerted on the particle to control it. We train a neural network following the strategy of reinforcement learning, and compare its performance on the two tasks with the performance of the optimal control obtained from the linear-quadratic-Gaussian (LQG) control theory \cite{LinearQuadraticControl}. Next, we extend the problem to an anharmonic setting by changing the potential to be quartic, and repeat the above procedure. For this case, an optimal control strategy is not known, and therefore we use suboptimal control strategies and Gaussian approximations and the local approximation of the linear control to derive several control protocols from a conventional point of view, and we compare their performances with the reinforcement learning control. We also compare the behaviour of the controls by looking at their outputs, and we discuss the properties of the underlying quantum systems to gain insights on the controllers' behaviour. \\

\section{Outline}\label{outline}
The present thesis is organized as follows.\\

In Chapter \ref{continuous quantum measurement}, we present a review of continuous measurement on quantum systems. We give the formulation of a general measurement, and formally derive the stochastic differential equations that govern the evolution of a quantum state subjected to continuous measurement, where the evolution is called a quantum trajectory. We discuss both jump and diffusive trajectories, and give the equation that is used in our investigated control problem.\\

In Chapter \ref{deep reinforcement learning}, we present a review of deep reinforcement learning. We start from the basics of machine learning and introduce deep learning with its motivation and uses, and introduce reinforcement learning, especially a particular type of reinforcement learning called Q-learning, which is used in our research. Finally we discuss the implementation of deep learning for a reinforcement learning problem, i.e. deep reinforcement learning.\\

In Chapter \ref{control quadratic potentials}, we describe the quadratic control problems that are introduced in the last section. We first analyse the problems to show that they can be solved by the standard LQG control, and then describe our problem setting and our learning system in detail, and we present the results of the reinforcement learning and those of the optimal control. We compare the results, and also directly compare the output from the deep learning system with that from the optimal control. We find that, both final performances and the control behaviours of the two are similar, which implies that the AI correctly learned the optimal control. There also exist small traces of imperfections concerning the AI's behaviour. We will make discussions on these results. \\

In Chapter \ref{control quartic potentials}, we describe the quartic anharmonic control problems. We follow the same line of reasoning as the quadratic case and show that this quartic case cannot be simplified in the same way as the quadratic one, and the system exhibits intrinsic quantum mechanical behaviour that cannot be modelled classically. We then discuss possible control strategies based on existing ideas and compare their performances with our trained reinforcement learning controllers, and organise the results and discussions in the same way as Chapter \ref{control quadratic potentials}. We find that when properly configured, the reinforcement learning controller could outperform all of our derived control strategies, which demonstrates the supremacy and the universality of reinforcement learning.\\

In Chapter \ref{summary}, we discuss the conclusions of this thesis and their implications, and we discuss the future perspectives.\\

Some technical details are discussed in appendices. Appendix A reviews the linear-quadratic-Gaussian (LQG) control theory that is used in Chapters \ref{control quadratic potentials} and \ref{control quartic potentials}. Appendix B explains the numerical methods implemented in our numerical simulation of the quantum systems. Appendix C presents detailed adopted techniques and configurations of our reinforcement learning algorithm.

	\chapter{Continuous Measurement on Quantum Systems}\label{continuous quantum measurement}
	In this chapter, we review the formulation of quantum measurement and its continuous limit. We review the indirect measurement model in Section \ref{measurement model}, and in Section \ref{continuous measurement} we discuss continuous measurement of the diffusive type, which describes the position measurement that we apply to the quantum system in our research in Chapter 4 and 5.\\

\section{General Model of Quantum Measurement}\label{measurement model}
In the postulates of quantum mechanics \cite{NielsenChuang}, a general measurement is described by a set of linear operators $\{M_m\}$, with $m$ denoting measurement outcomes. These operators act on the measured quantum system state space and satisfy the completeness condition\\
\begin{equation}\label{completeness}
\sum_{m}M^{\dagger}_{m}M_{m}=I
\end{equation}\\
such that the unconditioned post-measurement quantum state as a sum over measurement outcomes is trace-preserved:\\
\begin{equation}\label{quantum operation}
\rho'=\mathcal{E}\left(\rho\right)=\sum_{m}M_m\rho M^{\dagger}_m\quad\Rightarrow\quad\text{tr}\left(\rho'\right)=\text{tr}\left(\rho\sum_{m}M^{\dagger}_{m}M_{m}\right)=\text{tr}(\rho),
\end{equation}\\
where $\rho$ is the measured quantum state and $M_m\rho M^{\dagger}_m$ represents the state after a measurement outcome $m$ is observed. This condition of trace preservation ensures that the total probability of all measurement outcomes is one. To obtain a normalized state after a certain measurement outcome, it is divided by its outcome probability and becomes\\
\begin{equation}\label{postmeasurement state}
\rho_m=\dfrac{M_m\rho M^{\dagger}_m}{\text{tr}\left(M_m\rho M^{\dagger}_m\right)}\,,
\end{equation}\\
where the trace $\text{tr}\left(M_m\rho M^{\dagger}_m\right)$ is the probability of measurement outcome $m$ for state $\rho$.\\

The simplest and standard measurement is projection measurement $\{P_m\}$, satisfying Eq. (\ref{completeness}) and\\
\begin{equation}\label{projector1}
\forall m,\quad\forall n \in \mathbb{Z}^{+},\quad (P_m)^{n} = P_m,\quad P^{\dagger}_m = P_m
\end{equation}\\
and\\
\begin{equation}\label{projector2}
P_i P_j = \delta_{ij}P_i,\qquad 
\delta_{ij}=\left\{
\begin{array}{rl}
0 & \text{if } i \ne j;\\
1 & \text{if } i = j,
\end{array} \right.
\end{equation}
such that they are projectors. These projector properties ensure that after a measurement outcome $m$ is obtained, if you measure it again immediately, the measurement outcome must again be $m$ and the state is not changed. This is the simplest and basic quantum measurement we have. Now we show it is possible to extend the projection measurement to a general measurement $\{M_m\}$ as in equations (\ref{completeness}) to (\ref{postmeasurement state}) by using an indirect measurement scheme.
\begin{figure}[t]
	\centering
\begin{tikzpicture}
\draw[thick] (0,0) rectangle (1.2,2.3);
\node (U) at (0.6,1.15) {{\large $U$}};
\node (e) at (-1.5,0.4) {{\large $\rho_e$}};
\node (rho) at (-1.5,1.9) {{\large $\rho$}};
\draw[thick] (e) -- (0,0.4);
\draw[thick] (rho) -- (0,1.9);
\draw[thick] (2.7,0) rectangle (3.7,0.8);
\draw[thin] (3.62,0.2) arc (40:140:0.55);
\draw[thin,->] (3.2,0.05) -- (3.4,0.6);
\draw[thick] (1.2,0.4) -- (2.7,0.4);
\node (m) at (4.2,0.4) {{\large $m$}};
\node (rhom) at (3.75,1.9) {{\large $M_m\rho M^{\dagger}_m$}};
\draw[thick] (1.2,1.9) -- (rhom);
\end{tikzpicture}
\caption{Indirect measurement model. We let a meter interact with a measured quantum state, and we measure the meter to obtain a direct measurement result $m$.}
\label{fig:indirect measurement}
\end{figure}
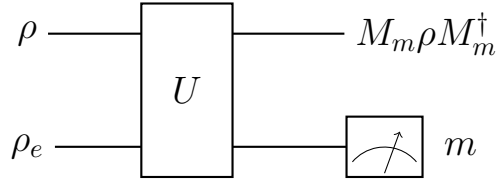\\

Suppose we want to measure a state $\rho$. We prepare a meter state $\rho_e$ which is known and not entangled with $\rho$, and let it interact with $\rho$ through a unitary evolution $U$, and then we measure the state of the meter using the projection measurement as schematically illustrated in figure \ref{fig:indirect measurement}. For a measurement outcome $m$ on the meter, the unnormalized post-measurement state of the initial $\rho$ becomes\\
\begin{equation}\label{indirect measurement1}
\tilde{\rho}_m=\text{tr}_e \left((I\otimes P_m)U(\rho \otimes \rho_e)U^{\dagger}(I\otimes P_m)\right)=\text{tr}_e \left((I\otimes P_m)U(\rho \otimes \rho_e)U^{\dagger}\right).
\end{equation}\\
For simplicity, we assume $\rho_e$ is pure, i.e. $\rho_e=|\psi_e\rangle\langle\psi_e|$, and we decompose $P_m$ into $P_m=\sum_{i}|\psi_i\rangle\langle\psi_i|$. The above result can be written as\\
\begin{equation}\label{indirect measurement2}
\tilde{\rho}_m=\sum\nolimits_{i}\langle\psi_i|U\left(|\psi_e\rangle\langle\psi_e| \otimes \rho\right) U^{\dagger}|\psi_i\rangle\,.
\end{equation}\\
If we define $M_{m,i}\equiv\langle\psi_i|U|\psi_e\rangle$, it becomes\\
\begin{equation}\label{indirect measurement3}
\tilde{\rho}_m=\sum_{i}M_{m,i}\,\rho M^{\dagger}_{m,i}\,,
\end{equation}\\
which can be considered as a measurement operator set $\{M_{(m,i)}\}$ with measurement outcomes $(m,i)$ as in equations (\ref{completeness}) to (\ref{postmeasurement state}), and we discard information on index $i$. If the projector $P_m$ only projects into one basis $|\psi_m\rangle\langle\psi_m|$, i.e. $i$ only has one choice, then it results in the measurement operator set $\{M_{m}\}$, and $\rho_m=\dfrac{M_m\rho M^{\dagger}_m}{\text{tr}\left(M_m\rho M^{\dagger}_m\right)}\,$. The completeness condition (\ref{completeness}) can be deduced from the completeness of projectors $\{P_m\}$. Conversely, for a given set of measurement operators $\{M_{m}\}$ that satisfy completeness condition (\ref{completeness}), there exists a unitary $U$ such that $\langle\psi_m|U|\psi_e\rangle=M_{m}$, and it allows us to implement the measurement $\{M_{m}\}$ through an indirect measurement with a direct projection measurement on the meter \cite{NielsenChuang}.\\

\section{Continuous Limit of Measurement}\label{continuous measurement}
\subsection{Unconditional State Evolution}
We consider the continuous limit of repeated measurements in infinitesimal time. When the measurement outcomes are not taken into account, the state evolution is deterministic, as $\rho\to\mathcal{E}(\rho)$ for each measurement done. We require this deterministic evolution to be continuous, that is\\
\begin{equation}\label{continuous}
0<\left\|\lim_{dt\to 0}\dfrac{\mathcal{E}(\rho)-\rho}{dt}\right\|<\infty,\quad \mathcal{E}(\rho)=\sum_{m}M_m\rho M^{\dagger}_m,
\end{equation}\\
where $\mathcal{E}$ necessarily depends on $dt$.\\

We first consider a binary measurement $\{M_0,M_1\}$ with two measurement outcomes. Due to the requirements $\sum_{m}M^{\dagger}_{m}M_{m}=I$ and $\left(\mathcal{E}(\rho)-\rho\right)\sim dt$, we set $M_0=I-\frac{R\,dt}{2}$ such that\\
\begin{equation}\label{binary default}
M_0 \rho M^{\dagger}_0=(I-R\,dt)\rho
\end{equation}\\
with $M_1$ satisfying\\
\begin{equation}\label{binary non-default}
M_1 = L_1\,\sqrt{dt},\quad L^{\dagger}_1L_1=R
\end{equation}\\
for the condition $\sum_{m}M^{\dagger}_{m}M_{m}=I$. In this way, all requirements for a continuous measurement are satisfied \cite{QuantumMeasurement}. Then similarly, we may add more operators $M_i$ into the operator set $\{M_m\}$, and they satisfy\\
\begin{equation}\label{multiple measurement outcomes}
M_i=L_i\,\sqrt{dt},\quad i=1,2,\cdots,m\ ,\quad M_0=I-\frac{dt}{2}\sum^{m}_{i=1}L^{\dagger}_iL_i,
\end{equation}\\
which produce the Lindblad equation\\
\begin{equation}\label{Lindblad}
\frac{d\rho}{dt}=-\frac{i}{\hbar}[H,\rho]+\sum_{i}\gamma_i\left(L_i\rho L^{\dagger}_i-\frac{1}{2}\{L^{\dagger}_iL_i,\rho\}\right),
\end{equation}\\
where a self-evolution term $[H,\rho]$ with Hamiltonian $H$ is taken into account, $\{\cdot,\cdot\}$ is the anticommutator, and $\gamma_i$ characterizes the strength of measurement. The above results can also be derived from the indirect measurement model by a repetition of a week unitary interaction between the state and the meter followed by a projection measurement on the meter \cite{OpenQuantumSystemsAngelRivas}.\\

\subsection{Quantum Trajectory Conditioned on Measurement Outcomes}\label{Quantum Trajectory section}
When the measurement outcomes of a continuous measurement are observed, the quantum state conditioned on the outcomes follows a quantum trajectory. This quantum trajectory can be considered as a stochastic process, and it is not necessarily continuous.\\

The probability to get a measurement outcome $i$ in an infinitesimal measurement in time $dt$ is\\
\begin{equation}\label{measurement outcome 1}
p_i(dt)=\text{tr}(L_i\rho L^{\dagger}_i)\,dt\,,
\end{equation}\\
which vanishes with $dt$. Therefore, in an infinitesimal length of time, measurement outcomes other than the outcome 0 can only appear with vanishingly small probabilities. We therefore take the limit that all measurement outcomes are sparse in time except for the outcome 0, that is, two or more of them do not occur in the same infinitesimal time interval in $dt$. If we denote the number of measurement outcome $i$ in $dt$ as $dN_i(t)$, they obey the following:\\
\begin{equation}\label{measurement events}
dN_i(t)= 0\ \text{or}\ 1,\quad dN_idN_j=\delta_{ij}\,dN_i,\quad E\left[dN_i(t)\right]=\text{tr}\left(L_i\rho(t)L^{\dagger}_i\right)dt\,.
\end{equation}\\
We now write the stochastic differential equation with these random variables to describe the state evolution, conditioned on these measurement outcomes \cite{QuantumMeasurement}. For a pure state $|\psi\rangle$, we have\\
\begin{equation}\label{jump evolution pure}
\begin{split}
d|\psi(t)\rangle&=\left[\sum_i dN_i(t)\left(\frac{L_i}{\sqrt{\langle L^{\dagger}_i L_i\rangle}}-1\right)+[1-\sum_i dN_i(t)]\left(\frac{M_0}{\sqrt{\langle M_0^{\dagger}M_0\rangle}}-1\right)\right]|\psi(t)\rangle\\
&=\left[\sum_i dN_i(t)\left(\frac{L_i}{\sqrt{\langle L^{\dagger}_i L_i\rangle}}-1\right)+1\cdot\left(\frac{1-\frac{dt}{2}\sum^{m}_{i=1}L^{\dagger}_iL_i}{\sqrt{\langle 1-dt\sum^{m}_{i=1}L^{\dagger}_iL_i\rangle}}-1\right)\right]|\psi(t)\rangle\\
&=\left[\sum_i dN_i(t)\left(\frac{L_i}{\sqrt{\langle L^{\dagger}_i L_i\rangle}}-1\right)-dt\left(\frac{1}{2}\sum^{m}_{i=1}L^{\dagger}_iL_i-\frac{1}{2} \sum^{m}_{i=1}\langle L^{\dagger}_iL_i\rangle\right)\right]|\psi(t)\rangle,
\end{split}
\end{equation}\\
where the term $[1-\sum_i dN_i(t)]$ is replaced by $1$ due to few non-zero events of $dN_i$. When the state Hamiltonian is taken into account, Eq. (\ref{jump evolution pure}) reduces to\\
\begin{equation}\label{jump evolution with Hamiltonian}
d|\psi(t)\rangle=\left[\sum_i dN_i(t)\left(\frac{L_i}{\sqrt{\langle L^{\dagger}_i L_i\rangle}}-1\right)-dt\left(\frac{i}{\hbar}H+\frac{1}{2}\sum^{m}_{i=1}L^{\dagger}_iL_i-\frac{1}{2} \sum^{m}_{i=1}\langle L^{\dagger}_iL_i\rangle\right)\right]|\psi(t)\rangle.
\end{equation}\\
This is called a nonlinear stochastic Schr\"odinger equation (SSE). For a general mixed state, the equation is\\
\begin{align}\label{jump evolution with Hamiltonian mixed}
d\rho=-dt\frac{i}{\hbar}(H_{\text{eff}}\rho-\rho H^{\dagger}_{\text{eff}})&+dt\sum^{m}_{i=1}\langle L^{\dagger}_iL_i\rangle\rho+\sum^{m}_{i=1}dN_i\left(\frac{L_i\rho L^{\dagger}_i}{\langle L^{\dagger}_iL_i\rangle}-\rho\right),\\ H_{\text{eff}}&:=H-\frac{i\hbar}{2}\sum_{i=1}^{m}L^{\dagger}_i L_i.
\end{align}\\
Note that the expectation value $\langle\cdot\rangle$ depends on the current state $\rho(t)$ or $|\psi(t)\rangle$, and it introduces nonlinearity regarding the state by the $\langle\cdot\rangle\rho$ and $\langle\cdot\rangle|\psi(t)\rangle$ terms.\\

Physically, when the operators $L_i$ are far from the identity and change the quantum state much, the rare non-zero $dN_i$ events are called quantum jumps, meaning that the state changes suddenly during its evolution. At another limit in which the operators $L_i$ are close to the identity and non-zero $dN_i$ occurs more frequently, the state can evolve smoothly. This is called the diffusive limit, and there are multiple ways to achieve this limit. If we have discrete measurement outcomes $0,1,2$ such that\\
\begin{equation}\label{diffusive limit discrete}
L_1=\sqrt{\frac{\Gamma}{2}}(I+l\,\hat{a}),\quad L_2=\sqrt{\frac{\Gamma}{2}}(I-l\,\hat{a}),\quad M_0=I-\frac{\Gamma l^2}{2}\hat{a}^{\dagger}\hat{a}\,dt\,,
\end{equation}
\begin{equation}\label{diffusive limit discrete condition}
l\to 0,\quad\Gamma\to\infty,\quad\Gamma l^2=\frac{\gamma}{2},
\end{equation}\\
where $\gamma$ is a constant. In this case, the frequency of measurement outcome $i=1,2$ becomes high:\\
\begin{equation}\label{frequent detection}
\begin{split}
E\left[\delta N_{1,2}(t)\right]=\text{tr}\left(L_{1,2}\,\rho L^{\dagger}_{1,2}\right)dt\cdot\frac{\delta t}{dt}&=\frac{\Gamma}{2}\left(1\pm l\langle\hat{a}+\hat{a}^{\dagger}\rangle+l^2\langle\hat{a}^{\dagger}\hat{a}\rangle\right)\delta t\\
&\approx\frac{\Gamma}{2}\left(1\pm l\langle\hat{a}+\hat{a}^{\dagger}\rangle\right)\delta t,
\end{split}
\end{equation}\\
where $\delta N_i(t)$ denotes the number of measurement outcome $i$ in a small time interval $\delta t$, and we assume that the state does not change much during this interval. Under this assumption, $\delta N_i(t)$ is a Poisson distribution for the interval $\delta t$, and therefore at $\Gamma\to\infty$, it is non-negligible and can be approximated to be Gaussian:\\
\begin{equation}\label{Wiener process}
\delta N_{1,2}(t)=\frac{\Gamma}{2}\left(1\pm l\langle\hat{a}+\hat{a}^{\dagger}\rangle\right)\delta t+\sqrt{\frac{\Gamma}{2}}\left(1\pm\frac{{l}}{2}\langle\hat{a}+\hat{a}^{\dagger}\rangle\right)\delta W_{1,2}(t),
\end{equation}
\begin{equation}\label{Wiener Gaussian}
\delta W_{1,2}(t)\sim\mathcal{N}\left(0,\sqrt{\delta t}^2\right),
\end{equation}\\
where $\Gamma\delta t$ is large, and $\delta W_i(t)$ is a Wiener increment satisfying $\delta W_i\delta W_j=\delta_{ij}\delta t$. To proceed, we first calculate $d\rho$:\\
\begin{equation}\label{d rho}
\begin{split}
d\rho=&-dt\frac{i}{\hbar}(H_{\text{eff}}\rho-\rho H^{\dagger}_{\text{eff}})+dt\sum^{2}_{i=1}\langle L^{\dagger}_iL_i\rangle\rho+\sum^{2}_{i=1}dN_i\left(\frac{L_i\rho L^{\dagger}_i}{\langle L^{\dagger}_iL_i\rangle}-\rho\right)\\
=&-dt\frac{i}{\hbar}[H,\rho]-dt{\frac{\Gamma}{2} }\left\{I+l^2\hat{a}^\dagger\hat{a},\rho\right\}+\Gamma\left(1+l^2\langle\hat{a}^{\dagger}\hat{a}\rangle\right)dt\,\rho\\
&+dN_1\left(\frac{\rho+l\hat{a}\rho+l\rho\hat{a}^{\dagger}+l^2\hat{a}\rho\hat{a}^{\dagger}}{1+l\langle\hat{a}+\hat{a}^{\dagger}\rangle+l^2\langle\hat{a}^{\dagger}\hat{a}\rangle}-\rho\right)+dN_2\left(\frac{\rho-l\hat{a}\rho-l\rho\hat{a}^{\dagger}+l^2\hat{a}\rho\hat{a}^{\dagger}}{1-l\langle\hat{a}+\hat{a}^{\dagger}\rangle+l^2\langle\hat{a}^{\dagger}\hat{a}\rangle}-\rho\right)\\
=&-dt\frac{i}{\hbar}[H,\rho]-dt{\frac{\Gamma l^2}{2} }\left\{\hat{a}^\dagger\hat{a},\rho\right\}+\Gamma l^2\langle\hat{a}^{\dagger}\hat{a}\rangle dt\,\rho\\
&+dN_1\left(l\hat{a}\rho+l\rho\hat{a}^{\dagger}+l^2\hat{a}\rho\hat{a}^{\dagger}-l\langle\hat{a}+\hat{a}^{\dagger}\rangle{\rho}+{l^2}\langle\hat{a}+\hat{a}^{\dagger}\rangle^2{\rho}-l^2\langle\hat{a}^{\dagger}\hat{a}\rangle\rho-l^2\langle\hat{a}+\hat{a}^{\dagger}\rangle(\hat{a}\rho+\rho\hat{a}^{\dagger})\right)\\
&+dN_2\left(-l\hat{a}\rho-l\rho\hat{a}^{\dagger}+l^2\hat{a}\rho\hat{a}^{\dagger}+l\langle\hat{a}+\hat{a}^{\dagger}\rangle{\rho}+{l^2}\langle\hat{a}+\hat{a}^{\dagger}\rangle^2{\rho}-l^2\langle\hat{a}^{\dagger}\hat{a}\rangle\rho-l^2\langle\hat{a}+\hat{a}^{\dagger}\rangle(\hat{a}\rho+\rho\hat{a}^{\dagger})\right)\\
=&-dt\frac{i}{\hbar}[H,\rho]-dt{\frac{\Gamma l^2}{2} }\left\{\hat{a}^\dagger\hat{a},\rho\right\}+\Gamma l^2\langle\hat{a}^{\dagger}\hat{a}\rangle dt\,\rho+(dN_1-dN_2)\left(l\hat{a}\rho+l\rho\hat{a}^{\dagger}-l\langle\hat{a}+\hat{a}^{\dagger}\rangle{\rho}\right)\\
&+(dN_1+dN_2)(l^2\hat{a}\rho\hat{a}^{\dagger}-l^2\langle\hat{a}^{\dagger}\hat{a}\rangle\rho-l^2\langle\hat{a}+\hat{a}^{\dagger}\rangle(\hat{a}\rho+\rho\hat{a}^{\dagger})+{l^2}\langle\hat{a}+\hat{a}^{\dagger}\rangle^2{\rho}),
\end{split}
\end{equation}
where we have expanded the denominators up to $O(l^2)$. To accumulate a total of $\dfrac{\delta t}{dt}\to\infty$ steps, we substitute Eq. (\ref{Wiener process}) into the above. It can be checked easily that most of the terms are cancelled and to the leading order in $l$ it becomes\\
\begin{equation}\label{delta rho}
\begin{split}
\delta\rho=&-\delta t\frac{i}{\hbar}[H,\rho]-\delta t\frac{\Gamma }{2}\sum^{m}_{i=1}\left[-2l^2\hat{a}\rho\hat{a}^{\dagger}+l^2\{\hat{a}^\dagger\hat{a},\rho\}\right]+\sqrt{\Gamma}l\sum_{i=1}^{m}\left[(\hat{a}-\langle\hat{a}\rangle)\rho+\rho(\hat{a}^{\dagger}-\langle\hat{a}^{\dagger}\rangle)\right]\delta W\\
=&\left[-\frac{i}{\hbar}[H,\rho]-\frac{\gamma }{4}\sum^{m}_{i=1}\left(\{\hat{a}^\dagger\hat{a},\rho\}-2\hat{a}\rho\hat{a}^{\dagger}\right)\right]\delta t+\sqrt{\frac{\gamma }{2}}\sum_{i=1}^{m}\left[(\hat{a}-\langle\hat{a}\rangle)\rho+\rho(\hat{a}^{\dagger}-\langle\hat{a}^{\dagger}\rangle)\right]\delta W,
\end{split}
\end{equation}
where the rule of calculation follows the It\^o calculus, and we have used a single Wiener increment $\delta W$ to represent the term $(\delta N_1 - \delta N_2)$. The above equation shows that our initial assumption that $\rho$ does not change much during a sufficiently small time interval $\delta t$ is true, as diverging quantities are cancelled and only non-diverging $\gamma$ terms before $\delta t$ and $\delta W$ remain, which scales with the length of a chosen time interval $\delta t$.\footnote{As can be seen from our derivation, the function before $\delta W$ should be evaluated at time $t$ but not at time $t+\dfrac{\delta t}{2}$. This point is crucial in stochastic calculus, and in this case it is called It\^o calculus. Another caveat is that the stochastic differential equations converge in $\sqrt{dt}$ rather than $dt$, which is different from usual differential equations and is important when we try to prove the above result from a rigorous mathematical point of view.} Rewriting $\delta t$ as $dt$, we obtain the final result\\
\begin{equation}\label{stochastic total equation}
d\rho=\left[-\frac{i}{\hbar}[H,\rho]-\frac{\gamma }{4}\sum^{m}_{i=1}\left(\{\hat{a}^\dagger\hat{a},\rho\}-2\hat{a}\rho\hat{a}^{\dagger}\right)\right]dt+\sqrt{\frac{\gamma }{2}}\sum_{i=1}^{m}\left[(\hat{a}-\langle\hat{a}\rangle)\rho+\rho(\hat{a}^{\dagger}-\langle\hat{a}^{\dagger}\rangle)\right]dW,
\end{equation}\\
where $dW\sim\mathcal{N}\left(0,\sqrt{dt}^2\right)$. For a pure state it is\\
\begin{equation}\label{stochastic total equation pure}
\begin{split}
d|\psi\rangle=&\left[-\frac{i}{\hbar}H-\frac{\gamma}{4}\sum_{i=1}^{m}\left(\hat{a}^{\dagger}\hat{a}-\hat{a}\langle\hat{a}+\hat{a}^{\dagger}\rangle+\frac{1}{4}\langle\hat{a}+\hat{a}^{\dagger}\rangle^2\right)\right]dt|\psi\rangle\\
&+\sqrt{\frac{\gamma}{2}}\sum_{i=1}^{m}\left(\hat{a}-\frac{1}{2}\langle\hat{a}+\hat{a}^{\dagger}\rangle\right)dW|\psi\rangle.
\end{split}
\end{equation}
\begin{figure}[t]
	\centering
	\includegraphics[width=0.5\linewidth]{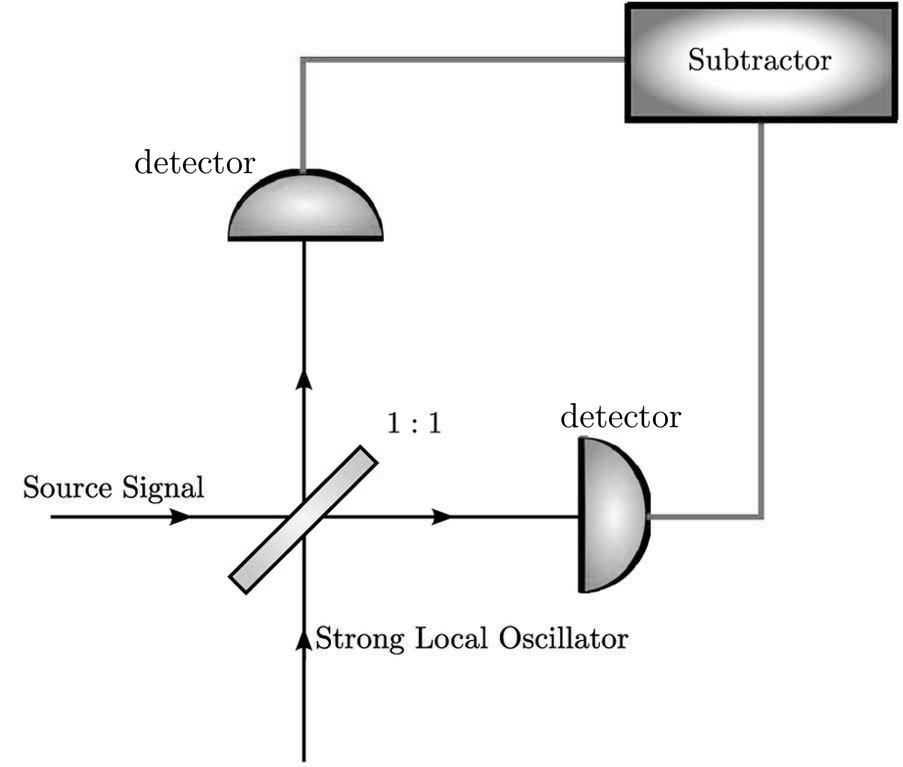}
	\caption{Setup of homodyne detection. An intense laser field (called a local oscillator) is superposed with a measured signal through a beam splitter. The two output fields are detected by two detectors which convert incident photons into electric currents. The electric currents are then fed into a balanced detector. Reproduced from Ref. \cite{ehek2015}.}
	\label{fig:homodyne}
\end{figure}\\

In real experiments, the above model describes a homodyne detection using a strong local oscillator, and $dN_1-dN_2$ is the observed signal as a differential photocurrent, which is illustrated in Fig. \ref{fig:homodyne}, which has a mean value of $\Gamma l \langle\hat{a}+\hat{a}^{\dagger}\rangle$ and a standard deviation of $\sqrt{\Gamma dt}$ regarding the measured signal. The random variable $dW$ represents the observed signal deviating from its mean value. This experimental setting explains why we need two measurement outcomes $dN_1$ and $dN_2$. If we only take one measurement outcome and block either branch of the light that goes out from the beam splitter, it comes back to the source signal and disturbs the Hamiltonian of the measured system drastically. \\

In the cases where measurement outcomes are not discrete but are real-valued, such as the direct position measurement, the measured system can also have a diffusive quantum trajectory, to which the above analysis does not apply. In such cases, we may assume the measurements are weak and performed repeatedly, so that they are averaged to obtain a total measurement, and then according to the central limit theorem, we can approximately describe the measurement effects and results as a Gaussian random process:\\
\begin{equation}\label{Gaussian measurement}
M_q=\left(\frac{\gamma dt}{\pi}\right)^\frac{1}{4}e^{-\frac{\gamma dt}{2}(\hat{x}-q)^2},
\end{equation}\\
where we have assumed that the measured physical quantity is the position of a particle. Note that the completeness condition is automatically satisfied by the property of the Gaussian integral:
\begin{equation}\label{Gaussian measurement completeness}
\int_{-\infty}^{\infty}M^{\dagger}_qM_q\, dq=1\,.
\end{equation}\\
Therefore it is a valid measurement. In this case, measurement outcomes are by themselves Gaussian and can be modelled as $\left(\langle\hat{x}\rangle+\dfrac{dW}{\sqrt{2\gamma}dt}\right)$, which is given in Ref. \cite{DIOSI1988419}. The deduction follows by explicitly calculating $M_q\rho M^{\dagger}_q$ by a straightforward expansion, and the final result is exactly the same as before, provided that $\hat{a}$ and $\hat{a}^{\dagger}$ are replaced by $\hat{x}$. The result for a pure state is\\
\begin{equation}\label{position measurement evolution}
d|\psi\rangle=\left[\left(-\frac{i}{\hbar}H-\frac{\gamma}{4}(\hat{x}-\langle\hat{x}\rangle)^2\right)dt+\sqrt{\dfrac{\gamma}{2}}(\hat{x}-\langle\hat{x}\rangle)dW\right]|\psi\rangle,
\end{equation}\\
where $dW$ is a Wiener increment as before. This is the equation that we use in the simulation of quantum systems in our research. If we express it in terms of the density matrix that admits a mixed state, the equation becomes\\
\begin{equation}\label{position measurement mixed}
d\rho=-\frac{i}{\hbar}[H,\rho]dt-\frac{\gamma}{4}[\hat{x},[\hat{x},\rho]]dt+\sqrt{\frac{\gamma}{2}}\{\hat{x}-\langle \hat{x}\rangle,\rho\}dW.
\end{equation}\\
In order to model incomplete information on measurement outcomes, we define a measurement efficiency parameter $\eta$, satisfying $0\le\eta\le1$, which represents the ratio of measurement outcomes that are obtained. In the above equations, measurement outcomes are represented by a Wiener increment $dW$, which can be considered as an accumulated value in a small time interval. Therefore, we rewrite the Wiener increment into a series of smaller Wiener increments which represent repeated weak measurement results, and zero out a portion of those results to obtain a total incomplete measurement outcome, i.e.,\\
\begin{equation}
dW=\sum_{i=1}^N dW_i,\qquad dW_i\sim\mathcal{N}\left (0,\frac{dt}{N}\right ),
\end{equation}\\
where we have discretized the time in units of $dt$ into $N$ steps to obtain $N$ Weiner increments. The original condition $dW\sim\mathcal{N}(0,{dt})$ is clearly satisfied. Then after removing a portion $1-\eta$ of the measurement results, the total $dW$ becomes\\
\begin{equation}
dW=\sum_{i=1}^{\eta N} dW_i,\qquad dW_i\sim\mathcal{N}\left (0,\frac{dt}{N}\right ),
\end{equation}\\
and therefore we have $dW\sim\mathcal{N}(0,{\eta\, dt})$, and the time-evolution equation is\\
\begin{equation}\label{position measurement incomplete information}
d\rho=-\frac{i}{\hbar}[H,\rho]dt-\frac{\gamma}{4}[\hat{x},[\hat{x},\rho]]dt+\sqrt{\frac{\gamma\eta}{2}}\{\hat{x}-\langle \hat{x}\rangle,\rho\}dW,\qquad dW\sim\mathcal{N}(0,{dt}),
\end{equation}\\
where we have rescaled $dW$ such that it is now a standard Wiener increment.
	\chapter{Deep Reinforcement Learning}\label{deep reinforcement learning}
	In this chapter, we briefly review deep reinforcement learning. It is constituted of two different subjects: (1) deep learning and (2) reinforcement learning. First, we begin with the general idea of machine learning in Section \ref{deep learning} and then introduce deep learning along with its motivations and uses. Then we review reinforcement learning, especially Q-learning, in Section \ref{reinforcement learning}, and discuss how it is implemented using deep learning technology. \\

In this thesis, we use the word ``machine learning" to refer to the general picture of artificial intelligence (AI) technology, and use the word ``deep learning" to refer to machine learning systems that specially use deep learning techniques. This chapter mainly discusses the AI technology relevant to the present research, and does not cover irrelevant deep learning or machine learning topics.\\
\section{Deep Learning}\label{deep learning}

\subsection{Machine Learning}
Generally speaking, learning usually refers to a process in which unpredictable becomes predictable, by building a model to correctly relate different pieces of relevant information. Machine learning aims to automatize this process. Although humans often achieve learning via a sequence of logical reasoning and validation, up to now, machines do not have a good common knowledge base to achieve creative logical reasoning to learn. To compensate for this deficiency, machine learning systems usually have a set of possible models beforehand, which represents conceivable relations among the different pieces of information that it is going to learn. The set of conceivable relations here is formally called the \textit{hypothesis space}. Then, it learns some provided example data, by looking for a model in its hypothesis space which fits the observed data best, and finally use the found model as the relation among the pieces of information it learns to give prediction on new data. As a result, the learned model is almost always only an approximate solution to the underlying problem. Nevertheless, it still works well enough in cases where a given problem cannot be modelled precisely but can be approximated easily.\\

To formally give a definition of machine learning, according to Tom M.~Mitchell \cite{MachineLearningBook}, ``A computer program is said to learn from experience $E$ with respect to some class of tasks $T$ and performance measure $P$, if its performance at tasks in $T$, as measured by $P$, improves with experience $E$." \\

\subsection{Feedforward Neural Networks}
As discussed, the setting of the hypothesis space of a machine learning system is crucial for the performance. Because different problems have different properties, before the emergence of deep learning, researchers considered various approximate models to describe the corresponding real-life problems, including text-voice transform, language translation, image recognition, etc., and therefore the researchers specialized in different machine learning tasks usually worked separately. However, the deep neural network as a general hypothesis space set outperformed all previous research results in 2012 \cite{ImageNetClassification}, and started a deep learning boom. Below we introduce the deep neural network model, or precisely, the deep feedforward neural network following the line of thoughts of the last section. \\

When we attempt to model a relation between two quantities, the simplest guess is the linear relation. Although real-world problems are typically high-dimensional and more complex, we may hold on to this linearity even in a multidimensional setting, and assume\\
\begin{equation}\label{Linear}
	\boldsymbol{y}=\textbf{M}\boldsymbol{x}+\boldsymbol{b}\, ,
\end{equation}\\
where we model the relation between $\boldsymbol{y}$ and $\boldsymbol{x}\,$. Here $\boldsymbol{y}$ and $\boldsymbol{x}$ are vectors, $\textbf{M}$ is a matrix. $\boldsymbol{b}$~is an additional bias term as a small compromise starting from the linear guess. $\textbf{M}$ and $\boldsymbol{b}$ are learned by fitting $\boldsymbol{y}$ and $\boldsymbol{x}$ pairs into existing training data pairs $\{(\boldsymbol{x},\boldsymbol{y})_i\}$. This process of learning is called \textit{linear regression} \cite{DeepLearningBook}.\\

Obviously, the simple linear (or affine) model above would not work for realistic complex problems, as it cannot model nonlinear relations. Therefore, we apply a simple nonlinear function $\sigma$ after the linear map, which is called an \textit{activation function}. This name is an analogy to the activation function controlling firings of neurons in neuroscience. For simplicity, this function is a scalar function and is applied to a vector in an element-wise manner, acting on every component of the vector separately. Then, the function may be constructed as\\
\begin{equation}\label{multilayer perceptron}
	\boldsymbol{y}=f(\boldsymbol{x})=\textbf{M}_2\cdot\sigma(\textbf{M}_1\boldsymbol{x}+\boldsymbol{b}_1)+\boldsymbol{b}_2\, .
\end{equation}\\

\begin{figure}[htb]
	\centering
	\subfloat[ReLU\quad$\sigma(x)=\max(0,x)$]{\label{ReLU}
		\begin{minipage}[c]{0.35\linewidth}
			\centering
			$\sigma(x)$\\
			\includegraphics[width=\linewidth]{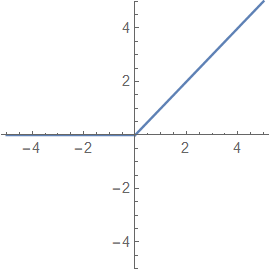}
	\end{minipage}}$x$\qquad\qquad
	\subfloat[Sigmoid\quad$\sigma(x)=\frac{1}{1+e^{-x}}$]{\label{sigmoid}
		\begin{minipage}[c]{0.35\linewidth}
			\centering
			$\sigma(x)$\\
			\includegraphics[width=\linewidth]{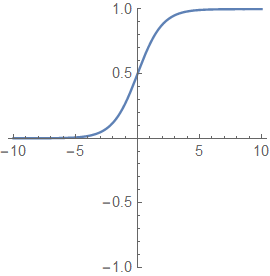}
	\end{minipage}}$x$
	\caption{two frequently used activation functions in deep learning}
	\label{activation functions}
\end{figure}

In practice, there is almost no constraint on the activation function, as long as it is nonlinear. The most commonly used two functions are the ReLU (Rectified Linear Unit) and the sigmoid, which are shown in figure \ref{activation functions}.\\

The ReLU function simply zeros all negative values and keeps all positive values, and the sigmoid is a transition between zero and one, which is also known as the standard logistic function. Because most of the time we use the ReLU as the activation function $\sigma$, we assume using the ReLU in the following context unless otherwise mentioned. \\

An immediate result which can be drawn is that the function $f$ in Eq. (\ref{multilayer perceptron}) is universal, in the sense that it can approximate an arbitrary continuous mapping from $\boldsymbol{x}$ to $\boldsymbol{y}\,$, provided that the parameters $\textbf{M}_1,\boldsymbol{b}_1,\textbf{M}_2,\boldsymbol{b}_2$ have sufficiently many dimensions and are complicated enough.\\

The above argument can be shown easily for the ReLU as $\sigma\,$, and then similarly for other activation functions. For simplicity we first consider one-dimensional $x,y\,$. First we note that the ReLU just effectively bends a line, and $\textbf{M}$ and $\boldsymbol{b}$ can be used to replicate, rotate and shift the straight line ${x}={x}\,$; then it can be realized that the above function in Eq.~(\ref{multilayer perceptron}) is constituted of three successive processes: (1) copying the line ${x}={x}\,$, plus customizable rotation and shift by $\textbf{M}_1,\boldsymbol{b}_1\,$, (2) bending each resultant line by $\sigma
$, (3) rotating, shifting and summing all lines by $\textbf{M}_2,\boldsymbol{b}_2\,$. Therefore, with correctly picked $\textbf{M}_1,\boldsymbol{b}_1,\textbf{M}_2,\boldsymbol{b}_2\,$, this function can be used to construct arbitrary lines that are piece-wise linear with finitely many bend points. Thus, it can approximate arbitrary functions from $x$ to $y$ with arbitrary precision, provided that parameters $\textbf{M}_1,\boldsymbol{b}_1,\textbf{M}_2,\boldsymbol{b}_2$ are appropriately chosen. For the case of higher dimensional $\boldsymbol{x}$ and $\boldsymbol{y}$, instead of bended lines, the function $f$ constructs polygons in the $\boldsymbol{x}$ space and the universality follows similarly. This argument of universality also holds for other types of nonlinear functions besides the ReLU, and can be shown by similar constructive arguments.\\

Now, we consider using the function in Eq. (\ref{multilayer perceptron}) as our hypothesis space for machine learning. Given data points $\{(\boldsymbol{x},\boldsymbol{y})_i\}$, if we follow the universality argument and fit the parameters in Eq. (\ref{multilayer perceptron}) to make $f$ reproduce the relation from $\boldsymbol{x}$ to $\boldsymbol{y}$ over the whole dataset $\{(\boldsymbol{x},\boldsymbol{y})_i\}$, then, as can also be seen from the universality argument, $f$ becomes the nearest-neighbour interpolation of data points $\{\boldsymbol{x}_i\}$. For any unseen data point $\boldsymbol{x}$ absent from the training set $\{(\boldsymbol{x},\boldsymbol{y})_i\}$, to predict its corresponding $\boldsymbol{y}\,$, $f$ finds its nearest neighbours in the data set $\{\boldsymbol{x}_i\}$ and predict its $\boldsymbol{y}$ as a linear interpolation of those neighbours' $\boldsymbol{y}$ values. This is a direct consequence of the polygon argument in the above paragraph, and implies that using this parametric function as the hypothesis space is still simple and naive, and that it cannot easily learn complex relations between $\boldsymbol{x}$ and $\boldsymbol{y}$ unless we have numerous data points in the training set to represent all possibilities of $\boldsymbol{x}$. Thus, we need further improvement so that the function can learn complex relations more easily.\\

As mentioned earlier, the nonlinear function $\sigma$ between linear mappings has its biological analogue as the activation function. This is because in neuroscience, the activation of a single neuron is influenced by its input linearly if its input is above an activation threshold, and if the input is below the threshold, there is no activation. This phenomenon is exactly modelled by the ReLU function following a linear mapping, where the linear mapping is connected to input neurons. Although each individual neuron functions simply, when they are connected, they may show extremely complex behaviour collectively. Motivated by this observation, we choose to apply the functions $f_i(\boldsymbol{x}):=\sigma(\textbf{M}_i\boldsymbol{x}+\boldsymbol{b}_i)$ sequentially and put them into the form $f_i\circ f_j \circ f_k\cdots\circ f_l$ to build a deep neural network. In this case, the output vector value of every $f_i$ represents a layer of neurons, with each scalar in it representing a single neuron, and every layer is connected to the previous layer through the weight matrix $\textbf{M}\,$. Note that the dimension of every $f_i$ may not be equal. This artificial network of neurons is called a feedforward neural network, since its information only goes in one direction and does not loops back to previous neurons. It can be written as follows:\\
\begin{equation}\label{feedforward neural network}
	\boldsymbol{y}=f(\boldsymbol{x})=\textbf{M}_n\left(f_{n-1}\circ f_{n-2} \circ f_{n-3}\cdots\circ f_1\left(\boldsymbol{x}\right)\right)+\boldsymbol{b}_n\,,
\end{equation}\\
where\\
\begin{equation}\label{feedforward neural network2}
	f_i(\boldsymbol{x})=\sigma\left(\textbf{M}_i\boldsymbol{x}+\boldsymbol{b}_i\right),\qquad\sigma=\max(0,x)\,.
\end{equation}\\
Intuitively speaking, although one layer of $f_i$ only bends and folds the line a few times, successive $f_i\,$s can fold on existing foldings and finally make the represented function more complex but with a certain regular shape. This equation (Eq. \ref{feedforward neural network}) is the deep feedforward neural network that we use in deep learning as our hypothesis space. For completeness, we recapitulate and define some relevant terms below.\\

The neural network is a function $f$ that gives an output $\boldsymbol{y}$ when provided with an input $\boldsymbol{x}$ as in Eq. (\ref{feedforward neural network}). The \textit{depth} of a deep neural network refers to the $n$ number in Eq. (\ref{feedforward neural network}), which is the number of linear mappings involved. $f_i$ stands for the $i$-th layer, and the activation values of the $i$-th layer is the output of $f_i\,$, and the \textit{width} of layer $f_i$ refers to the dimension of its output vector, i.e. the number of scalars (or neurons) involved. These involved scalars are called \textit{units} or \textit{hidden units} of the layer. When there is no constraint on matrix $\textbf{M}\,$, all units of adjacent layers are connected by generally non-zero entries in matrix $\textbf{M}$ and this case is called \textit{fully connected}. Usually, we call the $n$-th layer as the \textit{top} layer and the first layer as the \textit{bottom}. Concerning output $\boldsymbol{y}\,$, in Eq. (\ref{feedforward neural network}) it is a general real-valued vector, but if we know there exists some preconditions on the properties of $\boldsymbol{y}$, we may add an \textit{output function} at the top of the network $f$ to constrain $\boldsymbol{y}$ so that the preconditions are satisfied. This is especially useful for image classification tasks, where the neural network is supposed to output a probability distribution over discrete choices. For the classification case, the partition function is used as the output function to change the unnormalized output into a distribution. \\

\subsection{Training a Neural Network}

In the last section we defined our feedforward neural network $f$ with parameters $\{\textbf{M}_i,\boldsymbol{b}_i\}_{i=1}^{n}\,$. Different from the cases of Eq. (\ref{Linear}) and (\ref{multilayer perceptron}), it is not directly clear how to find appropriate parameters $\{\textbf{M}_i,\boldsymbol{b}_i\}_{i=1}^{n}$ to fit $f$ to a given dataset $\{(\boldsymbol{x},\boldsymbol{y})_i\}$. As the first step, we need to have a measure to evaluate how well a given $f$ fits into the dataset. For this purpose, a \textit{loss function} is used, which measures the difference between $f(\boldsymbol{x}_i)$ and $\boldsymbol{y}_i$ on a dataset $S\equiv\{(\boldsymbol{x},\boldsymbol{y})_i\}$, and a larger loss implies a lower performance. In the simplest case, the L2 loss is used:\\

\begin{equation}\label{L2 loss}
	L=\frac{1}{|S|}\sum_{(\boldsymbol{x}_i,\boldsymbol{y}_i)\in S}||{f}(\boldsymbol{x}_i)-\boldsymbol{y}_i||^2\,,
\end{equation}\\
where $||\cdot||$ is the usual L2 norm on vectors. This loss is also termed \textit{mean squared error} (MSE), i.e. the average of the squared error $||{f}(\boldsymbol{x}_i)-\boldsymbol{y}_i||^2$. It is widely used in machine learning problems as a fundamental measure of difference.\\

With a properly chosen loss function, the original problem of finding a $f$ that best fits the dataset $S$ reduces to finding a $f$ that minimizes the loss, which is an optimization problem in the parameter space $\{\textbf{M}_i,\boldsymbol{b}_i\}_{i=1}^{n}\,$. This optimization problem is clearly non-convex and hard to solve. Therefore, instead of looking for a global minimum in the parameter space, we only look for a local minimum which hopefully has a low enough loss to accomplish the learning task well. This is done via \textit{gradient descent}. Namely, we calculate the gradient of the loss with respect to all the parameters, and then modify all the parameters following the gradient to decrease the loss using a small step size, and then we repeat this process. Denoting the parameters by $\boldsymbol{\theta}\,$, the iteration process is given as below:\\
\begin{equation}\label{gradient descent}
	\boldsymbol{\theta}'=\boldsymbol{\theta} -\epsilon\,\nabla_{\boldsymbol{\theta}}L\left(\boldsymbol{\theta},\{(\boldsymbol{x},\boldsymbol{y})_i\}\right)\,,
\end{equation}\\
where $\epsilon$ is the iteration step size and is called the \textit{learning rate}. It is clear that, with a small enough learning rate, the above iteration indeed converges to a local minimum of $L$. This process of finding a solution is called \textit{training} in machine learning, and before training we initialize the parameters randomly. In practice, although Eq. (\ref{L2 loss}) uses the whole training set $S$ to define $L\,$, during training we only sample a minibatch of data points from $S$ to evaluate the gradient, and this is called \textit{stochastic gradient descent} (SGD). The sampling method significantly improves efficiency. \\

As can be seen, both training and evaluation of a neural network requires a great amount of matrix computation. In a typical modern neural network, the number of parameters is on the order of tens of millions and the required amount of computation is huge. Therefore, the potential of this neural network strategy did not attract much attention until the technology of GPU (Graphic Processing Unit)-based parallelized computation becomes available in recent years \cite{CUDA}, which makes it possible to train modern neural networks in hours or a few days, which would previously take many months on CPUs. This development of technology makes large-scale deep learning possible and is one important reason for the deep learning boom in recent years.\\

In real cases, the iteration in training process usually does not follow Eq. (\ref{gradient descent}) exactly. This is because this iteration strategy can cause the iteration to go forth and back inside a valley-shaped region, or be disturbed by local noise on gradients, or be blocked by barriers in the searched parameter space, etc. To alleviate these problems, some alternative algorithms have been developed as improved versions of the basic gradient descent, including Adam \cite{Adam}, RMSprop \cite{RMSprop}, and gradient descent with momentum \cite{DeepLearningBook}. Basically, these algorithms employ two strategies. The first one is to give the iteration step an inertia, so that the iteration step is not only influenced by the current gradient, but also by all previous gradients, and the effect of previous gradients decays exponentially every time a step is done. This is called the \textit{momentum} method, and the so-called \textit{momentum term} is one minus the decay coefficient, usually set to 0.9 , which represents how much the inertia is preserved per step. This training method is actually ubiquitous in deep learning. The second strategy is normalization of parameter gradients, such that the average of the iteration step for each parameter becomes roughly constant and not proportional to the magnitude of the gradient. In sparse gradient cases, this strategy dramatically speeds up training. RMSprop and Adam adopt this strategy.\\

To summarize, we first train a neural network to fit a given dataset $S=\{(\boldsymbol{x},\boldsymbol{y})_i\}$ by minimizing a predefined loss. Then we use the neural network to predict $\boldsymbol{y}$ of new unseen data points $\boldsymbol{x}\,$. Concerning practical applications, fully connected neural networks as described above are commonly used for regression tasks, of which the target is to fit a real-valued function, which is often multivariable. For image classification, motivated by the fact that a pixel in an image is most related to nearby pixels, we put the units of a neural network layer also into a pixelized space, and connect a unit only to adjacent units between layers. To extract nonlocal information as an output, we downsample the pixelized units. This structure is called the \textit{convolutional neural network}, and is the current state-of-the-art method for image classification tasks \cite{ImagenetSota}. Many other neural network structures exist, but we leave them here since they are not directly relevant to our study. Although deep neural networks work extremely well for various tasks, so far we do not precisely know the reason, which is still an important open question nowadays. \\

\section{Reinforcement Learning}\label{reinforcement learning}
\subsection{Problem Setting}
In a reinforcement learning (RL) task, we do not have a training dataset beforehand, but we let the AI interact with an environment to accumulate experience, and learn from the accumulated experience with a target set to maximize a predefined reward. Because the AI learns by exploring the environment to perform better and better, this learning process is called reinforcement learning, in contrast to supervised and unsupervised learning in which case the AI learns from a pre-existing dataset. \\

\begin{figure}[htb]
	\centering
	\includegraphics[width=0.8\linewidth]{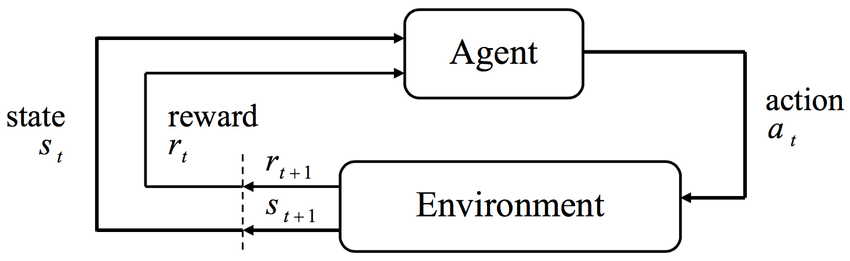}
	\caption{Setting of reinforcement learning, where an AI agent interacts with an environment. Subscripts $t$ and $t+1$ denote successive time steps. Reproduced from Ref. \cite{reinforcementLearningFigure}.}
	\label{fig:agent-environment-interaction-in-reinforcement-learning}
\end{figure}

Reinforcement learning is important when we do not understand an environment, but we can simulate the environment for an AI to interact with and gain experience from. Examples of reinforcement learning tasks include video games \cite{AtariEnvironment}, modern e-sports games \cite{Starcraft2}, chess-like games \cite{chess-like}, design problems \cite{MaterialDesign} and physical optimization problems in quantum control \cite{ErrorCorrectionReinforcement, GateOptimization}. In all situations the environment that the AI interact with is modelled by a Markov decision process (MDP) or a game in game theory, where there is a state representing the current situation of the environment, and the AI inputs the state and outputs an action which influences future evolution of the environment state. The goal of the AI is maximize the expected total reward in the environment, such as scores in games, winning probability in chess, and fidelity of quantum gates. This setting is illustrated in figure \ref{fig:agent-environment-interaction-in-reinforcement-learning}. Note that the evolution of the environment state and actions of the AI are discrete in time steps. \\

\subsection{Q-learning}\label{Q learning}
\begin{figure}[hbt]
	\centering
	\includegraphics[width=0.09\linewidth]{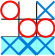}
	\caption{A tic-tac-toe game example, reproduced from Ref. \cite{tic-tac-toe}.}
	\label{fig:tic-tac-toe-game}
\end{figure}

There are many learning strategies for a reinforcement learning task. The most basic one is brute-force search, which is to test all possible action choices under all circumstances to find out the most beneficial strategy, in the sense of total expected reward. This strategy can be used to solve small-scale problems such as tic-tac-toe (figure \ref{fig:tic-tac-toe-game}).
\\

However, this strategy is applicable to tic-tac-toe only because the space of the game state is so small that it can be easily enumerated. In most cases, brute-force search is not possible, and we need better methods, often heuristic ones, to achieve reinforcement learning. In this section we discuss the mostly frequently used method, Q-learning \cite{Q-learning}, which is used in our research in the next few chapters.\\

The Q-learning is based on a function $Q^\pi(s,a)$ which is defined to be the expected future reward at a state $s$ when taking an action $a$, provided with a certain policy $\pi$ to decide future actions. In addition, expected rewards in future steps are discounted exponentially by $\gamma$ according to how far they are away from the present:\\

\begin{equation}\label{Q function}
	Q^\pi(s_t,a_t)=r(s_t,a_t)+\mathbb{E}_{{(a_{t+i}\sim\pi(s_{t+i}),\, s_{t+i})}_{i=1}^{\infty}}\left[\sum_{i=1}^{\infty}\gamma^i\, r(s_{t+i},a_{t+i})\right],\quad 0<\gamma<1\,,
\end{equation}\\
where $r(s,a)$ is the reward, and the expectation is taken over future trajectories $(s_{t+i},a_{t+i})_{i=1}^{\infty}\,$. Note that the environment evolution $(s_{t},a_{t})\mapsto s_{t+1}$ is solely determined by the environmental property and cannot be controlled.\\

This $Q$ function has a very important recursive property, that is\\

\begin{equation}\label{Q recursive}
	Q^{\pi}(s_t,a_t)=r(s_t,a_t)+\gamma\,\mathbb{E}_{a_{t+1}\sim\pi{(s_{t+1})},\,s_{t+1}}\left[Q^{\pi}(s_{t+1},a_{t+1})\right]
\end{equation}\\
which can be shown directly from Eq. (\ref{Q function}) with non-divergent $Q\,$s. If the action policy $\pi^*$ is optimal, $Q^{\pi^*}$ then satisfies the following Bellman Eq. (\ref{Bellman equation}) \cite{OptimalControlTheoryIntroduction}:\\

\begin{equation}\label{Bellman equation}
	Q^{\pi^*}(s_t,a_t)=r(s_t,a_t)+\gamma\,\mathbb{E}_{s_{t+1}}\left[\max_{a_{t+1}}Q^{\pi^*}(s_{t+1},a_{t+1})\right]\,,
\end{equation}\\
which is straightforward to show by following Eq. (\ref{Q recursive}), using the fact that policy $\pi^*$ takes every action to maximize $Q\,$; otherwise it cannot be an optimal policy as $Q$ would be increased by taking a maximum. \\

Then we look for such a $Q$ function. This function can be obtained by the following iteration:\\

\begin{equation}\label{Q iteration}
	Q'(s_t,a_t)=r(s_t,a_t)+\gamma\,\mathbb{E}_{s_{t+1}}\left[\max_{a_{t+1}}Q(s_{t+1},a_{t+1})\right]\,.
\end{equation}\\

After sufficiently many iterations, $Q'$ converges to $Q^{\pi^*}\,$ \cite{Q-learning}. This is due to the discount factor $0<\gamma<1$ and a non-diverging reward $r(s,a)\,$, which makes iterations of the above equation drop the negligible final term $\mathbb{E}_{s_{t+1}}\left[\max_{a_{t+1}}Q(s_{t+1},a_{t+1})\right]$ as it would be multiplied by $\gamma^n$ after $n$ iterations and diminishes exponentially, and the remaining value of $Q'$ would be purely determined by reward $r(s,a)\,$. Since this converged $Q'$ is not influenced by the initial $Q$ choice at the start of iterations, this $Q'$ must be unique, and therefore it is also the $Q^{\pi^*}$ in Eq. (\ref{Bellman equation}), so this converged $Q'$ is both unique and optimal. This is rigorously proved in Ref. \cite{Q-learning}. After obtaining the $Q^{\pi^*}\,$, we simply use $\max_{a_{t}}Q^{\pi^*}(s_{t},a_{t})$ to decide the action $a_{t}$ for a state $s_{t}\,$, and this results in the optimal policy $\pi^*\,$.\\

An important caveat here is that, the optimality above cannot be separated from the discount factor $\gamma\,$, which puts exponential discount on future rewards. This $\gamma$ is necessary for convergence of $Q\,$, but it also results in a ``time horizon" beyond which the optimal policy does not consider future rewards. However, the goal of solving the reinforcement learning problem is to actually maximize total accumulated reward, which corresponds to $\gamma=1\,$. The ideal case is that $\pi^*$ converges for $\gamma\to 1\,$; however, this is not always true, and different $\gamma\,$s represent strategies that have different amounts of foresight. In addition, a large $\gamma$ such as 0.9999 often make the learning difficult and hard to proceed. Therefore in practice, Q-learning usually does not achieve the absolute optimality that corresponds to $\gamma=1$, except for the case where the reinforcement learning task is well-bounded so that $\gamma\to1$ does not cause any problem.\\

\subsection{Implementation of Deep Q-Network Learning}\label{deep reinforcement learning implementation}
The Q-learning strategy that uses a deep neural network to approximate $Q$ function is called \textit{deep Q-network} (DQN) learning. It simply constructs a deep feedforward neural network to represent the $Q$ function, with input $s$ and several outputs as evaluated $Q$ values for each action choice $a\,$. Note that now action $a$ is a choice from a finite set and is not a continuous variable. The training loss is defined to minimize the absolute value of the \textit{temporal difference} error (TD error), which comes from Eq. (\ref{Q iteration}):\\

\begin{equation}\label{TD loss}
	L=Q(s_t,a_t)-\left(r(s_t,a_t)+\gamma\,\max_{a_{t+1}}Q(s_{t+1},a_{t+1})\right)\,,
\end{equation}\\
where $(s_t,a_t,r,s_{t+1})$ is a sampled piece of experience of the AI interacting with the environment. Due to the nature of sampling during training, we ignore evaluation of the expectation of $s_{t+1}$ in the equation. \\

\begin{figure}[htb]
	\centering
	\includegraphics[width=0.7\linewidth]{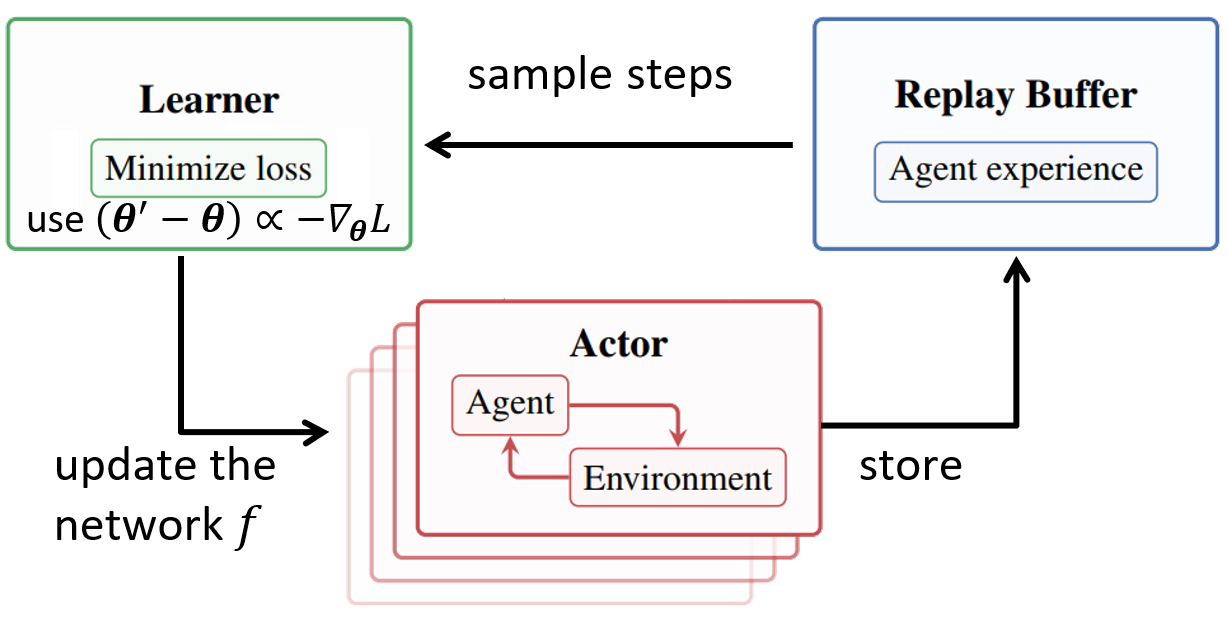}
	\caption{Reinforcement learning system, reproduced from Ref. \cite{TCLoss}.}
	\label{fig:reinforcementlearningsystem}
\end{figure}

Up to now we have obtained all the necessary pieces of information to implement deep reinforcement learning. First, we initialize a feedforward neural network with random parameters. Second, we use this neural network as our $Q$ function and use strategy $\max_{a_{t}}Q(s_{t},a_{t})$ to take actions (which may be modified). Third, we sample from what the AI has experienced, i.e. $(s_t,a_t,r,s_{t+1})$ data tuples, and calculate the error, (Eq. \ref{TD loss}), and use gradient descent to minimize the absolute error. Fourth, we repeat the second and third steps until the AI performs well enough. This system can be divided into three parts, and diagrammatically it is shown in figure \ref{fig:reinforcementlearningsystem}. Note that the second and the third steps above can be parallelized and executed simultaneously.
\\

In real scenarios, a great deal of technical modifications are applied to the above learning procedure in order to make the learning more efficient and to improve the performance. These advanced learning strategies include separation of the trained $Q$ function and the other $Q$ function term used in calculating the loss \cite{DQN}, taking random actions on occasion \cite{DQN}, prioritized experience sampling strategy \cite{prioritizedSampling}, double Q-networks to separately decide actions and $Q$ values \cite{DoubleDQN}, duel network structure to separately learn the average $Q$ value and $Q$ value change due to each action \cite{DuelDQN}, using random variables inside network layers to induce organized random exploration in the environment \cite{NoisyDQN}. These are the most important recent developments, and they improve the performance and stability of deep Q-learning, which is discussed in Appendix \ref{experiment details appendix}. These techniques are incorporated together in Ref. \cite{RainbowDQN} and the resultant algorithm is called \textit{Rainbow DQN}. We follow this algorithm in our reinforcement learning researches in the next chapters. The hyperparameter settings and details of our numerical experiments are provided in the corresponding chapters and Appendix \ref{experiment details appendix}.

	\chapter{Control in Quadratic Potentials}\label{control quadratic potentials}
	In this chapter, we discuss one particle control with continuous position measurement in one-dimensional quadratic potentials, including a harmonic potential and an inverted harmonic potential, and the target of control is to keep the particle at low energy around the center. We first analyse the system in Section \ref{quadra Gaussian approx} to simplify it, and show that it has an optimal solution of control in Section \ref{quadra optimal control} with reference to the Linear-Quadratic-Gaussian (LQG) control theory which is discussed in Appendix \ref{LQG appendix}. We explain our experimental setting of the simulated quantum system, train a neural network to control the system, and present the results in Section \ref{quadra experiments}. Next, we compare the resulting performance with the optimal control in Section \ref{quadra comparison}, and then consider different levels of handicaps on controllers, comparison them with suboptimal control strategies, and discuss what is precisely learned by the AI. Finally, in Section \ref{quadra conclusion} we present the conclusions of this chapter. \\

\section{Analysis of Control of a Quadratic Potential}\label{analysis of quadratic}
\subsection{Gaussian Approximation}\label{quadra Gaussian approx}
The state of a particle in quadratic potentials under position measurement is known to be well approximated by a Gaussian state \cite{LinearStochasticSchrodingerEquation, CoherentStatesByDecoherence}. We discuss this result in detail in this section, and derive a sufficient representation of the evolution equation of a particle only in terms of the first and second moments of its Wigner distribution in $(x,p)$ phase space. This significantly simplifies the problem, and reduces it to an almost classical situation.\\

By the term \textit{Gaussian state}, we mean that the one-particle state has a Gaussian shape distribution as its Wigner distribution in phase space. We show that a Gaussian-shaped Wigner distribution always keeps its Gaussian property when evolving in a quadratic potential. First, we consider the problem in the Heisenberg picture and evaluate the time evolutions of operators $\hat{x}$ and $\hat{p}$:\\
\begin{equation}\label{differential equation for x p}
\frac{d}{dt}\hat{x}=\frac{i}{\hbar}[H,\hat{x}],\qquad \frac{d}{dt}\hat{p}=\frac{i}{\hbar}[H,\hat{p}],
\end{equation}
\begin{equation}
H=\frac{k}{2}\hat{x}^2+\frac{\hat{p}^2}{2m},\qquad k\in\mathbb{R},\quad m \in\mathbb{R}^+,
\end{equation}\\
where $k$ can be both positive and negative. We do not assume a particular sign of $k$ in the following calculations. We substitute $H$ into Eq. (\ref{differential equation for x p}) and obtain\\
\begin{equation}
\frac{d}{dt}\hat{x}=\frac{\hat{p}}{m},\qquad \frac{d}{dt}\hat{p}=-k\hat{x},
\end{equation}\\
which constitute a system of differential equations as\\
\begin{equation}
\frac{d}{dt}
\left(\begin{array}{c}
\hat{x}\\
\hat{p}
\end{array}\right) = 
\left(\begin{array}{cc}
0 & \frac{1}{m}\\
-k & 0
\end{array}\right)
\left(\begin{array}{c}
\hat{x}\\
\hat{p}
\end{array}\right).
\end{equation}\\
This is solved by eigendecomposition:\\
\begin{equation}
\left(\begin{array}{c}
\hat{x}(t)\\
\hat{p}(t)
\end{array}\right) = e^{tA}\left(\begin{array}{c}
\hat{x}(0)\\
\hat{p}(0)
\end{array}\right),\qquad A:= \left(\begin{array}{cc}
0 & \frac{1}{m}\\
-k & 0
\end{array}\right)
\end{equation}\\
\begin{equation}
A=Q\Lambda Q^{-1}=\left(\begin{array}{cc}
	-\frac{1}{\sqrt{-mk}} & \frac{1}{\sqrt{-mk}}\\
	1 & 1
	\end{array}\right)\left(\begin{array}{cc}
	-\sqrt{\frac{-k}{m}} & 0\\
	0 & \sqrt{\frac{-k}{m}}
	\end{array}\right)\left(\begin{array}{cc}
	-\frac{\sqrt{-mk}}{2} & \frac{1}{2}\\
	\frac{\sqrt{-mk}}{2} & \frac{1}{2}
	\end{array}\right)
\end{equation}\\
\begin{equation}
M:=e^{tA}=Qe^{t\Lambda}Q^{-1}=Q\left(\begin{array}{cc}
e^{-t\sqrt{\frac{-k}{m}}} & 0\\
0 & e^{t\sqrt{\frac{-k}{m}}}
\end{array}\right)Q^{-1},
\end{equation}\\
\begin{equation}\label{operator evolution}
\left(\begin{array}{c}
\hat{x}(t)\\
\hat{p}(t)
\end{array}\right) =M\left(\begin{array}{c}
\hat{x}(0)\\
\hat{p}(0)
\end{array}\right).
\end{equation}\\
For simplicity, we define\\
\begin{equation}
\lambda:=\sqrt{\frac{-k}{m}}\equiv i\sqrt{\frac{k}{m}},
\end{equation}\\
and then the matrix $M$ can be written as\\
\begin{equation}\label{operator evolution matrix}
M=\frac{1}{2}\left(\begin{array}{cc}
e^{t\lambda}+e^{-t\lambda} & \frac{1}{\sqrt{-mk}}(e^{t\lambda}-e^{-t\lambda})\\
\sqrt{-mk}(e^{t\lambda}-e^{-t\lambda}) & e^{t\lambda}+e^{-t\lambda}
\end{array}\right),
\end{equation}\\
which is a symplectic matrix.\\

Therefore, a free evolution in a quadratic potential equivalently transforms the $\hat{x}$ and $\hat{p}$ operators through a symplectic transformation. Next we show that this results in the same symplectic transform of its Wigner distribution in terms of phase-space coordinates $(x,p)$. To show this, we use the characteristic function definition of Wigner distribution \cite{GaussianQuantumInformation}, as follows:\\
\begin{equation}\label{Wigner definition1}
W(\textbf{x})=\int_{\mathbb{R}^2}\frac{d^2\boldsymbol{\xi}}{(2\pi)^2}\exp(-i\textbf{x}^T \boldsymbol{\Omega}\boldsymbol{\xi})\chi(\boldsymbol{\xi}),
\end{equation}\\
\begin{equation}\label{Wigner definition2}
\textbf{x}=\left(\begin{array}{c}
x\\
p
\end{array}\right),\quad
\boldsymbol{\Omega}=\left(\begin{array}{cc}
0 & 1\\
-1 & 0
\end{array}\right),\quad
\textbf{x},\boldsymbol{\xi}\in\mathbb{R}^2,
\end{equation}\\
\begin{equation}\label{Wigner definition3}
\chi(\boldsymbol{\xi})= \text{tr}(\rho D(\boldsymbol{\xi})),\qquad D(\boldsymbol{\xi}):=\exp({i\hat{\textbf{x}}^T\boldsymbol{\Omega}\boldsymbol{\xi}}),\quad \hat{\textbf{x}}:= \left(\begin{array}{c}
\hat{x}\\
\hat{p}
\end{array}\right),
\end{equation}\\
where $\rho$ is a quantum state, $\chi$ is called the Wigner characteristic function and $D$ is the Weyl operator. The Wigner distribution is essentially a Fourier transform of the Wigner characteristic function, and both the characteristic function $\chi$ and the Wigner distribution $W$ contain complete information about the state $\rho$. Now we consider the time evolution of $\rho$, or equivalently, the time evolution of $\hat{\textbf{x}}$. According to Eqs. (\ref{operator evolution}) and (\ref{operator evolution matrix}) and the symplectic property of $M$, we have\\
\begin{equation}
\begin{split}
\chi(\boldsymbol{\xi},t)=\text{tr}(\rho D(\boldsymbol{\xi},t))&=\text{tr}(\rho \exp({i\hat{\textbf{x}}^T M^T\boldsymbol{\Omega}\boldsymbol{\xi}}))\\
&=\text{tr}(\rho \exp({i\hat{\textbf{x}}^T M^T\boldsymbol{\Omega}MM^{-1}\boldsymbol{\xi}}))\\
&=\text{tr}(\rho \exp({i\hat{\textbf{x}}^T \boldsymbol{\Omega}M^{-1}\boldsymbol{\xi}}))\\
&=\chi(M^{-1}\boldsymbol{\xi},0).
\end{split}
\end{equation}\\
Note that the matrix $M$ has a determinant equal to 1, i.e. $|M|=1$, and therefore it is always invertible. Then the Wigner distribution is\\
\begin{equation}
\begin{split}
W(\textbf{x},t)&=\int_{\mathbb{R}^2}\frac{d^2\boldsymbol{\xi}}{(2\pi)^2}\exp(-i\textbf{x}^T \boldsymbol{\Omega}\boldsymbol{\xi})\chi(\boldsymbol{\xi},t)\\
&=\int_{\mathbb{R}^2}\frac{d^2\boldsymbol{\xi}}{(2\pi)^2}\exp(-i\textbf{x}^T \boldsymbol{\Omega}\boldsymbol{\xi})\chi(M^{-1}\boldsymbol{\xi},0)\\
&=\int_{\mathbb{R}^2}\frac{d^2(M\boldsymbol{\xi})}{(2\pi)^2|M|}\exp(-i\textbf{x}^T \boldsymbol{\Omega}M\boldsymbol{\xi})\chi(\boldsymbol{\xi},0)\\
&=\int_{\mathbb{R}^2}\frac{d^2(M\boldsymbol{\xi})}{(2\pi)^2}\exp(-i\textbf{x}^T{M^{-1}}^T \boldsymbol{\Omega}\boldsymbol{\xi})\chi(\boldsymbol{\xi},0)\\
&=W(M^{-1}\textbf{x},0),
\end{split}
\end{equation}\\
where we have changed the integration variable $\boldsymbol{\xi} \gets M\boldsymbol{\xi}$ in the third line and used the condition of unbounded integration area $\mathbb{R}^2$ in the last line.\\

We see that a Wigner distribution $W(\textbf{x},0)$ simply evolves to $W(M\textbf{x},t)$ after time $t$. This is merely a linear transformation of phase space coordinates, and therefore the distribution as a whole remains unaltered, with only the position, orientation and width of the distribution possibly being changed. The shape of distribution never changes. Thus, a Gaussian distribution always stays Gaussian. This fact also holds true for Hamiltonians including the terms $(\hat{x}\hat{p}+\hat{p}\hat{x})$ and $\hat{x}$ and $\hat{p}$, which can be proved similarly. \\

Then, if at some instants we do weak position measurements that can be approximated to be Gaussian on the state $\rho$, as in Section \ref{Quantum Trajectory section}, the Gaussianity of the Wigner distribution on position coordinate $x$ increases, and along with the rotation and movement of the distribution in phase space, the Gaussianity of the whole distribution monotonically increases, and as a result it is always rounded to be approximately Gaussian in the long term. The main idea is that, non-Gaussianity never emerges by itself in quadratic potentials and under position measurement.\\

\subsection{Effective Description of Time Evolution}
In this case, the state can be fully described by its means and covariances as a Gaussian distribution in phase space. Those quantities are\\
\begin{equation}\label{mean and covariance matrix}
\begin{split}
\langle\hat{x}\rangle,\quad\langle\hat{p}\rangle,\quad &V_x:=\langle\hat{x}^2\rangle-\langle\hat{x}\rangle^2,\quad V_p:=\langle\hat{p}^2\rangle-\langle\hat{p}\rangle^2,\\
C&:=\frac{1}{2}\langle\hat{x}\hat{p}+\hat{p}\hat{x}\rangle-\langle\hat{x}\rangle\langle\hat{p}\rangle,
\end{split}
\end{equation}\\
i.e., totally five real values.\\

Therefore, when describing the time evolution of the state under continuous measurement, we may only describe the time evolution of the above five quantities instead, which is considerably simpler. We now derive their evolution equations.\\

We use the evolution equation for a state under continuous position measurement Eq. (\ref{position measurement incomplete information}) as a starting point:\\
\begin{equation}
d\rho=-\frac{i}{\hbar}[H,\rho]dt-\frac{\gamma}{4}[\hat{x},[\hat{x},\rho]]dt+\sqrt{\frac{\gamma\eta}{2}}\{\hat{x}-\langle \hat{x}\rangle,\rho\}dW,
\end{equation}\\
\begin{equation}
dW\sim\mathcal{N}(0,dt),\quad \gamma>0,\quad\eta\in[0,1].
\end{equation}\\
Recall that $dW$ is a Wiener increment and $\gamma$ and $\eta$ represent the measurement strength and efficiency respectively, and we use It\^o calculus formulation. We can now evaluate the time evolution of the quantities in Eq. (\ref{mean and covariance matrix}).\\
\begin{equation}\label{calculate x}
\begin{split}
d\langle\hat{x}\rangle=\text{tr}(\hat{x}\,d\rho)&=\text{tr}\left(\frac{i}{\hbar}\rho[H,\hat{x}]dt-\frac{\gamma}{4}[\hat{x},[\hat{x},\rho]]\hat{x}dt+\sqrt{\frac{\gamma\eta}{2}}\{\hat{x}-\langle \hat{x}\rangle,\rho\}\hat{x}\,dW\right)\\
&=\text{tr}\left(\rho\frac{\hat{p}}{m}dt-0+\sqrt{\frac{\gamma\eta}{2}}(2\hat{x}^2\rho-2\langle \hat{x}\rangle\hat{x}\rho)\,dW\right)\\
&=\frac{\langle \hat{p}\rangle}{m}dt+\sqrt{2\gamma\eta}(\langle\hat{x}^2\rangle-\langle \hat{x}\rangle^2)\,dW\\
&=\frac{\langle \hat{p}\rangle}{m}dt+\sqrt{2\gamma\eta}V_x\,dW.
\end{split}
\end{equation}\\
\begin{equation}\label{calculate p}
\begin{split}
d\langle\hat{p}\rangle=\text{tr}(\hat{p}\,d\rho)&=\text{tr}\left(\frac{i}{\hbar}\rho[H,\hat{p}]dt-\frac{\gamma}{4}[\hat{x},[\hat{x},\rho]]\hat{p}dt+\sqrt{\frac{\gamma\eta}{2}}\{\hat{x}-\langle \hat{x}\rangle,\rho\}\hat{p}\,dW\right)\\
&=\text{tr}\left(\rho(-k\hat{x})dt-\frac{\gamma}{4}\rho(\hat{x}\hat{x}\hat{p}-2\hat{x}\hat{p}\hat{x}+\hat{p}\hat{x}\hat{x})+\sqrt{2\gamma\eta}\left(\rho\frac{\hat{x}\hat{p}+\hat{p}\hat{x}}{2}-\rho\langle \hat{x}\rangle\hat{p}\right)\,dW\right)\\
&=-k{\langle \hat{x}\rangle}dt+\sqrt{2\gamma\eta}C\,dW.
\end{split}
\end{equation}\\
\begin{equation}\label{calculate Vx1}
\begin{split}
dV_x&=\text{tr}(\hat{x}^2\,d\rho)-d(\langle \hat{x}\rangle^2)=\text{tr}(\hat{x}^2\,d\rho)-2\langle \hat{x}\rangle d\langle \hat{x}\rangle-(d\langle \hat{x}\rangle)^2\\
&=\text{tr}(\hat{x}^2\,d\rho)-2\langle \hat{x}\rangle \left(\frac{\langle \hat{p}\rangle}{m}dt+\sqrt{2\gamma\eta}V_x\,dW\right)-2\gamma\eta V_x^2\,dt.
\end{split}
\end{equation}\\

Since it is too lengthy, we calculate $\text{tr}(\hat{x}^2\,d\rho)$ separately. In following calculations we need to use symmetric properties of $\rho$ as a Gaussian state. First we use $\text{tr}((\hat{x}-\langle\hat{x}\rangle)^3\rho)=0$, which means that the skewness of a Gaussian distribution is zero. It leads to
\begin{equation}\label{skewness Gaussian}
\text{tr}(\hat{x}^3\rho)=3\langle\hat{x}^2\rangle\langle\hat{x}\rangle-2\langle\hat{x}\rangle^3=3V_x\langle\hat{x}\rangle+\langle\hat{x}\rangle^3.
\end{equation}\\
\begin{equation}\label{calculate Vx2}
\begin{split}
\text{tr}(\hat{x}^2\,d\rho)&=\text{tr}\left(\frac{i}{\hbar}\rho[H,\hat{x}^2]dt-\frac{\gamma}{4}[\hat{x},[\hat{x},\rho]]\hat{x}^2dt+\sqrt{\frac{\gamma\eta}{2}}\{\hat{x}-\langle \hat{x}\rangle,\rho\}\hat{x}^2\,dW\right)\\
&=\text{tr}\left(\rho\frac{\hat{x}\hat{p}+\hat{p}\hat{x}}{m}dt+\sqrt{2\gamma\eta}\left(\rho{\hat{x}^3}-\rho\langle \hat{x}\rangle\hat{x}^2\right)\,dW\right)\\
&=\frac{\langle\hat{x}\hat{p}+\hat{p}\hat{x}\rangle}{m}dt+\sqrt{2\gamma\eta}\left(3V_x\langle\hat{x}\rangle+\langle\hat{x}\rangle^3-\langle \hat{x}\rangle (V_x+\langle \hat{x}\rangle^2)\right)\,dW\\
&=\frac{\langle\hat{x}\hat{p}+\hat{p}\hat{x}\rangle}{m}dt+2\sqrt{2\gamma\eta}V_x\langle\hat{x}\rangle\,dW,
\end{split}
\end{equation}\\
\begin{equation}\label{calculate Vx3}
\begin{split}
dV_x&=\text{tr}(\hat{x}^2\,d\rho)-2\langle \hat{x}\rangle \left(\frac{\langle p\rangle}{m}dt+\sqrt{2\gamma\eta}V_x\,dW\right)-2\gamma\eta V_x^2\,dt\\
&=\left(\frac{2C}{m}-2\gamma\eta V_x^2\right)dt.
\end{split}
\end{equation}\\

Next we calculate $V_p$ in a similar manner.\\
\begin{equation}\label{calculate Vp1}
\begin{split}
dV_p&=\text{tr}(\hat{p}^2\,d\rho)-d(\langle \hat{p}\rangle^2)\\
&=\text{tr}(\hat{p}^2\,d\rho)-2\langle \hat{p}\rangle \left(-k{\langle \hat{x}\rangle}dt+\sqrt{2\gamma\eta}C\,dW\right)-2\gamma\eta C^2\,dt.
\end{split}
\end{equation}\\
We need to use the following symmetric property:\\
\begin{equation}\label{calculate Vp2}
\begin{split}
&\text{tr}((\hat{p}-\langle\hat{p}\rangle)(\hat{x}-\langle\hat{x}\rangle)(\hat{p}-\langle\hat{p}\rangle)\rho)=0\\
\Rightarrow\quad&\text{tr}\left(\frac{\hat{x}\hat{p}^2+\hat{p}^2\hat{x}}{2}\rho\right)=\text{tr}((\hat{p}\hat{x}\hat{p})\rho)=2C\langle\hat{p}\rangle+V_p\langle\hat{x}\rangle+\langle\hat{p}\rangle^2\langle\hat{x}\rangle,
\end{split}
\end{equation}\\
\begin{equation}\label{calculate Vp3}
\begin{split}
\text{tr}(\hat{p}^2\,d\rho)&=\text{tr}\left(\frac{i}{\hbar}\rho[H,\hat{p}^2]dt-\frac{\gamma}{4}[\hat{x},[\hat{x},\rho]]\hat{p}^2dt+\sqrt{\frac{\gamma\eta}{2}}\{\hat{x}-\langle \hat{x}\rangle,\rho\}\hat{p}^2\,dW\right)\\
&=\text{tr}\left(-k\rho(\hat{x}\hat{p}+\hat{p}\hat{x})dt-\frac{\gamma}{4}\rho(\hat{x}\hat{x}\hat{p}^2-2\hat{x}\hat{p}^2\hat{x}+\hat{p}^2\hat{x}\hat{x})\right .\\
&\qquad\qquad\qquad\qquad\qquad\left .+\sqrt{2\gamma\eta}\left(\rho\frac{(\hat{x}\hat{p}^2+\hat{p}^2\hat{x})}{2}-\rho\langle \hat{x}\rangle\hat{p}^2\right)\,dW\right)\\
&=-k\langle\hat{x}\hat{p}+\hat{p}\hat{x}\rangle dt+\text{tr}\left(-\frac{\gamma}{4}\rho(2i\hbar\, \hat{x}\hat{p}-2i\hbar\, \hat{p}\hat{x})\right)+\sqrt{2\gamma\eta} 2C\langle\hat{p}\rangle\,dW\\
&=-k\langle\hat{x}\hat{p}+\hat{p}\hat{x}\rangle dt+\frac{\gamma}{2}\hbar^2 dt+\sqrt{2\gamma\eta} 2C\langle\hat{p}\rangle\,dW,
\end{split}
\end{equation}\\
\begin{equation}\label{calculate Vp4}
\begin{split}
dV_p&=\text{tr}(\hat{p}^2\,d\rho)-2\langle \hat{p}\rangle \left(-k{\langle \hat{x}\rangle}dt+\sqrt{2\gamma\eta}C\,dW\right)-2\gamma\eta C^2\,dt\\
&=(-2kC-2\gamma\eta C^2+\frac{\gamma}{2}\hbar^2)dt.
\end{split}
\end{equation}\\

Finally, we calculate the covariance $C$.\\
\begin{equation}\label{calculate C1}
\begin{split}
dC=\frac{1}{2}\text{tr}\left((\hat{x}\hat{p}+\hat{p}\hat{x})\,d\rho\right)&-d(\langle\hat{x}\rangle\langle\hat{p}\rangle)\\
=\frac{1}{2}\text{tr}\left((\hat{x}\hat{p}+\hat{p}\hat{x})\,d\rho\right)&-\langle\hat{x}\rangle d\langle\hat{p}\rangle-\langle\hat{p}\rangle d\langle\hat{x}\rangle-d\langle\hat{x}\rangle d\langle\hat{p}\rangle\\
=\frac{1}{2}\text{tr}\left((\hat{x}\hat{p}+\hat{p}\hat{x})\,d\rho\right)&-\langle\hat{x}\rangle(-k{\langle \hat{x}\rangle}dt+\sqrt{2\gamma\eta}C\,dW)\\
&- \langle\hat{p}\rangle\left(\frac{\langle p\rangle}{m}dt+\sqrt{2\gamma\eta}V_x\,dW\right)-2\gamma\eta V_xC\,dt.
\end{split}
\end{equation}\\
Here we need the following symmetry:\\
\begin{equation}\label{calculate C2}
\begin{split}
&\text{tr}((\hat{x}-\langle\hat{x}\rangle)(\hat{p}-\langle\hat{p}\rangle)(\hat{x}-\langle\hat{x}\rangle)\rho)=0\\
\Rightarrow\quad&\text{tr}\left(\frac{\hat{p}\hat{x}^2+\hat{x}^2\hat{p}}{2}\rho\right)=\text{tr}((\hat{x}\hat{p}\hat{x})\rho)=2C\langle\hat{x}\rangle+V_x\langle\hat{p}\rangle+\langle\hat{x}\rangle^2\langle\hat{p}\rangle.
\end{split}
\end{equation}\\
\begin{equation}\label{calculate C3}
\begin{split}
\frac{1}{2}\text{tr}\left((\hat{x}\hat{p}+\hat{p}\hat{x})\,d\rho\right)={}&\frac{1}{2}\text{tr}\left(\frac{i}{\hbar}\rho[H,(\hat{x}\hat{p}+\hat{p}\hat{x})]dt-\frac{\gamma}{4}[\hat{x},[\hat{x},\rho]](\hat{x}\hat{p}+\hat{p}\hat{x})dt\right.\\
&\quad+\left.\sqrt{\frac{\gamma\eta}{2}}\{\hat{x}-\langle \hat{x}\rangle,\rho\}(\hat{x}\hat{p}+\hat{p}\hat{x})\,dW\right)\\
={}&\frac{1}{2}\text{tr}\left(\rho\left(\frac{2\hat{p}^2}{m}-2k\hat{x}^2\right)dt-0+\sqrt{2\gamma\eta}\left(2\hat{x}\hat{p}\hat{x}\rho-(\hat{x}\hat{p}+\hat{p}\hat{x})\langle \hat{x}\rangle\rho\right)\,dW\right)\\
={}&\left(\frac{V_p+\langle \hat{p}\rangle^2}{m}-k(V_x+\langle \hat{x}\rangle^2)\right)dt\\
&\quad+\frac{1}{2}\left(\sqrt{2\gamma\eta}\left(4C\langle\hat{x}\rangle+2V_x\langle\hat{p}\rangle+2\langle\hat{x}\rangle^2\langle\hat{p}\rangle-2C\langle \hat{x}\rangle-2\langle\hat{p}\rangle\langle\hat{x}\rangle^2\right)\,dW\right)\\
={}&\left(\frac{V_p+\langle \hat{p}\rangle^2}{m}-k(V_x+\langle \hat{x}\rangle^2)\right)dt+\sqrt{2\gamma\eta}\left(C\langle\hat{x}\rangle+V_x\langle\hat{p}\rangle\right)\,dW,\\
\end{split}
\end{equation}\\
\begin{equation}\label{calculate C4}
\begin{split}
dC&=\frac{1}{2}\text{tr}\left((\hat{x}\hat{p}+\hat{p}\hat{x})\,d\rho\right)-\langle\hat{x}\rangle(-k{\langle \hat{x}\rangle}dt+\sqrt{2\gamma\eta}C\,dW)\\
&\qquad\qquad- \langle\hat{p}\rangle\left(\frac{\langle \hat{p}\rangle}{m}dt+\sqrt{2\gamma\eta}V_x\,dW\right)-2\gamma\eta V_xC\,dt\\
&=\left(\frac{V_p}{m}-kV_x-2\gamma\eta V_xC\right)dt.
\end{split}
\end{equation}\\

The results are summarized as follows:\\
\begin{equation}\label{mean and covariance evolution}
\begin{split}
d\langle\hat{x}\rangle&=\frac{\langle \hat{p}\rangle}{m}dt+\sqrt{2\gamma\eta}\,V_x\,dW,\\
d\langle\hat{p}\rangle&=-k{\langle \hat{x}\rangle}dt+\sqrt{2\gamma\eta}\,C\,dW,\\
dV_x&=\left(\frac{2C}{m}-2\gamma\eta V_x^2\right)dt,\\
dV_p&=(-2kC-2\gamma\eta C^2+\frac{\gamma}{2}\hbar^2)dt,\\
dC&=\left(\frac{V_p}{m}-kV_x-2\gamma\eta V_xC\right)dt.
\end{split}
\end{equation}\\

Our results coincide with the results presented in Ref. \cite{HarmonicOscillatorControl}, and we have also verified that our results are correct through numerical calculation. From the above equations, we can see that only the average position and momentum are perturbed by the stochastic term $dW$, and the covariances form a closed set of equations and evolve deterministically. Therefore, we can calculate their steady values:\\
\begin{equation}\label{covariances solution}
\begin{split}
dV_x=dV_p&=dC=0\cdot dt,\qquad V_x,V_p>0.\\
\Rightarrow C=\frac{-k+\sqrt{k^2+\gamma^2\eta\hbar^2}}{2\gamma\eta}&,\  V_x=\sqrt{\frac{C}{m\gamma\eta}},\  V_p=2C\sqrt{m\gamma\eta C}+k\sqrt{\frac{mC}{\gamma\eta}}.
\end{split}
\end{equation}\\

We observe that a state always evolves into a steady shape in numerical simulation, and due to this convergence we assume that the covariances are simply fixed as the above values. Then, the degrees of freedom considerably decrease, and the only remaining ones are the two real quantities $\langle\hat{x}\rangle$ and $\langle\hat{p}\rangle$, which are the means of the Gaussian distribution in phase space.\\

Equivalently, we may say that the degrees of freedom of the state is represented by a displacement operator $D(\alpha)$, which displaces the state from the origin of phase space, i.e. $\langle\hat{x}\rangle=\langle\hat{p}\rangle=0$. We denote the state centered at the origin of phase space by $\rho_0$, and we have \\
\begin{equation}\label{displaced means}
\langle\hat{x}\rangle_{D\rho_0D^\dagger}=\sqrt{\frac{\hbar}{2m\omega}}(\alpha+\alpha^*),\quad\langle\hat{p}\rangle_{D\rho_0D^\dagger}=i\sqrt{\frac{\hbar m\omega}{2}}(\alpha^*-\alpha),
\end{equation}\\
\begin{equation}\label{displacement definition}
D(\alpha)=e^{\alpha\hat{a}^\dagger-\alpha^*\hat{a}^\dagger},\quad \alpha\in\mathbb{C},
\end{equation}
\begin{equation}\label{annihilator definition}
\hat{a}:=\sqrt{\frac{m\omega}{2\hbar}}(\hat{x}+\frac{i}{m\omega}\hat{p}),\quad\omega:=\sqrt{\frac{|k|}{m}},
\end{equation}\\
where $\hat{a}$ is the annihilation operator. It has the following properties:\\ 
\begin{equation}\label{annihilation operator properties}
[\hat{a},\hat{a}^\dagger]=1,\quad \hat{x}=\sqrt{\frac{\hbar}{2m\omega}}(\hat{a}^\dagger+\hat{a}),\quad \hat{p}=i\sqrt{\frac{\hbar m\omega}{2}}(\hat{a}^\dagger-\hat{a}),
\end{equation}\\
and\\
\begin{equation}\label{annihilation operator displaced}
D^\dagger(\alpha)\hat{a}D(\alpha)=\hat{a}+\alpha,\quad D^\dagger(\alpha)\hat{a}^\dagger D(\alpha)=\hat{a}^\dagger+\alpha^*,
\end{equation}\\
\begin{equation}\label{number operator displaced}
\hat{n}:=\hat{a}^\dagger\hat{a},\quad D^\dagger(\alpha)\hat{n}D(\alpha)=\hat{n}+(\alpha\hat{a}^\dagger+\alpha^*\hat{a}) + |\alpha|^2,
\end{equation}\\
where $\hat{n}$ is the number operator. In the above calculations we do not assume the sign of $k$, but here it is necessary to use the positive coefficient $|k|$ to give an appropriate definition for the operators, and therefore $\omega$ is always positive.\\

For the state $\rho_0$, we have $\langle\hat{x}\rangle_{\rho_0}=\langle\hat{p}\rangle_{\rho_0}=0$, and therefore $\langle\alpha\hat{a}^\dagger+\alpha^*\hat{a}\rangle_{\rho_0}=0$. Then we have\\
\begin{equation}\label{number operator calculation}
\langle\hat{n}\rangle_{D\rho_0D^\dagger}=\langle\hat{n}\rangle_{\rho_0} + |\alpha|^2.
\end{equation}\\

Therefore, we can use $|\alpha|^2$ to express the expectation value of the operator $\left(\frac{\hat{p}^2}{2m}+\frac{|k|}{2}\hat{x}^2\right)$ for state $D\rho_0D^\dagger$:
\begin{equation}\label{energy displaced}
\left\langle\frac{\hat{p}^2}{2m}+\frac{|k|}{2}\hat{x}^2\right\rangle_{D\rho_0D^\dagger}=\hbar\omega\left(\langle\hat{n}\rangle_{D\rho_0D^\dagger}+\frac{1}{2}\right)=\hbar\omega\left(\langle\hat{n}\rangle_{\rho_0}+\frac{1}{2}+|\alpha|^2\right),
\end{equation}\\
where $\langle\hat{n}\rangle_{\rho_0}$ is a constant determined by the covariances in Eq. (\ref{covariances solution}). We also have $\langle\hat{x}\rangle_{D\rho_0D^\dagger}$ and $\langle\hat{p}\rangle_{D\rho_0D^\dagger}$ to represent the real and imaginary parts of $\alpha$, so we obtain the following formula:\\
\begin{equation}\label{energy representation}
\left\langle\frac{\hat{p}^2}{2m}+\frac{|k|}{2}\hat{x}^2\right\rangle_{D\rho_0D^\dagger}=\hbar\omega\left(\langle\hat{n}\rangle_{\rho_0}+\frac{1}{2}\right)+\left(\frac{\langle\hat{p}\rangle^2_{D\rho_0D^\dagger}}{2m}+\frac{|k|}{2}\langle\hat{x}\rangle^2_{D\rho_0D^\dagger}\right).
\end{equation}\\

Now, if we want to evaluate $\left\langle\frac{\hat{p}^2}{2m}+\frac{|k|}{2}\hat{x}^2\right\rangle$ for the state $\rho$, we can just replace the operators $\hat{p}$ and $\hat{x}$ by the means $\langle\hat{p}\rangle$ and $\langle\hat{x}\rangle$. \\

As we can see, this system turns out to be very simple. This can be understood by the fact that, concerning free evolution, a non-negative Wigner distribution behaves in phase space exactly in the same way as the corresponding classical distribution unless the Hamiltonian contains terms that are more than quadratic or non-analytic, since its evolution equation would reduces to the Liouville equation \cite{WignerDistributionEvolution, StatisticalMechanicsGibbs}. In addition, the position measurement only shrinks and squeezes the distribution in the $x$ direction, which does not introduce negativity into the distribution \cite{SqueezedStates}, and therefore the distribution evolves almost classically. The only quantumness in this system is the measurement backaction by $dW$ and the uncertainty principle which introduces a constant term in $dV_p$ (see Eq. (\ref{mean and covariance evolution})).\\

\subsection{Optimal Control}\label{quadra optimal control}
As the system is simplified, we now consider control of this quadratic system. The system is summarized by the following:\\
\begin{equation}\label{equation sets of quadra}
\begin{split}
d\langle\hat{x}\rangle&=\frac{\langle \hat{p}\rangle}{m}dt+\sqrt{2\gamma\eta}\,V_x\,dW,\\
d\langle\hat{p}\rangle&=-k{\langle \hat{x}\rangle}dt+\sqrt{2\gamma\eta}\,C\,dW,\\
C&=\frac{-k+\sqrt{k^2+\gamma^2\eta\hbar^2}}{2\gamma\eta},\\ V_x=\sqrt{\frac{C}{m\gamma\eta}}&,\quad V_p=2C\sqrt{m\gamma\eta C}+k\sqrt{\frac{mC}{\gamma\eta}},
\end{split}
\end{equation}\\
where the only independent degrees of freedom are $\langle\hat{x}\rangle$ and $\langle\hat{p}\rangle$. We consider using an external force to control the system, which is just an additional term $F_\text{con}\hat{x}$ added to the total Hamiltonian $H$. Then the time evolution becomes\\
\begin{equation}\label{controlled quadra}
\begin{split}
d\langle\hat{x}\rangle&=\frac{\langle \hat{p}\rangle}{m}dt+\sqrt{2\gamma\eta}\,V_x\,dW,\\
d\langle\hat{p}\rangle&=(-k{\langle \hat{x}\rangle}-F_{\text{con}})dt+\sqrt{2\gamma\eta}\,C\,dW,\\
\end{split}
\end{equation}\\
where $F_{\text{con}}$ actually gives a force in the opposite direction of its sign. We can confirm that the equations concerning $V_x$, $V_p$ and $C$ are not changed explicitly, or by interpreting the additional $F_{\text{con}}\hat{x}$ term in the Hamiltonian as a shift of the operator $\hat{x}$ by an amount of $\frac{F_{\text{con}}}{k}$, which clearly does not affect the covariances.\\

When $k$ is larger than zero, the Hamiltonian $H=\frac{\hat{p}^2}{2m}+\frac{k}{2}\hat{x}^2$ represents a harmonic oscillator, and here we consider controlled cooling of this system. Since the system involves only one particle, cooling amounts to decreasing its energy from an arbitrarily chosen initial state. Because we assume continuous measurement on this system, the previous analysis and simplification apply.\footnote{We do not consider a measurement strength that varies in time.} The target of control is to minimize energy $\langle H\rangle$, which amounts to minimizing the functional $\left(\frac{\langle\hat{p}\rangle^2}{2m}+\frac{k}{2}\langle\hat{x}\rangle^2\right)$ with $k>0$ according to Eq. (\ref{energy representation}), under the above time-evolution equations (\ref{controlled quadra}). As we consider a general cooling task, the minimization should be considered as minimizing the time-averaged total energy. We call this minimized function as a \textit{loss}, which is sometimes also called a control score. It is denoted as $L$:\\
\begin{equation}\label{control loss}
L:=\lim\limits_{T\to\infty}\frac{1}{T}\int_{0}^{T}\left(\frac{\langle\hat{p}\rangle^2}{2m}+\frac{k}{2}\langle\hat{x}\rangle^2\right)dt.
\end{equation}\\
Here the infinite limit of time is not crucial. It is put here for translational invariance of the control in time, and sufficient long-term planning of the control.\\

When the system is noise-free and deterministic, that is,
\begin{equation}\label{deterministic controlled quadra}
\begin{split}
d\langle\hat{x}\rangle&=\frac{\langle \hat{p}\rangle}{m}dt,\\
d\langle\hat{p}\rangle&=(-k{\langle \hat{x}\rangle}-F_{\text{con}})dt,
\end{split}
\end{equation}
the $L$ above can converge to 0 due to the term $\left (\lim\limits_{T\to\infty}\frac{1}{T}\right )$ and therefore is not a proper measure of loss that we wish to minimize. Therefore, we use the above definition (\ref{control loss}) only when the system contains noise, and for a deterministic case we need to redefine it as\\
\begin{equation}\label{deterministic loss}
L:=\lim\limits_{T\to\infty}\int_{0}^{T}\left(\frac{\langle\hat{p}\rangle^2}{2m}+\frac{k}{2}\langle\hat{x}\rangle^2\right)dt,
\end{equation}\\
which is not essentially different from Eq. (\ref{control loss}) but is well-behaved when we try to minimize it. Since these two definitions are only different by some mathematical subtlety, we do not specifically distinguish them when it is unnecessary, and it is clear from the context which one is being considered.\\

As introduced in Section \ref{quantum control}, we now seek for the optimal strategy of controlling the variable $F_{\text{con}}$ such that $L$ is minimized. For the deterministic system Eq. (\ref{deterministic controlled quadra}), minimization of $L$ can be achieved in a simple manner, by borrowing some ideas from physics. First, we note that the time-evolution equations concerning $\langle\hat{x}\rangle$ and $\langle\hat{p}\rangle$ are effectively classical, which means that, we have a classical particle with position $x$ and momentum $p$ satisfying $x_{t=0}=\langle\hat{x}\rangle_{t=0}$ and $p_{t=0}=\langle\hat{p}\rangle_{t=0}$, and the time evolution of $(x,p)$ can be the same as that of $(\langle\hat{x}\rangle,\langle\hat{p}\rangle)$ of the underlying quantum system, which can also be seen from the Ehrenfest theorem concerning quadratic potentials \cite{QuantumText}. Then, when looking at the functional $L$ as expressed in Eq. (\ref{deterministic loss}), one may recall the action and the Hamilton principle, i.e., a classical trajectory of mechanical variables $(x,p)$ minimizes the total action which is the time integral of the Lagrangian, which is very similar to the form of Eq. (\ref{deterministic loss}). Therefore, if we construct a Lagrangian $\mathcal{L}$ defined with $x$ and $p$ such that minimization of $\left(\int\mathcal{L}\,dt\right)$ corresponds to minimization of the loss $L$, then we can obtain a trajectory of variables $(\langle\hat{x}\rangle,\langle\hat{p}\rangle)$ as the classical trajectory of $(x,p)$ that minimizes the loss $L$. After we obtain the desired trajectory, an external control is applied to keep the quantities $(\langle\hat{x}\rangle,\langle\hat{p}\rangle)$ such that they stay on the desired trajectory. This completes a simple derivation of the so-called linear-quadratic optimal control for our system.\\

Following this argument, we define $\mathcal{L}=\frac{m\dot{x}^2}{2}+\frac{kx^2}{2}=\mathcal{T}-\mathcal{V}$, where the classical kinetic energy is $\mathcal{T}=\frac{p^2}{2m}=\frac{m\dot{x}^2}{2}$ and the potential is $\mathcal{V}=-\frac{kx^2}{2}$. Note that a Lagrangian must be defined in the form of $\mathcal{L}(\boldsymbol{q},\dot{\boldsymbol{q}},t)$ so that for this functional the Hamilton principle holds \cite{ClassicalMechanics}. The action $\left(\int\mathcal{L}\,dt\right)$ is equal to the loss $L$, and therefore a classical particle travelling in the potential $\mathcal{V}$ has mechanical variables $(x,p)$ which minimize $L$ when they are substituted by $(\langle\hat{x}\rangle,\langle\hat{p}\rangle)$, as $\langle\hat{x}\rangle$ and $\langle\hat{p}\rangle$ are constrained by the same relation as $x$ and $p$, i.e. $\frac{d}{dt}\langle\hat{x}\rangle=\frac{\langle \hat{p}\rangle}{m}$ (Eq. \ref{deterministic controlled quadra}) and $\frac{d}{dt}{x}=\frac{p}{m}$, which makes sure that when $x$ is substituted by $\langle \hat{x} \rangle$, $p$ is substituted by $\langle\hat{p}\rangle$. From a viewpoint of optimization, we see that the minimization of $L$ is done under the constraint $\frac{d}{dt}\langle\hat{x}\rangle=\frac{\langle \hat{p}\rangle}{m}$, which is achieved by the same constraint $\frac{d}{dt}x=\frac{p}{m}$ of the classical Lagrangian. Therefore, all necessary conditions are indeed satisfied, and a trajectory of $(\langle\hat{x}\rangle,\langle\hat{p}\rangle)$ which minimizes $L$ must be a classical physical trajectory of $(x,p)$ for Lagrangian $\mathcal{L}$, and we should use the control to achieve such a trajectory.\\

Next, we consider what trajectories of $\mathcal{L}$ can be used to minimize $L$. Because of the unstable potential $\mathcal{V}=-\frac{kx^2}{2}$ which is high at the center and low at both sides, a classical particle would have a divergent total action $\left(\int^\infty\mathcal{L}\,dt\right)$ unless the particle precisely stops at the top of the potential with a zero momentum, in which case the action becomes non-divergent. Therefore, we specifically look at the conditions under which it can be non-divergent. In order to precisely stop at the top of the potential $\mathcal{V}$, as a first condition its velocity and position should have opposite signs, so that it moves towards the center, and as a second condition it needs to dissipate all its energy when exactly reaching the top, i.e. $\frac{p^2}{2m}-\frac{kx^2}{2}=0$. Therefore, the trajectory of the particle's $(x,p)$ satisfies\\
\begin{equation}\label{optimal trajectory}
p=-\sqrt{mk}\,x.
\end{equation}\\
This is the main result of our optimal control.\\

Then, whenever the above condition is not satisfied for our state with $(\langle\hat{x}\rangle,\langle\hat{p}\rangle)$, we apply control $F_\text{con}$ to influence the evolution of $\langle\hat{p}\rangle$ so that it changes to satisfy the condition. If $F_\text{con}$ is not bounded, we can modify $\langle\hat{p}\rangle$ in an infinitesimal length of time to satisfy the condition quickly, and then keep a moderate strength of $F_\text{con}$ to keep $\langle\hat{p}\rangle$ always satisfying it. This is the optimal control which minimizes $L$ if the system variables evolve according to Eq. (\ref{deterministic controlled quadra}), which is deterministic and does not include noise.\\ 

The important but difficult final step is to show that, when measurement backaction noise is included as in Eq. (\ref{controlled quadra}), the above control strategy is still optimal. This is called the \textit{separation theorem} in the context of control theory, and it is not straightforward to prove. Therefore, we resort to the standard Linear-Quadratic-Gaussian (LQG) control theory \cite{LinearQuadraticControl} and prove it in the context of control theory in Sec.~\ref{linear quadratic Gaussian appendix section} in appendices. Since the reasoning follows a different line of thoughts, we do not discuss it here further.\footnote{A more general and rigorous proof can be found in Ref.~\cite{FeedbackControlOfLinearStochasticSystems}.} \\

Regarding the case of $k<0$, in which the system amounts to an inverted pendulum, we consider the minimization of a loss defined as
\begin{equation}\label{inverted potential optimal control loss}
L=\int\left(\frac{\langle\hat{p}\rangle^2}{2m}+\frac{|k|}{2}\langle\hat{x}\rangle^2\right)dt
\end{equation}
so that when it is minimized, both the position and momentum are kept close to zero, and therefore the particle stays stable near the origin of the $x$ coordinate. This makes the problem the same as before and produces the same optimal trajectory condition, that is
\begin{equation}\label{optimal trajectory inverted}
p=-\sqrt{m|k|}\,x,
\end{equation}
and we use this as the conventional optimal control strategy for the inverted harmonic potential problem.\\

\section{Numerical Experiments}\label{quadra experiments}
In this section, we describe the settings of our numerical experiments of the simulated quantum control in quadratic potentials under continuous position measurement. Detailed settings concerning specific deep reinforcement learning techniques and the corresponding hyperparameters are given in Appendix \ref{experiment details appendix}. \\

\subsection{Problem Settings}
First, we describe our settings of the simulated quantum system and the control.\\
\subsubsection{Loss Function}
As discussed in previous sections, we set the targets of the control to be minimizing the energy of the particle and keeping the particle near the center respectively for the harmonic oscillator and the inverted harmonic potentials. Because the problem is stochastic, the minimized quantity is actually the expectation of the loss, written as the following:\\
\begin{equation}\label{harmonic optential original loss}
E[L_1] = E\left [\lim\limits_{T\to\infty}\frac{1}{T}\int_{0}^{T}\left(\frac{\langle\hat{p}\rangle^2}{2m}+\frac{k}{2}\langle\hat{x}\rangle^2\right)dt\right ],
\end{equation}\\
\begin{equation}\label{inverted potential original loss}
E[L_2] = E\left [\frac{1}{T}\int_{0}^{T}g\left (\langle\hat{x}\rangle,\langle\hat{p}\rangle\right )\,dt\right ],
\end{equation}
where $E[\cdot]$ denotes the expectation value over trajectories of its stochastic variables, and $g$ is a function which judges whether the particle is away from the center (falling out) or is near the center (staying stable). The function $g$ is 1 when the particle is away, and is 0 otherwise. In our numerical simulation of the particle, we stop our simulation when the particle is already away and take $g=1$ afterwards. Therefore, we use a large $T$ rather than taking its infinite limit; otherwise it approaches 1. Concerning controls, we use deep reinforcement learning to learn to minimize these two quantities as reviewed in Chapter \ref{deep reinforcement learning}. However, the optimal control discussed in the last section only applies to loss functions of quadratic forms, and does not directly apply to Eq. (\ref{inverted potential original loss}). In order to obtain a control strategy for the inverted potential problem, we need to artificially define a different loss function to use optimal control theory. This is done in Eq. (\ref{inverted potential optimal control loss}). We use the optimal control derived from the artificially defined loss there and compare its performance with the deep reinforcement learning that directly learns the original loss. \\

\subsubsection{Simulation of the Quantum System}
Although we have a set of equations (\ref{controlled quadra}) which effectively describes the time evolution of the system, we still decide to numerically simulate the original quantum state in its Hilbert space. This is because we want to confirm that reinforcement learning can directly learn from a numerically simulated continuous-space quantum system without the need of simplification, and at the same time numerical error and computational budget are still acceptable. After this is confirmed, we may carry our strategy to other problems that are more difficult and cannot be simplified. Another reason for simulating the original quantum state is that, we input the quantum state directly to the neural network to make it learn from the state as well, and therefore we need the state.\\

To simulate the quantum system, we express the state in terms of the energy eigenbasis of the harmonic potential $V=\frac{|k|}{2}\hat{x}^2$. This simulation strategy is precise and efficient, because the state is Gaussian and thus can be expressed in terms of squeezing and displacement operators, which result in exponentially small values in high energy components of the harmonic eigenbasis. We set the energy cutoff of our simulated space at the 130-th excited state, and whenever the component on the 120-th excited state exceeds a norm of $10^{-5}$, we judge that the numerical error is going to be high and we stop the simulation. We call this as \textit{failing}, because the controller fails to keep the state stable around the center; otherwise the high energy components would not be large. This is used as our criterion to judge whether a control is successful or not.\\

To reduce the computational cost, we only consider pure states, and the time-evolution equation is Eq. (\ref{position measurement evolution}), i.e.,
\begin{equation*}
d|\psi\rangle=\left[\left(-\frac{i}{\hbar}H-\frac{\gamma}{4}(\hat{x}-\langle\hat{x}\rangle)^2\right)dt+\sqrt{\dfrac{\gamma}{2}}(\hat{x}-\langle\hat{x}\rangle)dW\right]|\psi\rangle,
\end{equation*}
where the Hamiltonian $H$ is
\begin{equation}\label{Hamiltonian quadra}
H=\frac{\hat{p}^2}{2m}+\frac{k}{2}\hat{x}^2+F_{\text{con}}\hat{x}.
\end{equation}
Also, due to the time-evolution equations of $\langle\hat{x}\rangle$ and $\langle\hat{p}\rangle$ (Eq.~(\ref{controlled quadra})), incomplete information on measurement outcomes, i.e. $\eta<1$, does not change the optimal control strategy, since the strategy merely depends on $k$ and $m$. Thus, we have the fact that the optimal control is the same for both a pure state with complete information and a mixed state with partial measurement information. As shown in Eq.~(\ref{controlled quadra}), the difference between incomplete and complete measurement information in this problem is only at the size of the additive noise, which does not essentially affect the system behaviour. Therefore, we do not experiment on mixed states with incomplete measurement results for simplicity.\\

\begin{figure}[tb]
	\centering\subfloat[]{
		\includegraphics[width=0.48\linewidth]{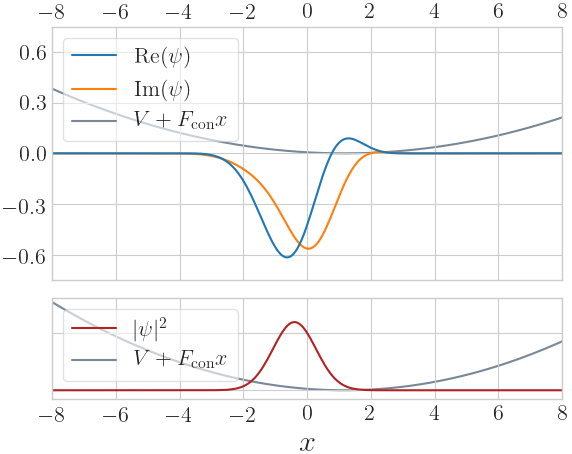}}
	\subfloat[]{
		\includegraphics[width=0.48\linewidth]{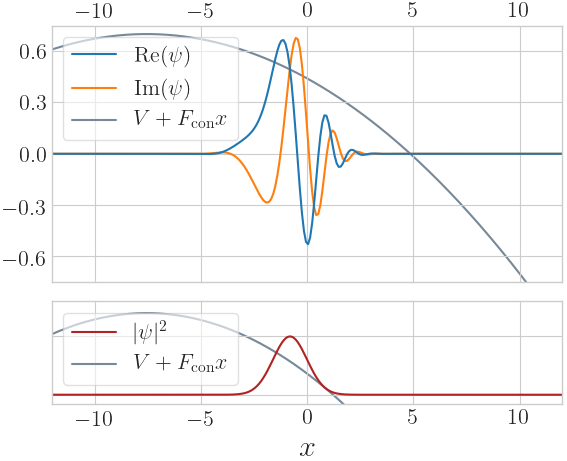}}
	\caption{Examples of controlled wavefunctions $\psi$ in the problems of cooling a harmonic oscillator (a) and stabilizing an inverted oscillator (b), plotted in $x$ space, together with schematic plots of the controlled potential (grey) and the probability distribution density (red). The real part and the imaginary part of the wavefunctions are plotted in blue and orange, respectively. The units of the horizontal axis is $\sqrt{\frac{\hbar}{m\omega}}$.}
	\label{fig:sampleharmonic}
\end{figure}

To numerically simulate the time-evolution equation, we discretize the time into time steps and do iterative updates to the state. The implemented numerical update scheme is a mixed explicit-implicit 1.5 order strong convergence scheme for It\^o stochastic differential equations with additional 2nd and 3rd order corrections of deterministic terms. It is nontrivial and is described in Section \ref{numerical update rule appendix} of Appendix \ref{numerical simulation appendix}. We numerically verified that our method has small numerical error, and specifically, the covariances of our simulated state differ from the calculated ones (see Eq. (\ref{equation sets of quadra})) by an amount of $10^{-4}\sim 10^{-6}$ when the state is stable, and differ by an amount of $10^{-3}\sim 10^{-4}$ when the simulated system fails, which is always below $10^{-2}$. Therefore, we believe that our numerical simulation of this stochastic system is sufficiently accurate. An example of the simulated system is plotted in Fig.~\ref{fig:sampleharmonic}.\\

To initialize the simulation, the state is simply set to be the ground state of the harmonic eigenbasis at the beginning, and then it evolves under the position measurement and control. The state leaves the ground state because of the measurement backaction, and it continuously gains energy if no control force is applied. Thus to keep the state at a low energy, it is necessary to make use of the control. When the total simulation time exceeds a preset threshold $t_{\text{max}}$ or when the simulation fails, we stop the simulation and restart a new one. We call one simulation from the start to the end as an \textit{episode}, following the usual convention of reinforcement learning. \\

\subsubsection{Constraints of Control Force}
In practice, the applied external control force $F_{\text{con}}$ must be finite and bounded. Because this external force as shown in Eq. (\ref{Hamiltonian quadra}) is equivalent to shifting the center of the potential by an amount of  $\dfrac{F_{\text{con}}}{k}$, we compare the width of the wavefunction and the distance of the potential shift, and keep them to be of the same order of magnitude. The wavefunctions in our experiments have standard deviations in the position space of around $0.67$ and $0.80$ in units of $\sqrt{\frac{\hbar}{m\omega}}$ for the harmonic oscillator problem and the inverted oscillator problem, and therefore the wavefunctions have widths of around $2.7$ and $3.2$. The allowed shifts of potentials for these two problems are correspondingly set to be $[-5,+5]$ and $[-10,+10]$\footnote{We do not use penalty on large control forces to prevent the divergence of control as done in the usual linear-quadratic control theory. This is because we also do not put penalty on any control choice of our neural network output and we want to compare the two strategies fairly.}, all in units of $\sqrt{\frac{\hbar}{m\omega}}$. In our numerical experiments, we find that the inverted problem is quite unstable. On the one hand, the noise is intrinsically unbounded as a Gaussian variable so it can overcome any finite control force; on the other hand, in the inverted potential, any deviation from the center makes the system harder to control since the particle tends to move away, and when it has deviated too much, the bounded control force may not be strong enough to work well. This is why we have set the allowed control force for the inverted problem to be larger. In the harmonic oscillator case, no matter how far the noise makes the particle deviate from the center, the particle always comes back again by oscillations, and a control force can always have a favourable effect on the particle to reduce its momentum as desired. This is why we have set the allowed control force for the harmonic oscillator problem to be small. In our experiments, we found that these allowed control force regions are sufficient to demonstrate the efficient cooling and control of the particle, as demonstrated in Section \ref{quadra performance}.\\

To make the control practical, we do not allow the controller to change its control force too many times during one oscillation period of the system, i.e. within one time period $\dfrac{2\pi}{\omega}$, where $\omega=\sqrt{\dfrac{|k|}{m}}$. This condition is imposed both for the harmonic and the inverted oscillator problems. We set that the controller can output 36 different control forces in one oscillation period $\dfrac{2\pi}{\omega}$, which amounts to 18 controls in half an oscillation period, i.e. the particle moving from one side to the other and changing the direction, or 9 controls in a one-quarter period. Our choice of this specific number here is only for divisibility regarding the time step of the simulated quantum system. A control force $F_{\text{con}}$ is applied to the system as a constant before the next control force is outputted from the controller. Therefore, the control forces on the system regarded as a function of time become a sequence of step functions, and we call each constant step in it as a \textit{control step}, in comparison to the time step of the numerical simulation.\\

\begin{figure}[tb]
	\centering
	\includegraphics[width=0.6\linewidth]{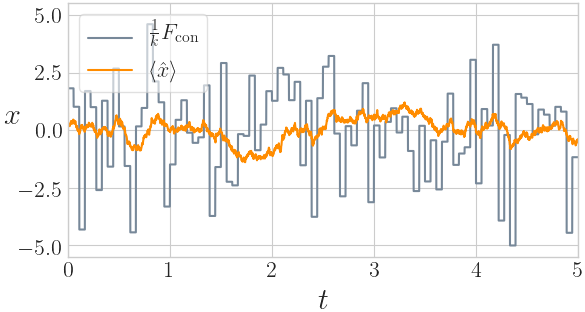}
	\caption{An example of the optimal control $\frac{1}{k}F_{\text{con}}$ and the average position $\langle\hat{x}\rangle$ in the cooling harmonic oscillator problem, plotted against time $t$ of the system. The units are consistent with Table~\ref{quadra parameter settings}. It can be seen that the control force $F_{\text{con}}$ is almost random, with a mean of zero, and $\langle\hat{x}\rangle$ fluctuates around zero.}
	\label{fig:harmoniccontrolsequence}
\end{figure}
Then, we need to specifically consider the implementation of the optimal controls. As control forces are bounded and discretized in time, we consider variations of the original continuous optimal control to satisfy these constraints, which is common in control theory. First, because each force is applied as a constant during the control step, the target of the control should be adapted and set to that the particle should be on the desired optimal trajectory at the end of a control step. We do so by solving for the control force $F_{\text{con}}$ using the time-evolution equation (\ref{deterministic controlled quadra}) and the current state $(\langle\hat{x}\rangle,\langle\hat{p}\rangle)$, with the target state $(\langle\hat{x}\rangle+d\langle\hat{x}\rangle,\langle\hat{p}\rangle+d\langle\hat{p}\rangle)$ satisfying the optimal trajectory condition in Eq.~(\ref{optimal trajectory inverted}). This is solved by simply taking $dt$ in Eq.~(\ref{deterministic controlled quadra}) to be the time of the control step and expanding $d\langle\hat{x}\rangle$ and $d\langle\hat{p}\rangle$ up to ${dt}^2$; this strategy is fairly accurate since we repeat it 18 times per one half of the oscillation period. The resulting control $F_{\text{con}}$ is linear with respect to the system variables $(\langle\hat{x}\rangle,\langle\hat{p}\rangle)$, and it is called a linear control. Finally, we bound $F_{\text{con}}$ by simply clipping it within the required bounds, ignoring the excessive values that are beyond the constraints. An example of the resulting control is shown in Fig.~\ref{fig:harmoniccontrolsequence}.\\

\subsubsection{Parameter Settings}
Many parameters of the simulated quantum system are redundant and can actually be rescaled to produce the same quantum evolution. Therefore, most of the parameters are arbitrary and we summarize our settings in Table \ref{quadra parameter settings}.\\
\begin{table}[hbt]
	\centering
	\begin{tabular}{cccccc}
		\toprule
		$\omega$ ($\omega_c$) & $m$ ($m_c$) & $k$ ($m_c\omega^2_c$) & $dt$ ($\frac{1}{\omega_c}$)  & $\eta$ & $\gamma$ ($\frac{m_c\omega_c^2}{\hbar}$)  \\[8pt]
		$\pi$  & $\dfrac{1}{\pi}$ & $\pm\pi$ & $\dfrac{1}{720}, \dfrac{1}{1440}$ & 1 & $\pi,2\pi$  \\ \bottomrule
	\end{tabular}\\[10pt]

	\begin{tabular}{cccc}
	\toprule
	 $n_{\text{max}}$ & $F_{\text{con}}$ ($\sqrt{\hbar m_c\omega_c^{3}}$)  & $N_\text{con}$  & $t_{\text{max}}$ ($\frac{1}{\omega_c}$) \\[8pt]
	 130 & $[-5\pi,+5\pi]$, $[-10\pi,+10\pi]$ & 18  & 100 \\ \bottomrule
	\end{tabular}\\
\caption{Parameter settings of the simulation of quadratic potentials under the position measurement. Units are shown in the parenthesis. The physical quantities $m_c$ and $\omega_c$ are used as a reference, and the parameters without units are dimensionless. The values are separated by a comma to show the specific settings for the harmonic (left) and the inverted (right) oscillator problems. }
\label{quadra parameter settings}
\end{table}\\
In Table \ref{quadra parameter settings}, $\eta$ denotes the measurement efficiency, and because we only consider pure states, it is 1; $\gamma$ is the measurement strength, and we set it so that the size of the wavefunction is about the same as that of the ground-state wavefunction of a harmonic oscillator. For the harmonic oscillator $\gamma=\pi$, and for the inverted oscillator $\gamma=2\pi$, and even if $\gamma$ is changed, our numerical simulation is still expected to produce similar results. In both the harmonic and the inverted problems we simulate the time evolution of the state only till time $t_{\text{max}}$. Other parameters include $n_{\text{max}}$, the high-energy cutoff, $dt$, the simulation time step, the control forces $F_{\text{con}}$ that is shown in ranges, and $N_\text{con}$, the number of output controls from the controller per unit time $\frac{1}{\omega_c}$. The oscillation period of the harmonic system here is exactly $2\times\frac{1}{\omega_c}$, which is our time scale. For example, for each simulation episode we simulate for time $100\times\frac{1}{\omega_c}$, i.e., 50 oscillation periods of the harmonic oscillator.\\

\subsubsection{Evaluation of Performance}\label{performance evaluation}
In this subsection, we explain how we evaluate the performances of different controllers. The performances are evaluated according to our numerical simulation, using the controllers to output control forces at every control step. Because the goals for the harmonic oscillator and the inverted oscillator problems are different, we use different methods to evaluate them.\\

For the problem of cooling a harmonic oscillator, we run the numerical simulation with control for 1000 episodes\footnote{Sometimes we simulate for more episodes.}, and sample the expected energy of the state as the phonon number $\langle\hat{n}\rangle$ for 4000 times in these episodes, and calculate the sample mean and estimate the standard error of the mean. We first initialize the state and let it evolve under control for 20 periods of oscillations, i.e. time $40\times\frac{1}{\omega_c}$, and then sample its $\langle\hat{n}\rangle$ per time $15\times\frac{1}{\omega_c}$, till the end of the episode. This is to remove the effect of initialization and sample correlations during sampling. The value $\langle\hat{n}\rangle$ is evaluated at the control steps, which is for consistency with the optimal control target that aims to put the state on the desired trajectory at the end of control steps. In this way we numerically evaluate the expected energy of the particle being controlled, or, the expected value of the loss function as in Eq.~(\ref{harmonic optential original loss}).\\

For the inverted oscillator problem, we use a totally different measure of performance. We use the failure events to judge whether the controlled particle is near or away from the center, which means that, if the simulated system fails (i.e., the components on eigenbasis $n=120$ exceeds the norm of $10^{-5}$), then the criterion function $g$ in Eq.~(\ref{inverted potential original loss}) equals 1; otherwise $g$ equals 0. In our numerical simulation, failure events correspond to the center of the wavefunction staying near $x=8$ or $x=9$ in units of $\sqrt{\frac{\hbar}{m\omega}}$, while the control force is allowed to move the center of the potential to $x=10$ at most. For simplicity, we consider a failure event in some time interval as a Bernoulli distribution of ``to fail" and ``not to fail", and we use the failing probability in one episode as our measure of control performance. To evaluate a controller, we run simulations of 1000 episodes to estimate its failing probability, and the variance of the estimation is calculated based the formula of binomial distributions.\\

\subsection{Reinforcement Learning Implementation}\label{quadraReinforcementImplementation}
In this section we describe how neural networks are trained to learn control strategies. We use deep Q-learning, and since we mainly follow Section \ref{deep reinforcement learning implementation} which is already explained in detail, we only discuss our specific settings for the problems concerned here. \\

We use Pytorch \cite{Pytorch} as our deep learning library, and when there are settings not mentioned in this thesis, it means that we use the default ones. For example, we use the default random initialization strategy for neural networks provided by Pytorch, and we do not manually set random seeds for the main process. As required, we sample random numbers in the main process to set the random seeds of subprocesses, which interact with the environment as reinforcement learning actors to accumulate experiences. \\

\subsubsection{Neural Network Architecture}
Neural network architecture is mostly related to the structure of its input, and therefore we need to decide what is the input to the neural network first. We consider three cases. The first is to use the quintuple $\left (\langle \hat{x}\rangle,\langle \hat{p}\rangle,V_x,V_p,C\right )$ as the input, because we know that this quintuple completely describes a Gaussian state, and actually it is already more than enough to determine the optimal control. The second case is to use the raw values of the wavefunction as the input, by taking all real-part and imaginary-part values of its components on the chosen eigenbasis. The third case is to use the measurement outcomes as the input. One or two measurement outcomes are not sufficient to determine the properties of the current state, and therefore we need to input many outcomes that are sequential in time into the neural network. In this case, the neural network is either a convolutional network or a recurrent neural network; otherwise it cannot process so many values ordered sequentially. Roughly speaking, we need the neural network to learn appropriate coarse-graining strategy for such a large number of measurement outcomes. \\

For the first and the second cases, the neural network is 4-layer fully connected feedforward neural network with the numbers of internal units being (512, 256, 256+128, 21+1), where the final two layers are separated into two different branches of computation as suggested in Ref.~\cite{DuelDQN} (see Sec.~\ref{Duel DQN section}). The 21 outputs are used to predict the action values deviating from their mean for different control choices, and the last 1 output is used to predict the mean action value. For the third input case, the neural network is a 6-layer feedforward neural network, with its first three layers being 1-dimensional convolutional layers, and others being fully connected ones. The kernel sizes of the three convolutional layers and their strides are respectively (13,5), (11,4), (9,4), with the number of filters being (32, 64, 64)\footnote{Experimentally we found that a larger size of the layers does not necessarily perform better, probably due to larger noises. Also, we cannot apply batch normalization between the convolutional layers, because this is a reinforcement learning task and the training target is not static.}. The fully connected three layers have their numbers of hidden units as (256, 256+128, 21+1), which is similar to the network of the previous case. Following Ref.~\cite{NoisyDQN}, the last two layers in our networks are constructed as noisy layers (see Sec.~\ref{Noisy DQN}). We notice that measurement outcomes only are not sufficient to predict the state, and we also need previous control forces that have been exerted on the state. Thus, we input the force and the measurement outcome data in parallel as a time sequence as the network input. The input time sequence contains force and measurement outcome history in the last $6\times\frac{1}{\omega_c}$ time for the cooling harmonic oscillator problem and contains the last $4\times\frac{1}{\omega_c}$ time for the inverted oscillator problem, which is due to the different measurement strength $\gamma$ in the two problems.\\

As discussed in Section \ref{reinforcement learning}, to implement Q-learning we need to define the action choices of the control, which is supposed to be a discrete set. However, currently we only have an allowed interval of control force values. Therefore, we discretize this interval into equispaced 21 different values as our 21 different control force choices, and then we do Q-learning as introduced in Section \ref{reinforcement learning}, which is the same for the harmonic and the inverted oscillator problems.\\

\subsubsection{Training Settings}
As we have used many deep reinforcement techniques, the hyperparameter settings relevant for the specific techniques are put into Appendix \ref{experiment details appendix}. In this subsection we only describe the settings and strategies that are relevant to the discussion we have presented so far.\\

First, in order to optimize the neural network with the training loss, we use the \linebreak RMSprop algorithm \cite{RMSprop} with its initial learning rate set to $2\times10^{-4}$ and momentum set to 0.9, and the training minibatch size is 512. The $\gamma$ coefficient that serves as the time horizon of Q-learning in Eq.~(\ref{Q iteration}) is 0.99, which means that future rewards are discounted by 0.99 for each control step. As $\frac{1}{1-0.99}=100$ and we have 36 controls per oscillation, the time horizon of the controller is around $10\times\frac{1}{\omega_c}$. We assume that this is sufficient for learning the tasks, i.e. a control does not influence the behaviour of the state after 5 periods of oscillation.\\

The reward of reinforcement learning is set to the control loss times a negative sign. For the cooling harmonic oscillator problem, because the value of $\langle\hat{n}\rangle$ decreases to a small number when the state is cooled, we multiply the loss by 5, and we also shift it by 0.4 such that it is closer to zero and can be either negative or positive. However, for the second and the third input cases as the learning is more difficult, we find that such large loss values make the learning process noisy and unstable, and as a result $\langle\hat{n}\rangle$ never decreases to a small value and the learning does not proceed. Therefore, for the second and third input cases we only rescale the loss by 2, and we shift it by 1. In addition, we stop the simulation whenever $\langle\hat{n}\rangle>10$ for the second input case or $\langle\hat{n}\rangle>20$ for the third input case is satisfied. With these modifications, the learned loss values are always reasonably small and the learning can proceed. Since the learned states always have low energy, in the wavefunction input case we only use the first 40 energy eigenbasis of the state as input. \\

Regarding the inverted oscillator problem, experimentally we find that the learning is unstable if we give a negative reward only at the moment when the simulation fails. This is probably due to the stochasticity of the system, because the stochasticity may push a state at the edge of failure either to fail or to turn back, which makes the training loss very high due to the unexpected behaviours of the states. In this case, the reward is called to be sparse. To alleviate this problem, we add a small portion of negative $\langle\hat{n}\rangle$ into the reward to facilitate the learning, which is multiplied by a factor of 0.02, with the original reward of a failure event being $-10$.\\

We use a memory replay buffer to store accumulated experiences of our reinforcement learning actors, and its size is set to containing a few thousands or hundreds of $t_{\text{max}}$ episodes. Due to different sizes of the inputs of different input cases, the sizes of the replay buffers are made different since we have limited computer memory. The sizes for the three cases are respectively 6000, 4000 and 400 episodes. To compromise, when new experiences are stored into the memory replay, we take away the experiences that are unimportant first, i.e. with low training loss values, so that important pieces of memory are preserved. We also use prioritized sampling during training as in Ref.~\cite{prioritizedSampling}, which is described in more detail in Sec.~\ref{Prioritized Replay} and \ref{Prioritized replay setting}.\\

To encourage exploring different possibilities of control, we make the reinforcement learning actors use the $\epsilon$-greedy strategy to take actions \cite{DQN}, i.e., the an action is randomly taken with a small probability $\epsilon$, and with a probability $1-\epsilon$ it is taken to pursue the highest expected reward (see Sec.~\ref{target network section}). This probability $\epsilon$ is 40\% at the beginning of training and then decreases rapidly to 2\%, and after several stages of decrease, it is suppressed to 0.01\% around the end of training.\\

The number of simulated episodes for training is around 10000 for the problem of cooling a harmonic oscillator and is around 30000 for the problem of stabilizing an inverted oscillator. Actually the simulation of the inverted problem fails quickly during the first a few thousand episodes, within an episode time length of around 1 or $2\times\frac{1}{\omega_c}$, and only the later episodes last longer, which shows that this problem is indeed a difficult problem. \\

\section{Comparison of Reinforcement Learning and Optimal Control}\label{quadra comparison}
\subsection{Performances}\label{quadra performance}
\begin{table}[tb]
	\centering
	\begin{tabular}{cp{8em}p{7.8em}p{8.8em}}
		\toprule
		& &  cooling harmonic oscillators  &  stabilizing inverted oscillators \\ \midrule
		\multirow{3}{3.7em}{Network Input} & $\left (\langle \hat{x}\rangle,\langle \hat{p}\rangle,V_x,V_p,C\right )$ &  $0.3252\pm0.0024$  &   $85.4\%\pm1.1\%$ \\ \cmidrule{2-4}
		& wavefunction &  $0.3266\pm0.0040$  &  $73.3\%\pm1.4\%$ \\ \cmidrule{2-4}
		& measurement outcomes &  $0.4505\pm0.0053$  &  $0.0\%$\linebreak {\small ($7.2\%\pm1.2\%$ provided with examples)} \\ \midrule
		\multicolumn{2}{c}{optimal control} &  $0.3265\pm0.0032$  &  $89.8\%\pm1.0\%$ \\ \bottomrule
	\end{tabular}
	\caption{Performance results for the problem of cooling harmonic oscillators and stabilizing inverted oscillators. The numbers behind $\pm$ signs show the estimated standard deviations of the reported means. As discussed in Sec.~\ref{performance evaluation}, the values reported for the cooling problem are the average phonon numbers $\langle\hat{n}\rangle$ and for the inverted problem are the success rates in one episode, i.e., one minus the probability of failure.}
	\label{quadra results}
\end{table}
\subsubsection{Cooling Harmonic Oscillators}
We now compare the performances of trained neural networks as described above with the performances of our derived optimal controls. For the problem of cooling a harmonic oscillator, the training was completed within one day on a Titan X GPU with 10-process parallelization running in Python, for each of the three input cases. The resulting performances in terms of $\langle\hat{n}\rangle$ are listed in Table \ref{quadra results}, compared with the optimal control. We can see that the performances except for the measurement input case are within one standard error, and therefore we conclude that they perform approximately the same. For completeness, we analytically calculate the theoretical optimal performance for this problem if provided with unbounded control. Assuming that $\langle\hat{p}\rangle=-\sqrt{mk}\langle\hat{x}\rangle$ always holds\footnote{From the perspective of the LQG theory, this condition can be obtained by taking the small control cost limit.}, the time evolution of $\langle\hat{x}\rangle$ becomes an Ornstein–Uhlenbeck process and we can calculate the expectation $E\left[\langle\hat{x}\rangle^2\right]$ and evaluate the loss given in Eq.~(\ref{harmonic optential original loss}). The result is $\langle\hat{n}\rangle\approx0.2565$, which is actually smaller than the above values. This discrepancy comes from the finite time length of our allowed control steps. If the number of control steps $N_{\text{con}}$ in time $\frac{1}{\omega_c}$ increases to 72, the performance of our optimal control is evaluated to be $\langle\hat{n}\rangle\approx0.2739\pm0.0047$, which is closer to the theoretical value and shows that our optimal control strategy is indeed valid. Therefore, by comparing with the optimal control in Table \ref{quadra results}, we can argue that our reinforcement learning is successful. Note that actually $\langle\hat{n}\rangle$ can never be reduced to zero, because the position measurement squeezes the wavefunction in space and the term $\langle\hat{n}\rangle_{\rho_0}$ in Eq.~(\ref{energy representation}) is non-zero. \\

As for the measurement-outcome-based input case, the lower performance may result from the intrinsic stochasticity of the measurement outcomes that disturb the process of learning. In deep learning practices, injected random noise in training data is typically found to decrease the final performance, and therefore, the measurement input case may have been affected similarly. This is possibly due to the interplay among the noise and the deep network structure and the gradient-based optimization. There seems to be no straightforward solution, and in this case it may be difficult for the reinforcement learning to learn from measurement outcomes alone.\\

\subsubsection{Stabilizing Inverted Oscillators}
Regarding the inverted harmonic potential problem, as the system is considerably more stochastic, a neural network is trained for three days for each input case. The resulting performances represented as success rates are given in Table \ref{quadra results}, i.e., one minus the probability of a failure event occurring in an episode. Here we find that only the first input case achieves a performance comparable to the optimal control, and that the performance roughly decreases with the increased difficulty of learning for the three input cases. To confirm that the measurement-outcome-based network can really learn, we add 100 episodes that are controlled by the optimal control into its memory replay as examples, such that when the network is trained, it learns both its own experiences and the examples given by the optimal control. In this case, the network performed better, and at the end of training it achieved a success rate of $7.2\%$, which implies that this network is indeed possible to complete the given task, but with much more difficulty. \\

In order to make a fairer comparison between the optimal control and the reinforcement learning controller, we consider a discretized version of the optimal control that has control forces discretized in the same way as the reinforcement learning controller. In order to obtain discretized values, we set the outputs of the optimal control to their nearest neighbouring values among the equispaced 21 choices of forces as the neural network, and then use the forces as the control. The performance of this discertized controller is evaluated, and the result is $89.3\%\pm1.0\%$, within one standard deviation compared with the continuous controller, which demonstrates that the discretization does not significantly decrease the controller's performance, and the performance of the reinforcement learning is not restricted by the discretization.\\

Finally we consider an even more discretized control strategy, the bang-bang protocol. The bang-bang protocol uses one of the extrema as its output, and in our case it outputs either the maximal force to the left or the maximal force to the right. If the optimal control has an output to the left, its variation as a bang-bang protocol outputs the left maximal output force, and vice versa. We then test the performance of this bang-bang protocol variation. Its performance is $71.0\%\pm1.4\%$, which is significantly lower. This bang-bang protocol always has the right direction of the control force, but its strength is larger than the optimal control and therefore the controlled system is perturbed more strongly. This shows that the inverted oscillator system is quite sensitive, and increased noise and disturbance reduce the stability of the system. Obviously this bang-bang protocol is a suboptimal control strategy, and both the best reinforcement learning control and our optimal control have performances superior to it. \\

\subsection{Response to Different Inputs}\label{quadraInputResponse}
\begin{figure}[tb!]
	\centering
	\includegraphics[width=0.4\linewidth]{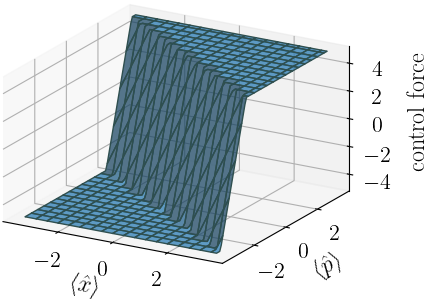}\qquad\quad
	\includegraphics[width=0.4\linewidth]{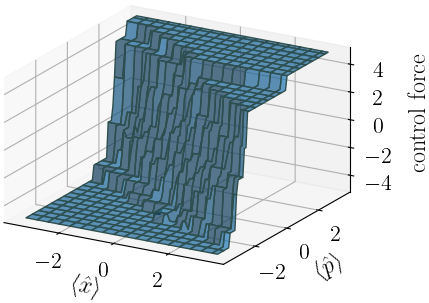}
	\caption{The control force plotted against the $\left (\langle \hat{x}\rangle,\langle \hat{p}\rangle\right )$ input for the problem of cooling a harmonic oscillator. The left panel shows the force of the optimal control, and the right one shows the control force of a trained reinforcement learning actor.}
	\label{fig:harmonic response small}
\end{figure}

\begin{figure}[tb!]
	\centering
	\includegraphics[width=0.4\linewidth]{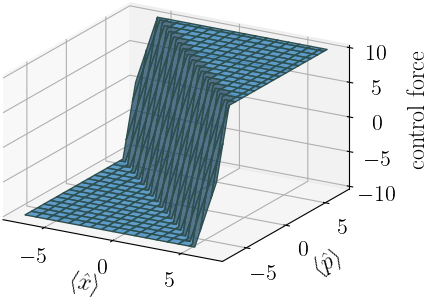}\qquad\quad
	\includegraphics[width=0.4\linewidth]{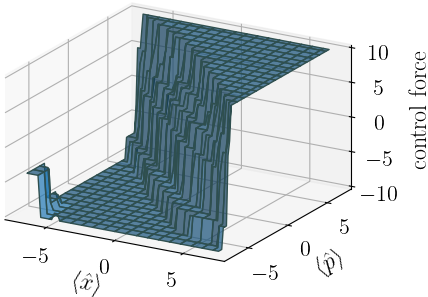}
	\caption{The control force plotted against the $\left (\langle \hat{x}\rangle,\langle \hat{p}\rangle\right )$ input for the problem of stabilization in an inverted harmonic potential. The left panel shows the the optimal control, and the right one shows our trained reinforcement learning actor.}
	\label{fig:inverted response small}
\end{figure}

To see what the reinforcement learning has learned in detail, we plot its output control force against different inputs of $\left (\langle \hat{x}\rangle,\langle \hat{p}\rangle\right )$, fixing $\left (V_x,V_p,C\right )$ at their average values. For the harmonic oscillator problem, the control force regarding $\left (\langle \hat{x}\rangle,\langle \hat{p}\rangle\right )$ is given in Fig.~\ref{fig:harmonic response small}, and for the inverted oscillator problem it is given in Fig.~\ref{fig:inverted response small}, and they are compared with the optimal controls. The leftmost and rightmost tails are the regions where the state is both staying away and moving away quickly, in which case it is almost destined to fail, and the central cliff of the plots is where the state stays most of the time. If the cliff becomes vertical, the control strategy reduces to the bang-bang protocol.\\

\begin{figure}[b!]
	\centering
	\includegraphics[width=0.9\linewidth]{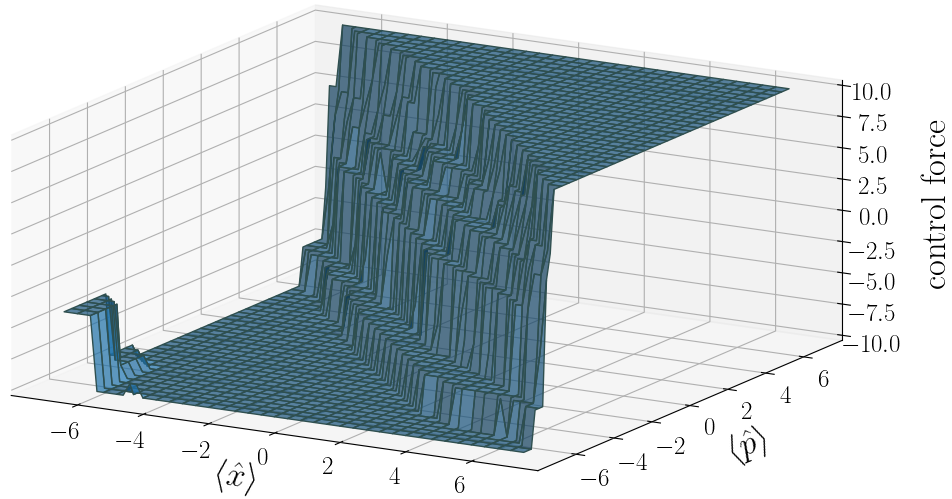}
	\caption{The control force of trained reinforcement learning plotted against $\left (\langle \hat{x}\rangle,\langle \hat{p}\rangle\right )$ input for the inverted harmonic potential problem. This is a finer and enlarged version of the right figure of Fig. \ref{fig:inverted response small}.}
	\label{fig:inverted response big}
\end{figure}

From the above figures we see that, the reinforcement learning controllers have learned qualitatively the same as the optimal control, and therefore we may say that they have learned the underlying properties of the system. However, some defects exist in their control decisions, which may be attributed to noise and insufficient training. Especially for the inverted oscillator problem, while the controller has 21 choices of control, it prefers outputting 4 or 5 choices around the central cliff in Fig. \ref{fig:inverted response small}, though all control choices are equispaced. This is shown more clearly in Fig. \ref{fig:inverted response big}. We suspect this to be an artefact of reinforcement learning. The reinforcement learning actor may discover when to use some of its control choices at the beginning, and then it converged to them rather than learning some other choices. This artefact may be suppressed by refinement of the training procedure, and the final performance may potentially increase. One reason for this artefact is that, in order to confirm that the performance of the neural network is lower than that of the optimal control, we need one thousand episodes of simulations, which amounts to $2\times10^6$ control steps, and therefore similarly, it is not easy for the reinforcement learning to discover that its strategy can still be improved. Also, the time horizon of control is set to be $\frac{1}{10}$ of an episode as discussed earlier, and therefore it may hard for the reinforcement learning to learn an optimal strategy on such long-time controls.\\ 

\section{Conclusion}\label{quadra conclusion}
In this chapter, we have shown that the problems of controlling a particle near the center of quadratic potentials with position measurement has optimal control solutions, and we have shown that a deep-reinforcement-learning-based controller can learn such a problem and control the system with a performance comparable to the optimal control. Also, the training of the reinforcement learning is within practical computational budgets, which is not specific to this quadratic case. We expect that similar reinforcement-learning-based controllers can be trained following the same line for other potentials where no optimal or reasonable control strategies are known, and we may carry the same settings to those other problems when possible. Therefore, naturally we move to the case of quartic potentials in the next chapter.\\

	\chapter{Control in Quartic Potentials}\label{control quartic potentials}
	In this chapter, we follow the same line as the last chapter to discuss controlling a quantum particle in a one-dimensional quartic potential, and the target of control is to keep the particle at a low energy. Unlike the last chapter, we do not consider inverted potential problems, because inverted anharmonic potentials are of less practical interest. We first show that the properties of the quartic system is totally different from the quadratic one and cannot be simplified in the same way in Section \ref{quartic breakdown}, and we derive some control strategies analogous to the optimal control of the quadratic case with some approximation in Section \ref{quartic controllers}. Our experimental settings to simulate quartic potentials are different from the quadratic case and we explain those settings in Section \ref{quartic experiments}. Based on the settings, we use deep reinforcement learning to control the system and present its control results in Section \ref{quartic performance section} and compare the results to our derived approximate control strategies. Finally, in Section \ref{quartic conclusion} we present the conclusions of this chapter.\\

\section{Analysis of a Quartic Potential}
In this section we show that a quartic potential cannot be simplified in the same way as a quadratic potential by revisiting the arguments presented in Section \ref{analysis of quadratic}. To obtain reasonable control strategies, we consider approximations to the time evolution of the system and present the corresponding derived controls, which serve as the comparison group as conventional control strategies. To make the properties of a quartic potential clear, we use figures to illustrate the behaviour of a particle inside a quartic potential before going into the next section.\\

\subsection{Breakdown of Gaussian Approximation and Effective Description}\label{quartic breakdown}
Following Section \ref{analysis of quadratic}, we consider the time evolutions of operators $\hat{x}$ and $\hat{p}$ in the Heisenberg picture for the Hamiltonian with a quartic potential:\\
\begin{equation}
H=\frac{\hat{p}^2}{2m}+\lambda\hat{x}^4,
\end{equation}
\begin{equation}\label{quartic operator evolution}
\frac{d}{dt}\hat{x}=\frac{i}{\hbar}[H,\hat{x}]=\frac{\hat{p}}{m},\qquad \frac{d}{dt}\hat{p}=\frac{i}{\hbar}[H,\hat{p}]=-4\lambda\hat{x}^3,
\end{equation}\\
which involves the third-order term of $\hat{x}$. Therefore, its time evolution is not a simple linear transformation between operators $\hat{x}$ and $\hat{p}$. It is nonlinear, and therefore the shape of its Wigner distribution is not preserved. Also, since the Hamiltonian is more than quadratic, the time-evolution equation of the Wigner distribution is different from the classical Liouville equation in phase space, and it is known that there exists no trajectory description of the Wigner distribution \cite{WignerAnharmonic}, which means that there is a non-classical effect.\\

We then consider the time evolution of the average momentum $\langle\hat{p}\rangle$. As shown in Eq. \nolinebreak(\ref{quartic operator evolution}), we have\\
\begin{equation}\label{quartic expected momentum evolution}
\frac{d}{dt}\langle\hat{p}\rangle=-4\lambda\langle\hat{x}^3\rangle,
\end{equation}\\
and if Gaussian approximation holds, $\langle\hat{x}^3\rangle$ can be calculated as in Eq. (\ref{skewness Gaussian}) exploiting the zero skewness property of the Gaussian. Therefore, we verify whether or not its zero skewness can always be kept zero as a Gaussian state in a quartic potential. The time evolution of skewness is\\
\begin{equation}
\begin{split}
d\left \langle\left (\hat{x}-\langle\hat{x}\rangle\right )^3\right \rangle&= \text{tr}\left (d\rho(\hat{x}-\langle\hat{x}\rangle)^3+\rho\, d(\hat{x}-\langle\hat{x}\rangle)^3\right )\\
&=\text{tr}\left (\frac{i}{\hbar}\rho\left [H,(\hat{x}-\langle\hat{x}\rangle)^3\right ] dt-3\rho\, (\hat{x}-\langle\hat{x}\rangle)^2 d\langle\hat{x}\rangle\right)\\
&=\text{tr}\left (\frac{i}{\hbar}\rho\left [\frac{\hat{p}^2}{2m},(\hat{x}-\langle\hat{x}\rangle)^3\right ] dt- \rho\,\dfrac{3\langle\hat{p}\rangle}{m} (\hat{x}-\langle\hat{x}\rangle)^2 dt\right )\\
&=\frac{3\left \langle(\hat{x}-\langle\hat{x}\rangle)\hat{p}(\hat{x}-\langle\hat{x}\rangle)\right\rangle}{m} dt- \dfrac{3\langle\hat{p}\rangle\langle(\hat{x}-\langle\hat{x}\rangle)^2\rangle}{m} dt\\
&=\frac{3\left \langle(\hat{x}-\langle\hat{x}\rangle)(\hat{p}-\langle\hat{p}\rangle)(\hat{x}-\langle\hat{x}\rangle)\right\rangle}{m} dt,
\end{split}
\end{equation}\\
which means that the skewness of the phase-space Wigner distribution in the $x$ direction is dependent on another skewness of it in the $x,p$ plane. We then calculate the time evolution of $\left \langle(\hat{x}-\langle\hat{x}\rangle)(\hat{p}-\langle\hat{p}\rangle)(\hat{x}-\langle\hat{x}\rangle)\right\rangle$:\\
\begin{equation}
\begin{split}
d\left \langle(\hat{x}-\langle\hat{x}\rangle)(\hat{p}-\langle\hat{p}\rangle)(\hat{x}-\langle\hat{x}\rangle)\right\rangle&=\text{tr} (d\rho(\hat{x}-\langle\hat{x}\rangle)(\hat{p}-\langle\hat{p}\rangle)(\hat{x}-\langle\hat{x}\rangle)\\
&\qquad\quad+\rho\, d(\hat{x}-\langle\hat{x}\rangle)(\hat{p}-\langle\hat{p}\rangle)(\hat{x}-\langle\hat{x}\rangle) )\\
&=\frac{2\left \langle(\hat{p}-\langle\hat{p}\rangle)(\hat{x}-\langle\hat{x}\rangle)(\hat{p}-\langle\hat{p}\rangle)\right\rangle }{m}dt\\
&\qquad\quad-4\lambda\left \langle\hat{x}^3(\hat{x}-\langle\hat{x}\rangle)^2\right \rangle dt+4\lambda \langle\hat{x}^3\rangle\left\langle(\hat{x}-\langle\hat{x}\rangle)^2\right \rangle dt\\
&=\frac{2\left \langle(\hat{p}-\langle\hat{p}\rangle)(\hat{x}-\langle\hat{x}\rangle)(\hat{p}-\langle\hat{p}\rangle)\right\rangle }{m}dt\\
&\qquad\quad-4\lambda\left \langle(\hat{x}^3-\langle\hat{x}^3\rangle)(\hat{x}-\langle\hat{x}\rangle)^2\right \rangle dt.
\end{split}
\end{equation}\\
The first term of the above is still a skewness, but the second term is rather nontrivial. This is because $\hat{x}^3-\langle\hat{x}^3\rangle\neq\left (\hat{x}-\langle\hat{x}\rangle\right )^3$. It appears to be non-zero, and we show this fact now. We evaluate the term $\left \langle(\hat{x}^3-\langle\hat{x}^3\rangle)(\hat{x}-\langle\hat{x}\rangle)^2\right \rangle$ for a Gaussian state:\\
\begin{equation}\label{5th moment 1}
\left \langle(\hat{x}^3-\langle\hat{x}^3\rangle)(\hat{x}-\langle\hat{x}\rangle)^2\right \rangle=\langle\hat{x}^5\rangle-2\langle\hat{x}^4\rangle\langle\hat{x}\rangle+\langle\hat{x}^3\rangle\langle\hat{x}\rangle^2-\langle\hat{x}^2\rangle\langle\hat{x}\rangle^3+\langle\hat{x}\rangle^5.
\end{equation}\\
Using the fact that a Gaussian distribution is central-symmetric and therefore has odd central moments being zero, similar to the skewness, we have $\langle(\hat{x}-\langle\hat{x}\rangle)^5\rangle=0$ and therefore\\
\begin{equation}
\langle\hat{x}^5\rangle=5\langle\hat{x}^4\rangle\langle\hat{x}\rangle-10\langle\hat{x}^3\rangle\langle\hat{x}\rangle^2+10\langle\hat{x}^2\rangle\langle\hat{x}\rangle^3-4\langle\hat{x}\rangle^5,
\end{equation}\\
and Eq (\ref{5th moment 1}) becomes\\
\begin{equation}\label{5th moment 2}
\left \langle(\hat{x}^3-\langle\hat{x}^3\rangle)(\hat{x}-\langle\hat{x}\rangle)^2\right \rangle = 3\langle\hat{x}^4\rangle\langle\hat{x}\rangle-9\langle\hat{x}^3\rangle\langle\hat{x}\rangle^2+9\langle\hat{x}^2\rangle\langle\hat{x}\rangle^3-3\langle\hat{x}\rangle^5.
\end{equation}\\
Using the fact that excess kurtosis is zero for a Gaussian distribution, i.e. $\dfrac{\langle(\hat{x}-\langle\hat{x}\rangle)^4\rangle}{V^2_x}\nolinebreak =\nolinebreak3$, where $V_x$ is the variance of $x$, we have\\
\begin{equation}\label{kurtosis Gaussian}
\langle\hat{x}^4\rangle=4\langle\hat{x}^3\rangle\langle\hat{x}\rangle-6\langle\hat{x}^2\rangle\langle\hat{x}\rangle^2+3\langle\hat{x}\rangle^4+3V_x^2,
\end{equation}\\
and Eq. (\ref{5th moment 2}) becomes\\
\begin{equation}\label{5th moment 3}
\begin{split}
\left \langle(\hat{x}^3-\langle\hat{x}^3\rangle)(\hat{x}-\langle\hat{x}\rangle)^2\right \rangle &= 3\langle\hat{x}^3\rangle\langle\hat{x}\rangle^2-9\langle\hat{x}^2\rangle\langle\hat{x}\rangle^3+6\langle\hat{x}\rangle^5+9V_x^2\langle\hat{x}\rangle\\
&= 3\left (3V_x\langle\hat{x}\rangle+\langle\hat{x}\rangle^3\right )\langle\hat{x}\rangle^2-9\langle\hat{x}\rangle^5+6\langle\hat{x}\rangle^5-9V_x\langle\hat{x}\rangle^3+9V_x^2\langle\hat{x}\rangle\\
&= 9V_x^2\langle\hat{x}\rangle.
\end{split}
\end{equation}\\
Therefore, whenever the mean $\langle\hat{x}\rangle$ of the Gaussian distribution is not zero, it gradually loses its Gaussianity in the quartic potential, and therefore the Wigner distribution cannot always be Gaussian.\\

Besides the failure of the Gaussian approximation, it is known that a quartic system is hard to analyse since it corresponds to the 1-dimensional $\phi^4$ theory \cite{anharmonicphi4}. Therefore we give up on analysing the system explicitly.\\

\subsection{Approximate Controls}\label{quartic controllers}
Although the state can no longer be described by only a few parameters, we still hope to obtain some reasonable control strategies for it. In this subsection we discuss possible control strategies.\\

If the system evolves deterministically, we can always find an appropriate control by simulating the state under the control and repeatedly refining the control to make the system evolve into a desired state. This local search method of control is actually widely used and it has many different variations and names as introduced in Chapter \ref{Introduction} \cite{QuantumOptimalControlTheory, CRAB, GRAPE, EvolutionaryQuantumControl}. In engineering, the famous ones include so-called differential dynamic programming (DDP) \cite{DDP} and the iterative linear-quadratic-Gaussian method (ILQG) \cite{ILQG}. However, these methods have strong limitations. First of all, since they search for possible controlled evolutions of a given state, if there are $N$ different given states, then they need to search $N$ times, using one for each, and if we do not know the given state in advance, we would need to do the search immediately after the state is given, which is time-consuming and inappropriate as a realistic control strategy. On the other hand, as these methods are based on local search, they can hardly be applied to systems that contain noise. This is because the methods investigate only one certain trajectory of the system and refine that investigated specific trajectory by back-and-forth iteration, which would result in a control that is inapplicable to other possibilities of noise. For a different random realization of noise, the controller does not have an idea about how to control the system and generally the control would fail. The control problem in this setting is called a stochastic optimal control (SOC) problem \cite{AICO}. To deal with this, the ILQG algorithm treats the effect of noise as small deviations from the expected trajectory and deals with the deviations perturbatively, but as already mentioned in the original paper, this method only handles a small noise that does not significantly affect the overall evolution of the system \cite{ILQG}. In our case, the quartic system has noise-induced behaviour, which makes an actual trajectory drastically different from the expected trajectory in the long term, and therefore this perturbative ILQG method can hardly apply. On the other hand, machine learning strategies such as the graphical-model-based approximate inference control (AICO) are often used to handle these stochastic control problems \cite{AICO}. Overall, if we want to cool down a particle in a quartic potential with position measurement and a random initial state, we actually have no existing well-established method that can exploit the properties of the system to control it, except for machine learning.\\

In this case, to obtain control strategies other than machine learning as a comparison group, we give up looking for usual globally optimal or locally optimal solutions and consider suboptimal controls and approximations of the system. As before, we use an external force to control the system, which is a linear term $F_{\text{con}}\hat{x}$ added to the system Hamiltonian that is parametrized by the control parameter $F_{\text{con}}$.
\subsubsection{damping controller}
As a first control strategy, we consider the time derivative of the system energy. It is\\
\begin{equation}
\begin{split}
dE=d\langle H\rangle=\text{tr}(d\rho\,H)&=\text{tr}(-\frac{i}{\hbar}[H+F_{\text{con}}\hat{x},\rho]H\,dt)=-\frac{F_{\text{con}}\langle\hat{p}\rangle}{m}dt,\\[6pt]
&\quad H=\frac{\hat{p}^2}{2m}+\lambda\hat{x}^4.
\end{split}
\end{equation}\\
Therefore, whenever $\langle\hat{p}\rangle>0$ is satisfied, we can use a positive $F_{\text{con}}$ to decrease the energy, and when $\langle\hat{p}\rangle<0$ is satisfied, we can use a negative $F_{\text{con}}$, which is the same for a classical system. This is called a steepest descent method. In numerical simulation, because we need to use finite time control steps, the action of this controller should be set to remove the particle's momentum at each control step, i.e. satisfying $\langle\hat{p}\rangle+d\langle\hat{p}\rangle=0$ with $dt$ taken to be the time of the control step. However, in numerical simulation we find that such a strategy would prevent the state from moving to the center and keep it at a high energy. Therefore, we use a damping strategy instead, i.e., $\langle\hat{p}\rangle+d\langle\hat{p}\rangle=(1-\zeta)\langle\hat{p}\rangle$ with $0\le\zeta<1$. We experimentally found that $\zeta=0.5$ results in the best performance in our problem setting and therefore we use the parameter setting $\zeta=0.5$. We call this the damping controller. Note that for a quadratic potential this strategy is not optimal, and it results in overdamping of the system. 
\subsubsection{quadratic controller}
As a second control strategy, we apply the usual linear-quadratic-Gaussian controller to the quartic system. This is because linear-quadratic-Gaussian controllers are often used in real situations where we have some quantities to minimize and do not care about the strict optimality of control \cite{LQGasApproximation}, and therefore, we just follow this strategy. To apply the linear-quadratic-Gaussian (LQG) theory we need to set a quadratic control target, define system variables, and linearise the system. After these procedures, the LQG theory produces a controller which is optimal on the transformed system, and we use this controller to control our original system. In analogy to the harmonic problem in Chapter \ref{control quadratic potentials}, we use $\langle\hat{x}\rangle$ and $\langle\hat{p}\rangle$ as our system variables, which actually only contain partial information of the system, and we define the minimized loss as $L=\int \left (\frac{k}{2}\langle\hat{x}\rangle^2+\frac{\langle\hat{p}\rangle^2}{2m}\right )dt$, where the parameter $k$ is searched and determined by testing the controller's performance. Following the arguments presented in Chapter \ref{control quadratic potentials}, with discretized control steps, the control force is determined such that $(\langle\hat{x}\rangle+d\langle\hat{x}\rangle,\langle\hat{p}\rangle+d\langle\hat{p}\rangle)$ satisfies the optimal trajectory condition $\langle\hat{p}\rangle=-\sqrt{km}\langle\hat{x}\rangle$ to the first order of $dt$, with $dt$ being the time of a control step. We call this derived controller as the quadratic controller, because it only considers observables that are sufficient for the quadratic problem case.
\subsubsection{Gaussian approximation controller}
As a fourth control strategy, we use a Gaussian approximation of the state and establish the correspondence between $(\langle\hat{x}\rangle,\langle\hat{p}\rangle)$ of the quantum state and $(x,p)$ of a classical particle, as done in the quadratic system case. By Eq. (\ref{quartic expected momentum evolution}), we know that the evolution of $\langle\hat{p}\rangle$ depends on $\langle\hat{x}^3\rangle$, which involves skewness and is not defined for a classical point-like particle. However, based on a Gaussian approximation, the state has zero skewness and $\langle\hat{x}^3\rangle=3V_x\langle\hat{x}\rangle+\langle\hat{x}\rangle^3$ as shown in Eq. (\ref{skewness Gaussian}). If we further assume that $V_x$ is a constant, we obtain a classical particle having $(x,p)$ corresponding to $(\langle\hat{x}\rangle,\langle\hat{p}\rangle)$, where $p$ evolves according to $\dfrac{dp}{dt}=-12\lambda V_xx-4\lambda x^3$, i.e. in a static potential $V=6\lambda V_x x^2+\lambda x^4$. Then by following the same Lagrangian argument in Section \ref{quadra optimal control}, we derive an optimal control protocol for the classical particle in this quartic potential. In this way we can go beyond the quadratic problem. Using Eq. (\ref{kurtosis Gaussian}), the loss under the above Gaussian approximation is\\
\begin{equation}\label{quartic Gaussian approx loss}
\begin{split}
L&=\int\left (\frac{\langle\hat{p}^2\rangle}{2m}+\lambda\langle\hat{x}^4\rangle\right )\,dt\\
&=\int\left (\frac{\langle\hat{p}\rangle^2+V_p}{2m}+\lambda(4\langle\hat{x}^3\rangle\langle\hat{x}\rangle-6\langle\hat{x}^2\rangle\langle\hat{x}\rangle^2+3\langle\hat{x}\rangle^4+3V_x^2)\right )\,dt\\
&=\int\left (\frac{\langle\hat{p}\rangle^2+V_p}{2m}+\lambda(6V_x\langle\hat{x}\rangle^2+\langle\hat{x}\rangle^4+3V_x^2)\right )\,dt,
\end{split}
\end{equation}\\
and we take $V_p$ and $V_x$ as constants. These above assumptions are expected to hold when the position measurement on the system is strong, so that the Gaussian measurement dominates the evolution of the shape of the state, and the quartic potential has little effect on the state. \\

To make the loss function as a classical action, we interpret the integrated term in Eq. (\ref{quartic Gaussian approx loss}) as a Lagrangian, with the potential term being $\mathcal{V}=-6\lambda V_x x^2-\lambda x^4$. Following the same argument as Section \ref{quadra optimal control}, the optimal trajectory should satisfy\\
\begin{equation}
p=-\sqrt{2m(6\lambda V_x+\lambda x^2)}x,
\end{equation}\\
so that the particle exactly stops at the top of the potential. We call this controller as our Gaussian approximation controller.\\

To implement it in numerical simulation, we proceed similarly as before to set the state $(\langle\hat{x}\rangle,\langle\hat{p}\rangle)$ on the optimal trajectory after applying a control step. Since the variance $V_x$ may actually change in time, at each step we calculate the current variance to determine the control.\\
\subsection{Behaviour of a Quartic Potential}\label{quartic potential demonstration}
Before proceeding into the next section, we use examples to demonstrate the typical behaviour of a quartic potential according to our numerical simulation. In our simulation, we initialize the state as a Gaussian wave packet near the center of the potential, and let it evolve in time. First we investigate the deterministic case with no position measurement imposed. The time evolution is plotted in Fig. \ref{wave evolution}. It can be seen that at the beginning the wave is localized in space and shows oscillatory behaviour inside the potential, with its center of mass moving forward and backward. However, as the wave slowly spreads and delocalize throughout time, the particle gradually spread in the bottom of the potential and delocalize, and the center of mass ceases to oscillate, as shown in Fig. \ref{quartic deterministic center of mass}.\clearpage
\begin{figure}[p]
	\begin{tikzpicture}
	\node[inner sep=0pt] (wave evolution1) at (0,0)
	{\centering
		\subfloat[]{
			\includegraphics[width=0.47\linewidth]{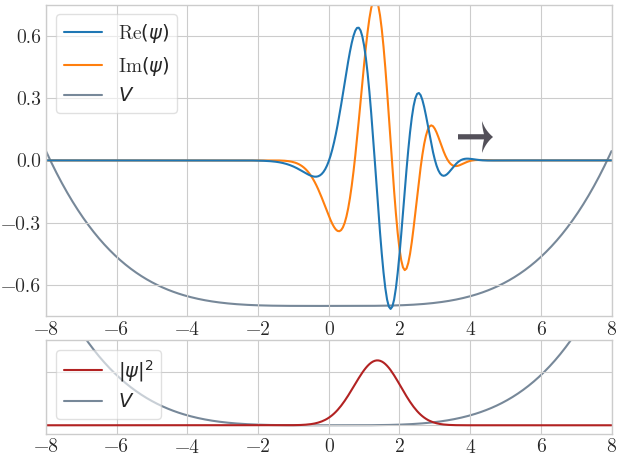}}};
	\node[inner sep=0pt] (wave evolution2) at (8.2,0)
	{\centering\subfloat[]{
			\includegraphics[width=0.47\linewidth]{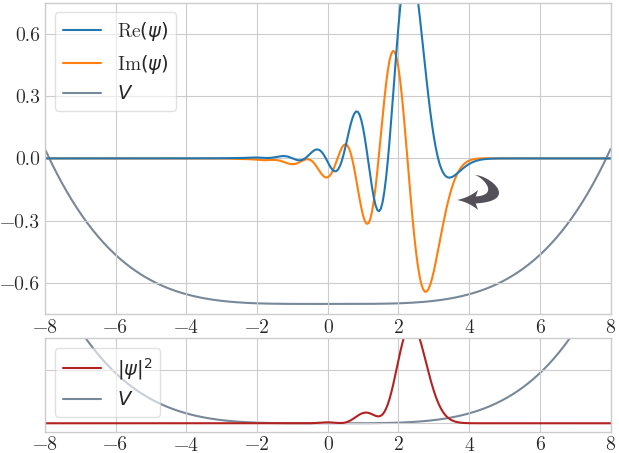}}};
	\draw[-stealth,thick] (wave evolution1.east) -- (wave evolution2.west);
		\node[inner sep=0pt] (wave evolution3) at (0,-6.8)
	{\centering\subfloat[]{
			\includegraphics[width=0.47\linewidth]{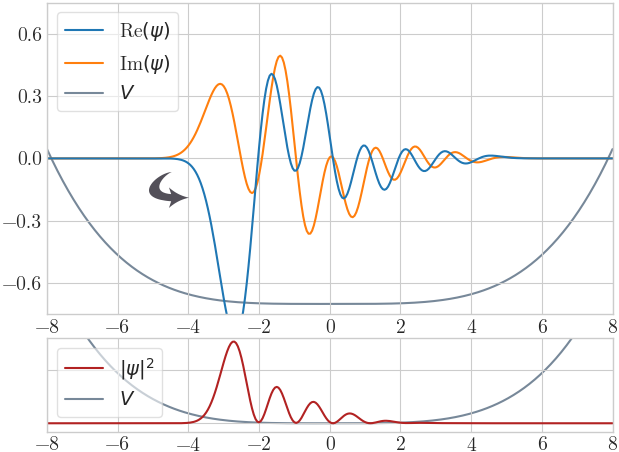}}};
	\draw[-stealth,thick] (wave evolution2.south west) -- (wave evolution3.north east);
			\node[inner sep=0pt] (wave evolution4) at (8.2,-6.8)
	{\centering\subfloat[]{
			\includegraphics[width=0.47\linewidth]{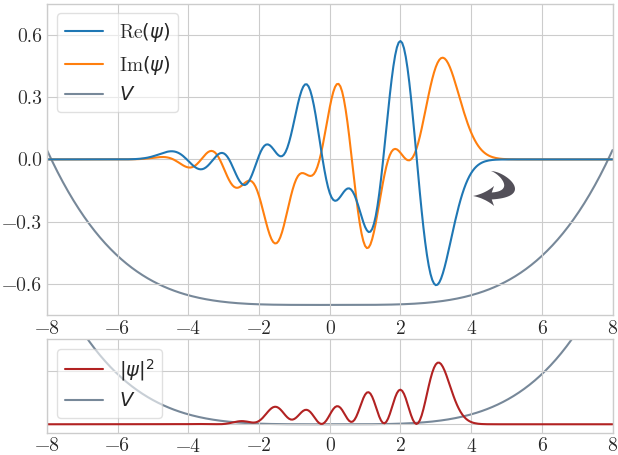}}};
	\draw[-stealth,thick] (wave evolution3.east) -- (wave evolution4.west);
				\node[inner sep=0pt] (wave evolution5) at (0,-13.8)
	{\centering\subfloat[]{
			\includegraphics[width=0.47\linewidth]{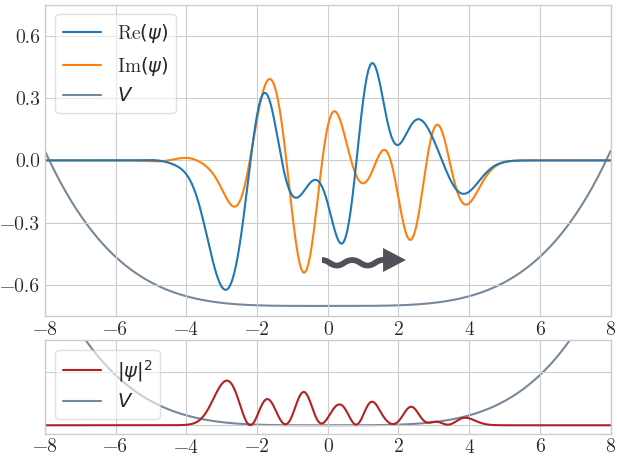}}};
	\draw[-stealth,thick] (wave evolution4.south west) -- (wave evolution5.north east);
					\node[inner sep=0pt] (wave evolution6) at (8.2,-13.8)
	{\centering\subfloat[]{
			\includegraphics[width=0.47\linewidth]{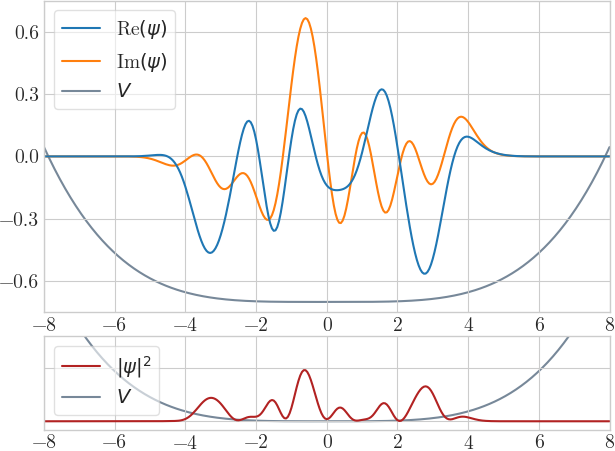}}};
	\draw[-stealth,thick] (wave evolution5.east) -- (wave evolution6.west);
	\end{tikzpicture}
	\caption{Deterministic time evolution in a quartic potential $V$ for a particle initialized as a Gaussian wave packet in (a), and panels (a) to (f) are ordered chronologically as indicated by the arrows. The grey arrows inside figures show the propagation directions of the density wave, and its movement of the center of mass becomes more and more obscured as time elapses. The particle gradually delocalize and loses its forward-backward oscillation in position space. In the plots, blue and orange curves show the real and imaginary parts of the wave function, read ones show the probability density distribution, and grey one show the quartic potential.}
	\label{wave evolution}
\end{figure}
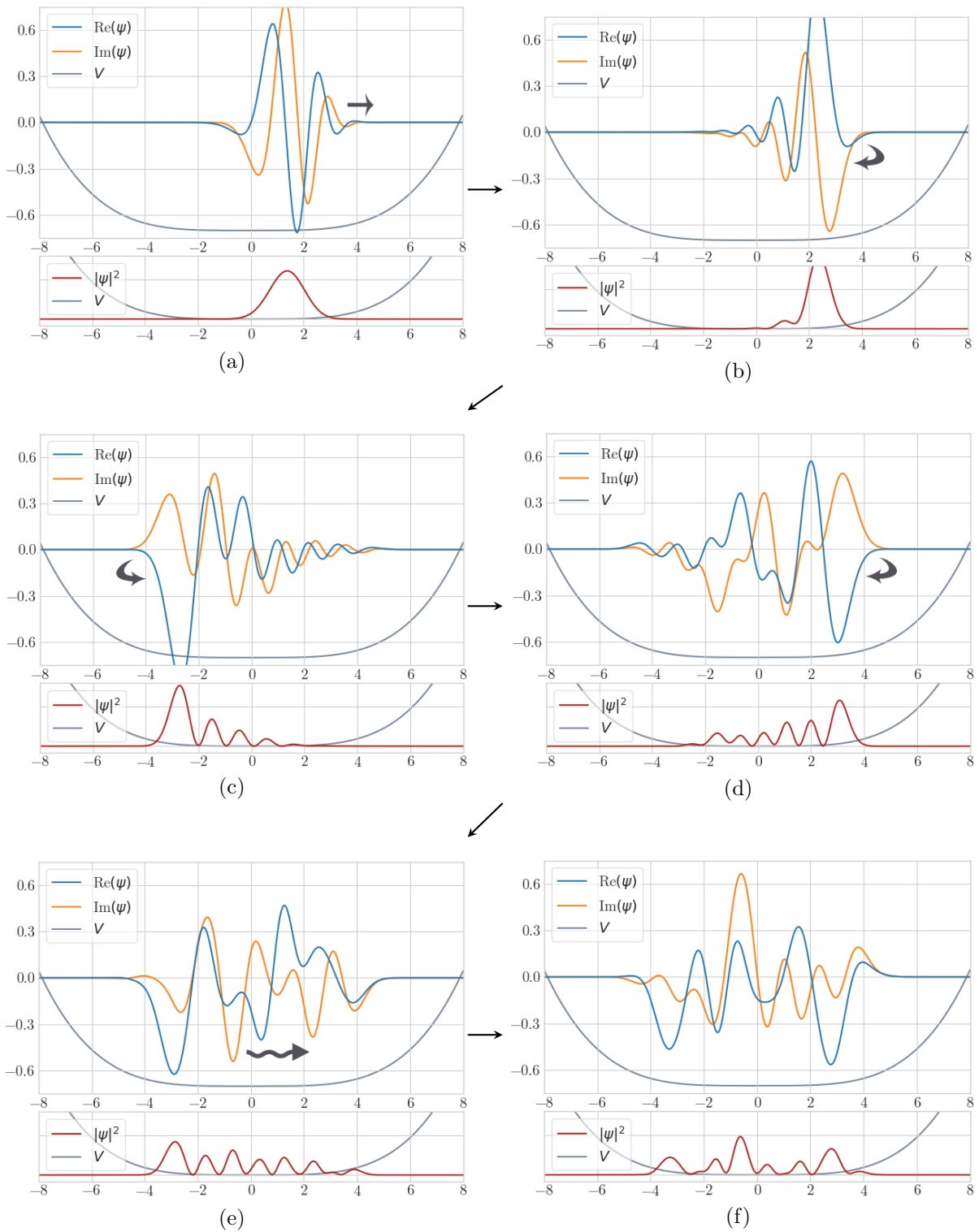\clearpage
\begin{figure}[t]
	\centering
	\includegraphics[width=0.6\linewidth]{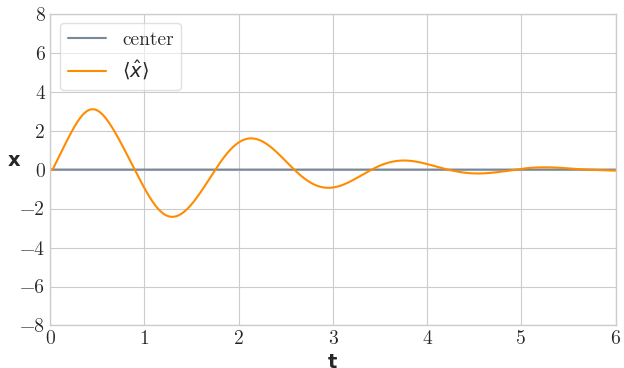}
	\caption{Time evolution of the average position $\langle\hat{x}\rangle$ of the wavefunction in Fig. \ref{wave evolution} (orange curve). The units of the time and position are consistent with those in the main text. As the state delocalizes, its center of mass gradually ceases to oscillate.}
	\label{quartic deterministic center of mass}
\end{figure}

The different behaviour from quadratic potentials is mainly due to the irregular potential shape. The bottom of the quartic potential is a plateau, and the wave function evolves almost freely. However, the two sides of the potential abruptly increase in value and therefore the wave front is strongly reflected at both ends and interfere with the tail of the wavefunction, which can be observed from the distribution density plots in Fig. \ref{wave evolution}(b) and \ref{wave evolution}(c). As the process repeats, the wavefunction becomes more and more delocalized and tend to resemble a wavefunction in an infinite-well potential, and its center-of-mass motion disappears. This is contrary to the quadratic case, where the shape of the wavefunction is preserved and no interference emerges.
\begin{figure}[b!]
	\centering
	\includegraphics[width=0.6\linewidth]{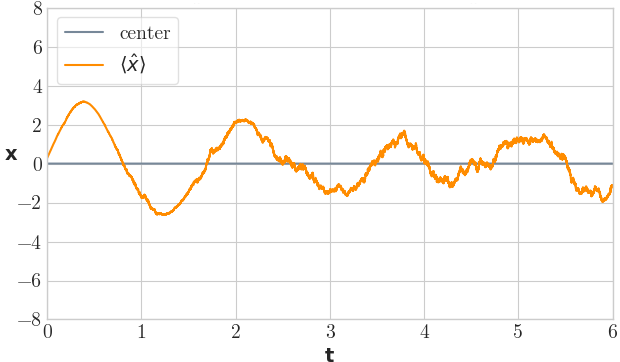}
	\caption{Time evolution of the average position $\langle\hat{x}\rangle$ of the wavefunction initialized as in Fig. \ref{quartic deterministic center of mass} but subject to continuous position measurement. This figure is to be compared with Fig. \ref{quartic deterministic center of mass}. At the beginning, the wavefunction is initialized with a low variance in position and is affected less by the measurement, but as it delocalizes, it experiences a stronger measurement effect and noise, and its center-of-mass oscillation is also increasingly more perturbed by the noise. Concerning the probability density distribution of the wavefunction, we can observe a rough ``envelope" which signifies the center of mass.}
	\label{quartic stochastic center of mass}
\end{figure}\\

On the other hand, when there is weak position measurement imposed, the properties of the system will change as illustrated in Fig. \ref{quartic stochastic center of mass}. The position measurement shrinks the wavefunction and makes the wavefunction imbalance at the central plateau of the potential, and therefore the center of the wavefunction slowly oscillates with noise, which is in contrast to the case of Fig. \ref{quartic deterministic center of mass}. Without measurement we observe no such long-term oscillatory behaviour. Due to this reason, it is indeed possible for those simple controllers described in Section \ref{quartic controllers} to successfully cool down the quartic system when the position measurement is imposed. Otherwise the situation would be extremely unclear, because without the measurement-induced effect both $\langle\hat{x}\rangle$ and $\langle\hat{p}\rangle$ tend to shrink to zero while the energy of the system is high.\\

\section{Numerical Experiments}\label{quartic experiments}
In this section we discuss our experimental setting for simulating the quartic system. Because the reinforcement learning implementation is almost the same as the quadratic case in Chapter \ref{control quadratic potentials}, we only briefly discuss the parameter settings and choices.
\begin{figure}[tb]
	\centering
	\subfloat[]{\label{quartic harmonic components}
		\includegraphics[width=0.48\linewidth]{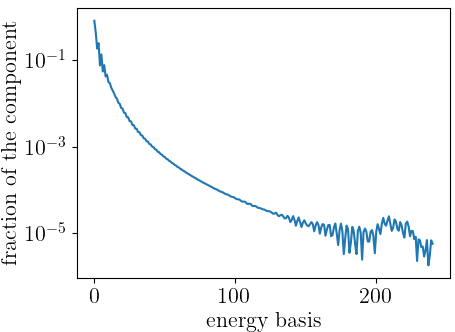}}\ 
	\subfloat[]{\label{quadratic harmonic components}
		\includegraphics[width=0.47\linewidth]{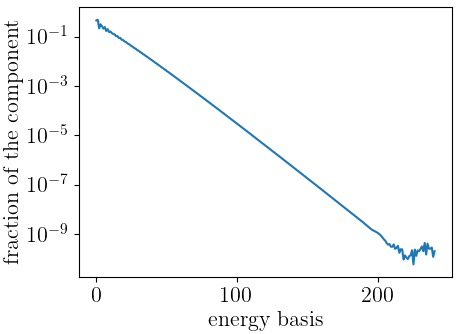}}
	\caption{Fraction of the component in the harmonic-oscillator eigenbasis for states in (a) quartic and (b) quadratic potentials. The states are subject to the continuous position measurement. The ordinate is plotted in a log scale.}
	\label{components in harmonic basis}
\end{figure}\\

\subsection{Simulation of the Quantum System}\label{quartic simulation}
The numerical simulation of a quartic potential is carried out in discretized real space. This is because we find that, as the wavefunction has become non-Gaussian, it is inefficient to simulate the wavefunction using the eigenbasis of a harmonic oscillator. We plot the norms of the wavefunction on high-energy components when it is simulated using the eigenbasis of a harmonic oscillator in Fig.~\ref{components in harmonic basis}(a), and compare it with a state evolving in a quadratic potential as plotted in Fig.~\ref{components in harmonic basis}(b). As the figure shows, for a state in a quadratic potential which is effectively Gaussian, the norms of its components on the high-energy part of the harmonic basis decrease exponentially, which is shown as the straight line in the log scale plot in Fig.~\ref{components in harmonic basis}(b). However, the components for a state in a quartic potential do not decrease exponentially as shown in Fig. \ref{components in harmonic basis}(a), and thus simulating the state using the harmonic basis is both inaccurate and inefficient. Therefore, we change to the usual finite-difference time-domain (FDTD) method, which discretizes both space and time to simulate the wave function. Our parameters and choices are described as follows.\\

We use a central difference method to evaluate the derivatives of the wavefunction in position space, i.e. to approximate the operators $\dfrac{\partial}{\partial x}$ and $\dfrac{\partial^2}{\partial x^2}$. The central difference method we use involves 9 points in total, and has $O(d^8)$ errors for the first- and second-order derivative evaluations with $d$ being the discretization grid size. The numerical evaluations of the derivatives are found to be fairly accurate by comparing the results with analytically calculated derivatives for Gaussian wave packets, with an error of around $10^{-5}$. Except for the discretization, for simplicity we keep most of our settings the same as those in the quadratic problems in Section~\ref{quadra experiments}, including the time step, the control step, the control force and the mass, and also the reference quantities $m_c$ and $\omega_c$ that define the units of mass and frequency. However, some other parameter settings need to be changed. As a starting point, we need to determine the quartic potential coefficient $\lambda$. To make the quartic system of a similar size as the quadratic system, we set the values of the harmonic potential and the quartic potential to be equal around position $x=3.5\left (\sqrt{\frac{\hbar}{m_c\omega_c}}\right )$\footnote{The expressions in parenthesis are physical units.}, and therefore we take the value $\lambda=\dfrac{\pi}{25}\left (\frac{m^2_c\omega^3_c}{\hbar}\right )$. Next, to avoid the numerical divergence problem that results from high energy modes on the discretized space grid and large potentials, we restrict our simulated space to be from $x=-8.5\left (\sqrt{\frac{\hbar}{m_c\omega_c}}\right )$ to $x=+8.5\left (\sqrt{\frac{\hbar}{m_c\omega_c}}\right )$, and we discretize it using a step of $d=0.1\left (\sqrt{\frac{\hbar}{m_c\omega_c}}\right )$. Note that because the quartic potential increases fast, the largest value of the simulated potential exceeds the harmonic potential at position $x=20\left(\sqrt{\frac{\hbar}{m_c\omega_c}}\right )$, and therefore a properly controlled and cooled wavefunction does not have enough energy to come close to the border of the space, which we have also confirmed in our numerical simulation.\\

Even with such a coarse grid and moderate potential values, the numerical simulation still diverges after tens of thousands of time steps. To make minimal additional modifications to handle this diverging problem, we add higher order correction terms concerning the Hamiltonian into our state-update equation of the numerical simulation. This is easily done by adding more Taylor-expansion terms $\sum_n\frac{1}{n!}(-{dt}\frac{i}{\hbar}H)^n|\psi\rangle$ with $dt$ being our simulation time step. In our case, we sum to the $5$-th order correction, and the numerical divergence disappears. It is important to note that such a correction does not necessarily increase the precision of the simulation, because when adding these additional terms, we do not consider their interactions with other non-Hamiltonian terms that would result in values of similar orders. On the other hand, the space discretization also limits our simulation precision. The additional higher order corrections of the Hamiltonian is only intended to alleviate the numerical divergence, not to increase an overall precision. To avoid large numerical errors, we claim that the simulation fails when its energy exceeds an amount of $20\ (\hbar\omega_c)$ or when the norm of the wavefunction exceeds $10^{-5}$ near the border of the space.\footnote{Specifically, we evaluate the value of the wavefunction on the 5th leftmost point and the 5th rightmost point on our space grid. This is because a point too close to the border is affected by the error coming from the finite-difference-based derivative estimation, and the wavefunction may not evolve correctly.} When such conditions are not violated, we numerically confirmed that our simulation of the wavefunction has an error below $0.05\%$\footnote{This is confirmed by simulating the same wavefunction using finer discretization steps and then comparing the simulation results. We only evaluated the deterministic evolution of the state, i.e., $\gamma=0$.} for simulation time $10\times\frac{1}{\omega_c}$, i.e., the time horizon of the reinforcement learning control.\\

The parameters used are summarized as follows:\\

\begin{table}[hbt]
	\centering
	\begin{tabular}{ccccccc}
		\toprule
		$\lambda$ $\left (\frac{m^2_c\omega^3_c}{\hbar}\right )$  &  $m$ ($m_c$)  &  $d$ $\left (\sqrt{\frac{\hbar}{m_c\omega_c}}\right )$  &  $x$ $\left (\sqrt{\frac{\hbar}{m_c\omega_c}}\right )$  &  $dt$ $\left (\frac{1}{\omega_c}\right )$   &  $\eta$  &  $\gamma$ $\left( \frac{m_c\omega_c^2}{\hbar}\right) $  \\[8pt]
		$\dfrac{\pi}{25}$  & $\dfrac{1}{\pi}$ & $0.1$ & $[-8, +8]$ & $\dfrac{1}{1440}$ & 1 & $0.01\pi$  \\ \bottomrule
	\end{tabular}\\[8pt]
	
	\begin{tabular}{ccc}
		\toprule
		$F_{\text{con}}$ ($(m_c\hbar)^\frac{1}{2}\omega_c^\frac{3}{2}$)  & $N_\text{con}$  & $t_{\text{max}}$ $\left (\frac{1}{\omega_c}\right )$ \\[8pt]
		$[-5\pi,+5\pi]$ & 18  & 100 \\ \bottomrule
	\end{tabular}\\
\caption{The parameter settings used in our quartic system simulations. The units are specified in parenthesis, and we use the same reference quantities $m_c$ and $\omega_c$ as Table \ref{quadra parameter settings}. When there is no unit, the quantity is dimensionless.}
\label{quartic parameter settings}
\end{table}
In the above Table \ref{quartic parameter settings}, the parameter $\lambda$ denotes the strength of our quartic potential, $m$ is the mass, $d$ is the discretization step of position space, $x$ represents the positions considered in our simulation, $dt$ is the time step, $\eta$ is measurement efficiency, and $\gamma$ is the measurement strength. This measurement strength is much smaller than the ones used in quadratic problems, because we want the wavefunction to be sufficiently non-local in our quartic potential. For the control part, $F_{\text{con}}$ is the external control force, $N_\text{con}$ is the number of control steps in a unit time $\frac{1}{\omega_c}$, and $t_{\text{max}}$ is the total time of a simulation episode. When we implement our derived conventional controllers as in Sec. \ref{quartic controllers}, we always discretize the control forces in the same way as the neural network controller before applying the forces, which is a strategy already mentioned in Sec.~\ref{quadra performance} concerning the discretized optimal control. \\

A final issue here is the initialization of the wavefunction. This is not a significant issue for the case of quadratic potentials, where the state quickly evolves into a Gaussian state with fixed covariances and follow simple time-evolution equations. However, as illustrated in Fig.~\ref{wave evolution}, we know that a freely evolving wavefunction in a quartic potential is typically non-Gaussian, and therefore, in order to initialize a ``typical" state in the quartic potential, we use a two-step initialization strategy. First, we initialize the state as a Gaussian wave packet as in Fig.~\ref{wave evolution}(a). Second, we let the state evolve in the potential for time $15\times\frac{1}{\omega_c}$ under the position measurement. If the resulting state has an energy below $18\ (\hbar\omega_c)$, we accept it as an initialized state that can be used in our simulation; otherwise we repeat the above initialization process to obtain other states. The initialization energy is dependent on the initial Gaussian wave packet, and we set its wave-number to be randomly chosen from the interval $[-0.4,+0.4]\ \left (\sqrt{\frac{m_c \omega_c}{\hbar}}\right )$, and we set its mean to be 0 and standard deviation to be $1\ \left (\sqrt{\frac{m_c\omega_c}{\hbar}}\right )$. The units are the same as those in Table~\ref{quartic parameter settings}, Fig.~\ref{wave evolution} and Fig.~\ref{quartic deterministic center of mass}. This initialization strategy produces initial energies of the states that range from $1$ to $18\ (\hbar\omega_c)$ with an average around $10\ (\hbar\omega_c)$. \\

When evaluating a controller's performance, since the oscillation behaviour is much slower compared with the quadratic case, we sample the energy of the controlled state per time $20\times\frac{1}{\omega_c}$ after the episode has started for time $40\times\frac{1}{\omega_c}$, i.e., using 3 samples in each simulated episode. Then we calculate the sample mean and the standard error.\\

\subsection{Implementation of Reinforcement Learning}
The reinforcement learning system is implemented basically in the same way as that described in Sec.~\ref{quadraReinforcementImplementation} for the quadratic problems. One difference here is that, we do not consider measurement-outcome-based controllers, since we know that they would result in lower performances as discussed in Sec.~\ref{quadra comparison}.\\

Similar to the case of quadratic problems, we consider two cases concerning the neural network inputs. While the second case is simply to input the wavefunction, the first case is no longer to input the quintuple $\left (\langle \hat{x}\rangle,\langle \hat{p}\rangle,V_x,V_p,C\right )$, because we know that these five values are not sufficient to describe the state. To proceed further, we input higher order moments of the phase space quasi-distribution of the state beyond the covariances. Namely, we input the third-order central moments $\langle \left (\hat{x}-\langle\hat{x}\rangle\right )^3\rangle$, $\langle \left (\hat{x}-\langle\hat{x}\rangle\right )\left (\hat{p}-\langle\hat{p}\rangle\right )\left (\hat{x}-\langle\hat{x}\rangle\right )\rangle$, $\langle \left (\hat{p}-\langle\hat{p}\rangle\right )\left (\hat{x}-\langle\hat{x}\rangle\right )\left (\hat{p}-\langle\hat{p}\rangle\right )\rangle$ and $\langle \left (\hat{p}-\langle\hat{p}\rangle\right )^3\rangle$ into the neural network, which are called skewnesses, and similarly, we input all forth- and fifth-order central moments. The resulting input contains 20 values in total, and this is used as the first input case for our quartic problem. For the wavefunction input case, as we know that the wavefunction near the borders of the simulated position space is less relevant and contains numerical error, we do not input wavefunction values near the border. Specifically, regarding the wavefunction amplitudes defined on spatial grids, we discard 15 wavefunction values on both the left and the right border, and use the rest as the neural network input, by separating the complex amplitudes into their real-valued parts and imaginary-valued parts. \\

The reward is defined to be $-2$ times the expected energy of the state in the quartic potential, shifted by $2$. Similar to the quadratic case in Sec.~\ref{quadraReinforcementImplementation}, the simulation is stopped when the energy is too high, with the threshold energy being $20\ (\hbar\omega_c)$. This keeps both the learned loss values reasonably small and the numerical simulation error of the system reasonably small, which is already mentioned in Sec.~\ref{quartic simulation}. The number of simulation episodes for training is set to be 12000, and the replay memory sizes of the two input cases are respectively set to contain 6000 and 2000 episodes that are of the length of $t_{\text{max}}$.\\

\section{Comparison of Reinforcement Learning and Conventional Control}\label{quartic performance section}
In this section we present the performances of the reinforcement learning controllers and the controllers described in Sec.~\ref{quartic controllers}. The results are summarized in Table. \ref{quartic results}. Unlike the quadratic case, we do not directly compare the behaviours of reinforcement learning controllers and conventional controllers as in Sec.~\ref{quadraInputResponse}, because the relevant inputs have become much more complicated and high-dimensional and can hardly be plotted.\\
\begin{table}[tb]
	\centering
	\begin{tabular}{cp{8em}p{6.5em}}
		\toprule
		& &  cooling quartic oscillators  \\ \midrule
		\multirow{2}{3.7em}{Network Input} &  \raggedright phase space distri-\linebreak bution moments \linebreak (up to the fifth) &  $0.7393\pm0.0003$ \tabularnewline \cmidrule{2-3}
		& wavefunction &  $0.7575\pm0.0009$ \\ \midrule
		\multirow{3}{3.7em}{Derived Controls} & damping control & $0.7716\pm0.0024$ \\ \cmidrule{2-3}
		& quadratic control & $0.8211 \pm 0.0031$ \\ \cmidrule{2-3}
		& Gaussian approximation & $0.7562\pm0.0010$ \\ \midrule
		\multicolumn{2}{l}{Ground State Energy} & $0.7177$ \\ \bottomrule
	\end{tabular}
	\caption{Performance results for the problem of cooling a quartic oscillator. The numbers after the $\pm$ signs show the estimated standard deviations for the above reported means. The values show the average energies of the controlled states in units of $\hbar\omega_c$.}
	\label{quartic results}
\end{table}

The damping controller uses a damping factor of 0.5, and the quadratic controller uses a parameter $k=2\times\lambda\cdot\frac{\hbar}{m_c\omega_c}$, which are determined by a rough grid search of the possible values. The ground-state energy is calculated in our discretized and bounded position space, by direct diagonalization of the system Hamiltonian.\footnote{For completeness, the first three excited states have energies that are given by 2.5718, 5.0463 and 7.8816 in units of $\hbar\omega_c$. It can be seen that the controlled states are very close to the ground state.} In Table \ref{quartic results}, it can be seen that the first input case of the reinforcement learning controller performs best, clearly better than all the other controllers by many times of standard deviations. Also we find that the energy variance of its controlled states is much smaller than other controllers, which means that the control is very stable, and results in an extremely small standard error as given in Table \ref{quartic results}. Compared with the first input case, the second input case performs comparatively worse, probably because the neural network cannot easily evaluate the relevant physical observables using the wavefunction, as physical observables are always quadratic in terms of the wavefunctions. This difficulty typically induces noise and small error in an AI's decisions, and can lower its final performance slightly. In fact, we indeed find a larger variance of the energies of the controlled states, which supports the above interpretation.\\

\section{Conclusion}\label{quartic conclusion}
In this chapter, we have shown that the problem of cooling a particle in a one-dimensional quartic potential can be learned and solved well by a deep-reinforcement-learning-based controller, and the controller can outperform all conventional non-machine-learning strategies that we have considered, which serves as an example of application of deep reinforcement learning to physics. We have also found that when relevant information of the input cannot be extracted by the neural network precisely and easily, the neural network typically performs slightly worse. This fact suggests that adaptation and modification of the neural network architecture for specific physical problems may increase the final performance of the AI, which is a possible direction of future research concerning applying deep learning to physics. \\

On the other hand, we find that this cooling problem for quartic potentials does not significantly demonstrate a superior performance of deep reinforcement learning. This is probably due to the fact that all the controllers have cooled the states to have energies that are very close to the ground state, and as can be observed experimentally, all the resulting wavefunctions have become very similar in shape, which suggests that a perturbative solution to the problem can probably work very well after all, and that the system eventually becomes quite static and simple. Therefore, the complexity of the properties of the quartic potential does not necessarily imply that conventional strategies cannot cool the system well, and in order to demonstrate significant superior performance of deep reinforcement learning, we actually need problems that are more complicated than this cooling task.
	\chapter{Conclusions and Future Perspectives}\label{summary}
	In this thesis, based on the formulation of continuous measurement and the deep reinforcement learning technology which are reviewed in Chapters \ref{continuous quantum measurement} and \ref{deep reinforcement learning}, we have investigated the control problems of one particle in one-dimensional quadratic and quartic potentials subject to continuous position measurement, and showed that the performance of deep-reinforcement-learning based control is comparable to or better than existing non-machine-learning strategies. Although this research is not the first application of deep reinforcement learning to quantum control, this research has a lot of important implications which we will summarise and discuss in this chapter.\\

In our research we have shown that when the reinforcement learning controller is correctly configured and sufficiently trained, with the recent developments of reinforcement learning technology it can achieve a performance that is satisfactory enough when compared with known control strategies. Even when we give the controller redundant information but not the directly relevant information as its input, such as the wavefunction or the measurement outcomes, the reinforcement-learning controller is still able to extract relevant physical information from its input to output a reasonable control, though it may need more training. This is especially useful when we do not know what physical quantities are relevant. Even when we do not understand how to simplify or interpret the problem, the AI still gives us an answer that can be verified, which is of great practical importance for controlling complex quantum systems that we do not understand well. Although previous researches also proceeded along this line, as discussed in Chapter \ref{Introduction}, they mainly focused on open-loop controls in which the state evolves deterministically and the AI only searches for one controlled trajectory of the state. However, for such open-loop control problems, there are already plenty of efficient search algorithms that can be used to find the trajectory and AI technology may not be necessary. On the other hand, in our problem setting the state evolves stochastically due to measurement backaction, and therefore the controller is a measurement-feedback closed-loop controller, since it reads the post-measurement state which is conditioned on the measurement. Measurement is useful in real circumstances as it acquires information of the state and purifies the state, and in order to accompany measurement, a closed-loop controller that works with measurement-induced stochasticity is necessary. This is where a reinforcement learning controller can be used, and explains the uniqueness and importance of our research. \\

In reality, AI-based real-time closed-loop controllers are indeed feasible by embedding the AI computation into a programmed microchip to work sufficiently fast \cite{embeddedDeepLearning}, such as the application-specific integrated circuit (ASIC) and field-programmable gate array (FPGA), which respond and operate in a time scale of microseconds and nanoseconds \cite{deeplearningChip1, deeplearningChip2}. In this way, AI-based controllers can operate around the same time scale as the underlying quantum systems and become realizable in experiments and real-world quantum devices, which may help the development of quantum technology in a direct manner. If we do not consider using the AI controller directly, we may as well design our own control methods based on the AI's learned strategy, similar to what is done in Ref. \cite{SpinControlDeepLearning} concerning spin-chain controls. \\

From a theoretical perspective, we may do reverse engineering on the AI controller to discover underlying properties of the controlled system. We may analyse the system under control to gain insights on how the system behaves and develop effective theories to describe it and explain AI's control. That is, we start from the answer of the control problem obtained from the AI and proceed to find the reason for such an answer. In this manner AI strategies can open up a new way of research on complex quantum systems and bring a lot of new possibilities. Although we have not done this for our quartic systems in Chapter \ref{control quartic potentials}, we expect that it is possible, and this reverse-engineering strategy may give us a lot of insights on the quartic system since we almost have no other methods or lines of research to follow to analyse the system behaviour. The same logic applies to other complex systems, and this is why we believe AI strategies bring us new possibilities of research on complex quantum systems. \\

Besides the above implications concerning physics, there are also new questions to answer from the perspective of deep reinforcement learning. In our research, we simulated the quantum system for $10^4$ episodes, i.e. around $10^5\sim 10^6$ oscillations, to train the reinforcement learning controller on the quantum control problem, which is of a considerable computational cost. It would be desirable if we can reduce the number of simulations by one or two orders of magnitude while achieving a similar performance of reinforcement learning, which asks for a better simulation or training strategy. This can be a future research direction, improving the implementation of reinforcement learning on physics; otherwise it would not be feasible to train reinforcement learning AI to control quantum systems that require a long time to simulate. Also, there are possibilities to suitably modify reinforcement learning systems for underlying physical problems to facilitate the learning. For example, we know that any expectation value of physical observables are of a quadratic form regarding the state vector $|\psi\rangle$, as $\langle\psi|\hat{O}|\psi\rangle$. However, the deep neural networks we used are only composed of linear mappings and rectifier units, which cannot express quadratic forms straightforwardly. Thus, a learning system taking the state vector as an input may work more efficiently if we design a new structure of its neural network to make it learn quadratic forms. Similar optimizations of the learning system can be many, and deep reinforcement learning has not been well adapted to physics yet.\\

On the whole, our research has demonstrated new possibilities of applying deep reinforcement learning to quantum physics and real-world applications. There are many new questions to answer and problems to solve concerning the interplay between deep leaning and physics, and in the future, AI technology may hopefully help both real-world problems and theoretical physical research.
	
	\renewcommand\chaptername{Appendix}
	\begin{appendices}
		\chapter{Linear-Quadratic-Gaussian Control}\label{LQG appendix}
		In this appendix we review the linear-quadratic-Gaussian (LQG) control theory, which enables us to conduct an optimal control in its specific LQG problem setting. We first discuss the linear-quadratic control problem without a Gaussian noise in Section \ref{linear quadratic regulator}, and then discuss a situation in which a Gaussian noise is added in Section \ref{linear quadratic Gaussian appendix section}, which completes the LQG theory. Our notations follow Ref. \cite{LinearQuadraticControl}. \\

\section{Deterministic Linear-Quadratic Control}\label{linear quadratic regulator}
The controller for the linear-quadratic control problem is called a \textit{linear-quadratic regulator} (LQR) \cite{LinearQuadraticControl}, and we introduce its problem setting and derivations in this section.\\

\subsection{Problem Setting and the Quadratic Optimal Cost}
As introduced in Chapter \ref{Introduction}, a control problem is concerned with a controlled system state, which is denoted as $x$, and a control parameter, which is denoted as $u$. Here $x$ and $u$ are vectors and we do not write them in boldfaces for simplicity. The linear-quadratic problem involves a linear system in the sense that the time evolution of $x$ is linear with respect to combined $x$ and $u$, i.e. \\
\begin{equation}\label{linear control equation}
dx=Fx\,dt+Gu\,dt,
\end{equation}\\
where $F$ and $G$ are matrices, and we have suppressed time dependences of all the variables $F$, $G$, $x$ and $u$. Then, the control problem considers minimization of a quadratic control cost\footnote{This quantity has many nomenclatures. It is also called a performance index or score or cost-to-go.} accumulated from the start of the control at time $t_0$ to the end time $T$, which is formulated as\\
\begin{equation}\label{quadratic cost equation}
V(x(t_0), u(\cdot), t_0)=\int_{t_0}^{T}\left (u^{\dagger}Ru+x^{\dagger}Qx\right )dt+x^{\dagger}(T)Ax(T),
\end{equation}\\
where $Q$ and $A$ are non-negative matrices and $R$ is a positive matrix, and therefore we have $V\ge0$. In the above equation, $V$ is dependent only on the initial condition of $x$, i.e., $x(t_0)$, because the time-evolution equation of $x$ is determined once we know the control trajectory $u(\cdot)$. From now on we always assume that the control stops at time $T$, while the starting time of control $t_0$ may be changed. \\

From Eq.~(\ref{linear control equation}) we can conclude that the controlled system $x$ is memoryless, and therefore for any consistent control strategy, the control $u$ must be a function that is only dependent on the current values of $x$ and $t$. Let us denote a possible optimal control strategy of the problem by $u^*$, which minimizes the control cost $V$. The minimized cost is then denoted by $V^*(x(t_0),t_0)$, where $u^*$ has been totally determined by $x$ and $t$ and therefore disappeared. Now we prove that $V^*$ must have the form of\\
\begin{equation*}
V^*(x(t),t)=x^{\dagger}(t)P(t)x(t),
\end{equation*}\\
where matrix $P(t)$ is independent of $x$, and is symmetric (self-conjugate) and non-negative. First, we show that for a constant $\lambda$, we have $V^*(\lambda x,t)=|\lambda|^2V^*( x,t)$. This can be shown as follows. First,\\
\begin{equation}\label{quadratic form 1}
\begin{split}
V^*(\lambda x,t)&=V(\lambda x,u^*_{\lambda x}(\cdot),t)\\
&\le V(\lambda x,\lambda u^*_{x}(\cdot),t)\\
&=|\lambda|^2V( x, u^*_{x}(\cdot),t)\\
&=|\lambda|^2 V^*( x,t),
\end{split}
\end{equation}\\
where we have used $u^*_{x}(\cdot)$ to denote the optimal control for an initial state $x$ and $u^*_{\lambda x}(\cdot)$ for an initial state $\lambda x$. Using the optimality of control $u^*_{\lambda x}(\cdot)$ regarding the initial state $\lambda x$ we proceed from the first line to the second line, and using the linearity of Eq. (\ref{linear control equation}) and the definition of $V$ in Eq. (\ref{quadratic cost equation}) we proceed from the second line to the third line. Note that we have suppressed the time argument of the initial condition. Similarly, we have\\
\begin{equation}\label{quadratic form 2}
|\lambda|^2 V^*( x,t)\le |\lambda|^2 V( x,\lambda^{-1}u^*_{\lambda x}(\cdot),t)=V( \lambda x,u^*_{\lambda x}(\cdot),t)=V^*(\lambda x,t),
\end{equation}\\
and therefore we have\\
\begin{equation}\label{quadratic form condition 1}
V^*(\lambda x,t)=|\lambda|^2V^*( x,t).
\end{equation}\\
Secondly, we show that $V^*(x_1,t)+V^*(x_2,t)=\frac{1}{2}\left[V^*(x_1+x_2,t)+V^*(x_1-x_2,t)\right]$. It is done similarly using the linearity of trajectory $x(\cdot)$ regarding the initial condition $x(t)$ and control $u(\cdot)$ and using the quadratic property of $V$:\\
\begin{equation}\label{quadratic form 3}
\begin{split}
V^*(x_1,t)+V^*(x_2,t)&=\frac{1}{4}\left[V^*(2x_1,t)+V^*(2x_2,t)\right]\\
&\le\frac{1}{4}\left[V((x_1+x_2)+(x_1-x_2),u^*_{x_1+x_2}+u^*_{x_1-x_2},t)\right.\\
&\qquad\left.+V((x_1+x_2)-(x_1-x_2),u^*_{x_1+x_2}-u^*_{x_1-x_2},t)\right].\\
\end{split}
\end{equation}\\
From Eq. (\ref{linear control equation}), we know that for an initial state $x_1(t_0)$ and control $u_1(\cdot)$ we can obtain a trajectory $x_1(\cdot)$, and if there is another initial state $x_2(t_0)$ and control $u_2(\cdot)$ and a resulting trajectory $x_2(\cdot)$, we can obtain a third valid trajectory $k_1x_1(\cdot)+k_2x_2(\cdot)$ with control $k_1u_1(\cdot)+k_2u_2(\cdot)$, which is the linearity. By Eq. (\ref{quadratic cost equation}) we have\\
\begin{equation}
\begin{split}
V((x_1+x_2),(u_1+u_2)(\cdot),t)&=\int_{t}^{T}\left [(u_1+u_2)^{\dagger}(s)R(u_1+u_2)(s)\right.\\
&\qquad\left.+(x_1+x_2)^{\dagger}(s)Q(x_1+x_2)(s)\right ]ds\\
&\qquad+(x_1+x_2)^{\dagger}(T)A(x_1+x_2)(T),\\[6pt]
V((x_1-x_2),(u_1-u_2)(\cdot),t)&=\int_{t}^{T}\left [(u_1-u_2)^{\dagger}(s)R(u_1-u_2)(s)\right.\\
&\qquad\left.+(x_1-x_2)^{\dagger}(s)Q(x_1-x_2)(s)\right ]ds\\
&\qquad+(x_1-x_2)^{\dagger}(T)A(x_1-x_2)(T),
\end{split}
\end{equation}
and therefore\\
\begin{equation}
\begin{split}
V((x_1+x_2),(u_1+u_2)(\cdot),t)&+V((x_1-x_2),(u_1-u_2)(\cdot),t)\\
&=2\int_{t}^{T}\left (u_1^{\dagger}Ru_1+x_1^{\dagger}Qx_1+u_2^{\dagger}Ru_2+x_2^{\dagger}Qx_2\right )ds\\
&\qquad+2x_1^{\dagger}(T)Ax_1(T)+2x_2^{\dagger}(T)Ax_2(T)\\
&=2V(x_1,u_1(\cdot),t)+2V(x_2,u_2(\cdot),t).
\end{split}
\end{equation}\\
Note that the cancellation of the $x_1(\cdot)$ and $x_2(\cdot)$ terms is possible due to the fact that they are the trajectories resulting from initial conditions $x_1$ and $x_2$ and controls $u_1$ and $u_2$, and therefore they are already fully determined. Using the above formula we can transform Eq. (\ref{quadratic form 3}) into\\
\begin{equation}\label{quadratic form 4}
\begin{split}
V^*(x_1,t)+V^*(x_2,t)&\le\frac{1}{2}\left[V(x_1+x_2,u^*_{x_1+x_2},t)+V(x_1-x_2,u^*_{x_1-x_2},t)\right]\\
&=\frac{1}{2}\left[V^*(x_1+x_2,t)+V^*(x_1-x_2,t)\right].
\end{split}
\end{equation}\\
Next we perform a change of variables:\\
\begin{equation}\label{quadratic form 5}
\begin{split}
\frac{1}{2}\left[V^*(x_1+x_2,t)+V^*(x_1-x_2,t)\right]&\le\frac{1}{4}\left[V^*(2x_1,t)+V^*(2x_2,t)\right]\\
&=V^*(x_1,t)+V^*(x_2,t).
\end{split}
\end{equation}\\
Therefore, we have obtained \\
\begin{equation}\label{quadratic form condition 2}
V^*(x_1,t)+V^*(x_2,t)=\frac{1}{2}\left[V^*(x_1+x_2,t)+V^*(x_1-x_2,t)\right],
\end{equation}\\
and together with Eq. (\ref{quadratic form condition 1}) we have the result that $V^*$ takes a quadratic form with respect to $x$, as shown in Ref. \cite{LinearQuadraticControl}. Therefore it can be written as \\
\begin{equation}\label{deterministic control cost}
V^*(x,t)=x^{\dagger}P(t)x,
\end{equation}\\
where we have denoted the moment to start the control as $t$ rather than $t_0$, and this is because we will do variations on this variable later. The matrix $P(t)$ above is symmetric and non-negative and independent of $x$. Its symmetric property can be assumed without loss of generality because we can equivalently rewrite it as $\frac{1}{2}\left (P+P^{\dagger}\right )$, and its non-negativity follows from the non-negativity of $V$. Note that everything here is implicitly dependent on the ending time $T$ of the control.\\

\subsection{The Hamilton-Jacobi-Bellman Equation and the Optimal Control}\label{quadratic optimal control appendix}
To solve for the matrix $P(t)$, we need to introduce the Hamilton-Jacobi-Bellman (HJB) equation. This equation is in essence similar to the Bellman equation for the Q-function in Section \ref{Q learning}, but it has no discount factor and deals with continuous space. To proceed, we first derive this HJB equation.\\

For an optimal control cost $V^*(x(t),t)$, by its optimality its control $u^*$ should minimize its own expectation of its control cost, and therefore for an infinitesimal time $dt$ it satisfies\\
\begin{equation}\label{HJB 1}
V^*(x(t),t)=\min_u\left \{V^*(x(t+dt),t+dt) +\ell(x(t),u) dt \right \},
\end{equation}\\
where $\ell$ is the integrated term in the control cost (i.e., $u^{\dagger}Ru+x^{\dagger}Qx$ in Eq. (\ref{quadratic cost equation})), and the state $x(t+dt)$ is dependent on the control $u$, introducing a non-trivial $u$ dependence into the minimization. By the Taylor-expansion of $V^*(x(t+dt),t+dt)$ up to the first order of $dt$, we obtain\\
\begin{equation}\label{HJB 2}
V^*(x(t+dt),t+dt)=V^*(x(t),t)+\partial_xV^*(x(t),t)\,dx+\partial_tV^*(x(t),t)\,dt,
\end{equation}\\
where $\partial_x$ and $\partial_t$ denote the differentiation with respect to the first and the second arguments of $V^*$. Therefore Eq. (\ref{HJB 1}) can be written as\\
\begin{equation}\label{HJB 3}
-\partial_tV^*(x(t),t)\,dt=\min_u\left \{\partial_xV^*(x(t),t)\,dx +\ell(x(t),u)dt \right \}.
\end{equation}
If we express the time-evolution equation of $x$ as $\dfrac{dx}{dt}=f(x,u)$, then we have the HJB equation as\\
\begin{equation}\label{HJB}
-\partial_tV^*(x(t),t)=\min_u\left \{\partial_xV^*(x(t),t)f(x,u) +\ell(x(t),u) \right \}.
\end{equation}\\
Sometimes this equation is also written as\\
\begin{equation}\label{HJB2}
\dot{V}(x,t)+\min_u\left \{\nabla V(x,t)\cdot f(x,u)+\ell(x,u)\right \}=0,
\end{equation}\\
where the superscript stars are omitted, and the superscript dot and nabla denote $\partial_t$ and $\partial_x$.\\

Because the optimal control cost $V^*(x,t)$ is defined on the whole space of $x$ and $t$ and by definition $V^*$ is the smallest possible control cost, we have the fact that as long as $V^*$ exists, it is unique. Therefore, if we have a function $V$ that satisfies the HJB equation (Eq. (\ref{HJB})), it must be the unique optimal control cost $V^*$, and the HJB equation becomes a sufficient and necessary condition for optimal control. In addition, the optimal control strategy $u^*$ is to take the minimization of $u$ in the above HJB equation, which is possibly not unique. This sufficient and necessary condition provided by the HJB equation is important because it makes ``guessing" the optimal control possible.\\

After we have obtained the HJB equation, we now turn to solve the matrix $P(t)$ in Eq. (\ref{deterministic control cost}). By straightforward substitution, we have\footnote{In usual control theory, complex values are not considered. However, for consistency with quantum mechanics, we have generalized the derivation to allow complex values. Therefore, one or two terms in our derivation may be different from those in standard textbooks.}\\
\begin{equation}\label{LQR optimal 1}
\begin{split}
-x^{\dagger}\dot{P}x&=\min_u\left \{x^{\dagger}P(Fx+Gu)+(Fx+Gu)^{\dagger}Px+u^{\dagger}Ru + x^{\dagger}Qx \right \}\\
&=\min_u\left \{x^{\dagger}(PF+F^{\dagger}P)x + x^{\dagger}Qx+x^{\dagger}PGu+u^{\dagger}G^{\dagger}Px+u^{\dagger}Ru \right \}.
\end{split}
\end{equation}
We can arrange the $u$ dependent terms into a complete square to eliminate them through minimization:\\
\begin{equation}\label{LQR optimal 2}
\begin{split}
-x^{\dagger}\dot{P}x&=\min_u\left \{x^{\dagger}(PF+F^{\dagger}P)x + x^{\dagger}Qx+(u^{\dagger}+x^{\dagger}PGR^{-1})R(R^{-1}G^{\dagger}Px+u) \right.\\
&\qquad\qquad\left.-x^{\dagger}PGR^{-1}G^{\dagger}Px \right \}\\
&=x^{\dagger}(PF+F^{\dagger}P)x + x^{\dagger}Qx-x^{\dagger}PGR^{-1}G^{\dagger}Px,
\end{split}
\end{equation}
where we have assumed that $P$ is symmetric, or hermitian, and that $R$ is positive. The minimization in the above equation is completed by forcing the condition $R^{-1}G^{\dagger}Px+u=0$, and therefore the optimal control $u^*$ is \\
\begin{equation}\label{LQR control law}
u^*=-R^{-1}G^{\dagger}Px,
\end{equation}\\
where the matrix $P$ is obtained from the differential equation (\ref{LQR optimal 2}) and matrices $G$ and $R$ are defined in Eq. (\ref{linear control equation}) and (\ref{quadratic cost equation}). The above equation concerning $P$ can be further simplified by removing the arbitrary variable $x$ on both sides. Then we obtain the matrix Riccati equation:\\
\begin{equation}\label{LQR finite time}
-\dot{P}=PF+F^{\dagger}P -PGR^{-1}G^{\dagger}P+ Q.
\end{equation}\\
Since this equation is a differential equation, we still need to find its boundary condition to solve it. Recall that there is a final control-independent term $x^{\dagger}(T)Ax(T)$ in the definition of $V$ in Eq. (\ref{quadratic cost equation}), we obtain the boundary condition for $P(t)$, i.e. $P(T)=A$. In cases $A=0$, we have $P(T)=0$. This completes the optimal control for finite-horizon control problems, i.e., $T<\infty$.\\

Next we consider the case of $T\to\infty$. Although it is still possible to proceed with time dependent matrices $F$, $G$, $R$ and $Q$, for simplicity we require them to be constant in the following discussion. \\

When all the matrices in the definition of the problem are time-independent, the property of the controlled system becomes completely time-invariant if $T$ also goes to infinity. In this case, the control strategy $u$ should be consistent in time, i.e. time-independent, and thus $P$ cannot change with time\footnote{This result can also be obtained by analysing the differential equation and take the limit that the time is infinitely long.}, and we have $\dot{P}=0$, which yields the continuous-time algebraic Riccati equation (CARE):\\
\begin{equation}\label{continuous Riccati equation}
PF+F^{\dagger}P -PGR^{-1}G^{\dagger}P+ Q=0.
\end{equation}\\
Mathematical software usually provides numerical methods to solve this algebraic Riccati equation, and this equation is of great importance in engineering and control theory. We summarise the results for the infinite-time control case below.\\
\begin{equation}
dx=Fx\,dt+Gu\,dt,
\end{equation}

\begin{equation}
V(x(t_0), u(\cdot))=\int_{t_0}^{\infty}\left (u^{\dagger}Ru+x^{\dagger}Qx\right )dt,\quad R>0,Q\ge0,
\end{equation}

\begin{equation}
V^*(x(t_0))=x^{\dagger}(t_0)Px(t_0),\qquad P\ge0,
\end{equation}

\begin{equation}
u^*=-R^{-1}G^{\dagger}Px,
\end{equation}

\begin{equation}
PF+F^{\dagger}P -PGR^{-1}G^{\dagger}P+ Q=0,
\end{equation}\\
where $u^*$ is the optimal control, and $V^*$ is the optimal cost.\\

From the above equations, we can see that a finite choice of the matrix $R$ is necessary to produce a non-divergent $u^*$. Otherwise, when solving for $u^*$, typically its value goes to infinity. This is reasonable since the condition $R=0$ implies that the control is for free and we can use as large control as we want, which clearly does not produce finite control strengths. However, we may look at the conditions where the control $u^*$ is precisely zero, and interpret the infinite control strengths as pushing the state $x$ to these stationary points where $u^*$ is zero. This justifies our treatment of the control forces in Chapter \ref{control quadratic potentials} where we have assumed $R=0$. It can be seen that in this case, the time evolution of the system, i.e. matrix $F$, is no longer relevant for this control problem, and the optimal control is determined by the loss.\\

For completeness, we present the results of discrete linear systems with discrete controls, i.e., $x_n$ and $u_n$, without proof. The proof follows the same line as the above continuous case, and interested readers are referred to Ref. \cite{LinearQuadraticControl}.\\
\begin{equation}
x_{t+1}=Fx_{t}+Gu_{t},
\end{equation}

\begin{equation}
V(x_{t_0}, u_{(\cdot)}))=\sum_{t=t_0}^{\infty}\left (u_t^{\dagger}Ru_t+x_{t+1}^{\dagger}Qx_{t+1}\right ),\quad R>0,Q\ge0,
\end{equation}

\begin{equation}
V^*(x(t_0))=x^{\dagger}(t_0)Px(t_0),\qquad P\ge0,
\end{equation}

\begin{equation}
u^*=-\left (G^{\dagger}SG+R\right )^{-1}G^{\dagger}SFx,\qquad S = P+Q,
\end{equation}

\begin{equation}\label{discrete Riccati equation}
S=F^{\dagger}\left[S-SG\left(G^{\dagger}SG+R\right)^{-1}G^{\dagger}S\right]F+Q.
\end{equation}\\
The last equation \ref{discrete Riccati equation} is called the discrete-time algebraic Riccati equation (DARE).

\section{Linear-Quadratic Control with Gaussian Noise}\label{linear quadratic Gaussian appendix section}
When an additive Gaussian noise is introduced into the time-evolution equation, the optimal control strategy in Eqs. (\ref{LQR control law}) and (\ref{continuous Riccati equation}) is not changed. In this section we prove this result.\\

\subsection{Definitions of the Optimal Control and Control Cost}
When a Gaussian noise introduced, the time-evolution equation (\ref{linear control equation}) becomes\\
\begin{equation}\label{linear control Gaussian}
dx=Fx\,dt+Gu\,dt + \hat{Q}^{\frac{1}{2}}\,dW,
\end{equation}\\
where $dW$ is a multi-dimensional Wiener increment with the variance $dt$ for each dimension, and $\hat{Q}$ is a semidefinite covariance matrix. \\

Without noise, as long as the system is controllable, it is always possible to use $u$ to control the state $x$ to become 0 and leave both $x$ and $u$ zero afterwards to stop the accumulation of control cost, which shows that there always exist control strategies with finite control costs even when the limit $T\to\infty$ is taken, and therefore the optimal cost $V^*$ must also have a finite value. However, this is not true when noise is present. When the system contains noise, expectation values of the state typically converge to some steady values, and they are not zero. Therefore, the expected control cost $V$ would be proportional to the control time $T$ in the long term, and we have $\lim\limits_{T\to\infty}V\to\infty$ and $\lim\limits_{T\to\infty}\frac{1}{T}V\to c$ where $c$ is some constant. Therefore, we can neither consider a minimization of $\lim\limits_{T\to\infty}V$ nor $\lim\limits_{T\to\infty}\frac{1}{T}V$ to deduce the optimal control, since $\lim\limits_{T\to\infty}\frac{1}{T}V$ is never affected by any short-time change of control $u$. In this case, in order to define the optimal control for an infinite time horizon $T$, we investigate the convergence of the optimal control strategy $u^*$ at the large limit of $T$, rather than the convergence of the control cost $V^*$. Therefore we start from a finite $T$ to deduce $V^*$ and $u^*$.\\

In the presence of a Wiener increment, the derivation of the HJB equation (\ref{HJB}) should be modified. Especially, the Taylor expansion of $V^*(x(t+dt),t+dt)$ should be taken to the second order of $dx$. Note that the control cost $V$ is an expectation now. The Taylor expansion is\\
\begin{equation}\label{control cost stochastic expansion}
\begin{split}
V^*(x(t+dt),t+dt)&=E\left[V^*+\partial_tV^*dt+\partial_xV^*dx+\frac{1}{2}\partial_x^2V^*dx^2\right]\\
&=V^*+\partial_tV^*dt+\partial_xV^*(Fx\,dt+Gu\,dt)+\frac{1}{2}\text{tr}\left(\partial_x^2V^*\hat{Q}\right)\,dt,
\end{split}
\end{equation}
and the HJB equation (\ref{HJB}) becomes\\
\begin{equation}\label{HJB stochastic}
-\partial_tV^*(x(t),t)=\min_u\left \{\partial_xV^*(x(t),t)(Fx+Gu)+\frac{1}{2}\text{tr}\left(\partial_x^2V^*(x(t),t)\hat{Q}\right) +(u^{\dagger}Ru+x^{\dagger}Qx) \right \}.
\end{equation}\\
As discussed in Section \ref{quadratic optimal control appendix}, if the solution $V^*$ to the above equation exists, it must be unique. Therefore, we try to make up such a solution by referring to the results obtained in the previous section.\\

\subsection{The LQG Optimal Control}\label{LQG optimal result appendix}
As explained above, we expect $V$ to be proportional to time $T$. Since Eq. (\ref{HJB stochastic}) is only different from the deterministic case by the term $\dfrac{1}{2}\text{tr}\left(\partial_x^2V^*(x(t),t)\hat{Q}\right)$, without any change of the optimal control, we expect this term to be a constant, which is indeed the case for a $V^*$ quadratic in $x$. Its value is $\dfrac{1}{2}\text{tr}\left(2P(t)\hat{Q}\right)=\text{tr}\left(P(t)\hat{Q}\right)$. Recall that $P(t)$ is constrained by the continuous algebraic Riccati equation and boundary condition and is independent of the control $u$. Therefore, the minimization in Eq. (\ref{HJB stochastic}) concerning $u$ is not affected by the term $\text{tr}\left(P(t)\hat{Q}\right)$. Assuming that $\partial_xV^*$ is the same as the deterministic case, we can proceed similarly and reaches\\
\begin{equation}\label{HJB stochastic makeup 1}
-\partial_tV^*(x(t),t) = x^{\dagger}(PF+F^{\dagger}P)x + x^{\dagger}Qx-x^{\dagger}PGR^{-1}G^{\dagger}Px + \text{tr}\left(P(t)\hat{Q}\right),
\end{equation}\\
where we have used the optimal control $u^*=-R^{-1}G^{\dagger}Px$ as before.\\

The above formula clearly suggests an additional term in $V^*$ to represents the effect of $\text{tr}\left(P\hat{Q}\right)$, so that the original quadratic term $x^{\dagger}P(t)x$ can be kept unaltered. It can be constructed as follows:\\
\begin{equation}\label{HJB stochastic makeup 2}
V^*(x,t)=x^{\dagger}P(t)x+\int_{t}^{T}\text{tr}\left(P(s)\hat{Q}\right)ds,
\end{equation}\\
where $P(t)$ satisfies the finite-time Riccati equation (\ref{LQR finite time}). For this constructed $V^*$, Eq.~(\ref{HJB stochastic makeup 1}) is clearly satisfied, and therefore the HJB equation (\ref{HJB stochastic}) is also satisfied, so that the optimal control $u^*$ is a valid control. As a final step, we confirm the convergence behaviour of $u^*$ at large $T$.\\

The convergence of $u^*$ at large $T$ by definition depends on the convergence of the matrix $P(t)$. However, this is already shown in Section \ref{quadratic optimal control appendix}. We have known that in the deterministic case $P(t)$ loses its dependency on $t$ at the limit $T\to\infty$, and the finite-time Riccati equation that governs the evolution of $P(t)$ is the same as the equation for the stochastic case here. Therefore at the large limit of $T$, $P(t)$ also loses its time-dependence in this stochastic problem, and $u^*$ converges. The $P$ matrix finally satisfies the continuous-time algebraic Riccati equation (\ref{continuous Riccati equation}), which is time-independent. This completes our proof.

		\chapter{Numerical Simulation of Stochastic Differential Equation}\label{numerical simulation appendix}
		To train the reinforcement learning agent, we simulate the controlled quantum system for hundreds of thousands of oscillation periods, and therefore both efficiency and consistency are crucial. In this appendix we introduce stochastic It\^o--Taylor expansion and give the update rule of our numerical approximation in Eq. (\ref{order 1.5 explicit}) and (\ref{partial implicit}). The contents and notations follow Ref. \cite{NumericalSimulationofStochasticDE}.\\

\section{Stochastic It\^o--Taylor Expansion}
Let $W(t)$ be a Wiener process with time argument $t\in [0,T]$ that is randomly chosen and satisfies \\ 
\begin{equation}
W(t+dt)-W(t)\sim\mathcal{N}(0,dt),
\end{equation}\\
and the quantity $X_t$ evolves according to the It\^o stochastic differential equation \\
\begin{equation}\label{defining equation of X_t}
dX_t = a(X_t)\,dt+b(X_t)\,dW,
\end{equation}\\
where $a$ and $b$ are smooth functions differentiable up to some necessary orders, when their derivatives appear in the following context.\\

We consider a function of the variable $X_t$ as $f(X_t)$. By the It\^o formula:\\
\begin{equation}\label{Ito formula}
df(X_t) = \left(a(X_t)\frac{\partial}{\partial x}f(X_t) + \frac{1}{2}b^2(X_t)\frac{\partial^2}{\partial x^2}f(X_t) \right)dt + b(X_t)\frac{\partial}{\partial x}f(X_t)\,dW_t,
\end{equation}\\
where the partial differentials are taken with respect to the argument of function $f(\cdot)$, and\\ $dW_t\equiv W(t+dt)-W(t)$. All time dependences are indicated as subscripts. The It\^o formula can be considered as a Taylor expansion of $f(\cdot)$ on its argument using the first and the second terms subject to the relation $dW^2=dt$.\\

First, we rewrite the stochastic differential equations in their integral forms formally:\\
\begin{equation}\label{integral form of stochastic equation}
X_t = X_{0}+\int_{0}^{t}a(X_s)\,ds+\int_{0}^{t}b(X_s)\,dW_s,
\end{equation}\\
which is just the previous stochastic differential equation on $dX_t$. Similarly we have\\
\begin{equation}\label{integral form of f}
f(X_t) = f(X_0)+\int_{0}^{t}\left(a(X_s)f'(X_s) + \frac{1}{2}b^2(X_s)f''(X_s) \right)ds + \int_{0}^{t}b(X_s)f'(X_s)\,dW_s,
\end{equation}\\
where we have written differentiation with respect to its argument as primes, which we will also write in the form of $f^{(n)}$. We then apply the above equation to $a(X_t)$ and $b(X_t)$, and substitute them into Eq. (\ref{integral form of stochastic equation}), obtaining \\
\begin{equation}\label{first stochastic expansion}
\begin{split}
X_t &= X_{0}+\int_{0}^{t}a(X_{s_1})\,d{s_1}+\int_{0}^{t}b(X_{s_1})\,dW_{s_1}\\
&=X_{0}\\
+&\int_{0}^{t}\left(a(X_0)+\int_{0}^{s_1}\left(a(X_{s_2})a'(X_{s_2}) + \frac{1}{2}b^2(X_{s_2})a''(X_{s_2}) \right)ds_2 + \int_{0}^{s_1}b(X_{s_2})a'(X_{s_2})\,dW_{s_2}\right)\,d{s_1}\\
+&\int_{0}^{t}\left(b(X_0)+\int_{0}^{s_1}\left(a(X_{s_2})b'(X_{s_2}) + \frac{1}{2}b^2(X_{s_2})b''(X_{s_2}) \right)d{s_2} + \int_{0}^{s_1}b(X_{s_2})b'(X_{s_2})\,dW_{s_2}\right)\,dW_{s_1}\\
&=X_{0}+\int_{0}^{t}a(X_0)\,d{s_1}+\int_{0}^{t}b(X_0)\,dW_{s_1} \\
&+\int_{0}^{t}\int_{0}^{s_1}\left(aa' + \frac{1}{2}b^2a'' \right)ds_2\,d{s_1} + \int_{0}^{t}\int_{0}^{s_1}a'b\ dW_{s_2}\,d{s_1}\\
&+\int_{0}^{t}\int_{0}^{s_1}\left(ab' + \frac{1}{2}b^2b'' \right)d{s_2}\,dW_{s_1} + \int_{0}^{t}\int_{0}^{s_1}bb'\ dW_{s_2}\,dW_{s_1}\\
&=X_{0}+a(X_0)\,t+b(X_0)(W_t-W_0) + R,
\end{split}
\end{equation}\\
where $R$ stands for the remaining terms, and $dW_{s_2}$ and $dW_{s_1}$ are the Wiener increments for the same Wiener process $W_t$. It can easily be realized that the terms $\left(aa' + \frac{1}{2}b^2a'' \right)$ and $a'b$ which depend on $X_{s_2}$ can be further expanded by Eq. (\ref{integral form of f}), and this results in integrations with a constant $f(X_0)$ plus a remaining factor that scales with orders of $t$. This recursive expansion produces the It\^o--Taylor expansion. For simplicity, we define two operators to represent this substitutive recursion:\\
\begin{equation}\label{differential operator L}
L^0:=a\,\frac{\partial}{\partial x}+\frac{1}{2}b^2\frac{\partial^2}{\partial x^2}, \qquad L^1=b\,\frac{\partial}{\partial x},
\end{equation}\\
where the functions $a$ and $b$ take the same arguments as the function that is differentiated by them. Then Eq. (\ref{integral form of f}) can be written into\\
\begin{equation}\label{integral f simpler}
f(X_t) = f(X_0)+\int_{0}^{t}L^0f(X_s)ds + \int_{0}^{t}L^1f(X_s)\,dW_s.
\end{equation}\\
To demonstrate the recursive expansion, we expand Eq. (\ref{first stochastic expansion}) for a second time:\\
\begin{equation}\label{second stochastic expansion}
\begin{split}
X_t &=X_{0}+a(X_0)\,t+b(X_0)(W_t-W_0)+\int_{0}^{t}\int_{0}^{s_1}L^0a(X_{s_2})\ ds_2\,d{s_1} + \int_{0}^{t}\int_{0}^{s_1}L^1a(X_{s_2})\ dW_{s_2}\,d{s_1}\\
&+\int_{0}^{t}\int_{0}^{s_1}L^0b(X_{s_2})\ d{s_2}\,dW_{s_1} + \int_{0}^{t}\int_{0}^{s_1}L^1b(X_{s_2})\ dW_{s_2}\,dW_{s_1}\\
&=X_{0}+a(X_0)\,t+b(X_0)(W_t-W_0)+\int_{0}^{t}\int_{0}^{s_1}L^0a(X_0)\ ds_2\,d{s_1} + \int_{0}^{t}\int_{0}^{s_1}L^1a(X_0)\ dW_{s_2}\,d{s_1}\\
&+\int_{0}^{t}\int_{0}^{s_1}L^0b(X_0)\ d{s_2}\,dW_{s_1} + \int_{0}^{t}\int_{0}^{s_1}L^1b(X_0)\ dW_{s_2}\,dW_{s_1}\\
&+\int_{0}^{t}\int_{0}^{s_1}\int_{0}^{s_2}L^0L^0a(X_{s_3})\ d{s_3}\,ds_2\,d{s_1} + \int_{0}^{t}\int_{0}^{s_1}\int_{0}^{s_2}L^1L^0a(X_{s_3})\ dW_{s_3}\,ds_2\,d{s_1} \\
&+ \int_{0}^{t}\int_{0}^{s_1}\int_{0}^{s_2}L^0L^1a(X_{s_3})\ d{s_3}\,dW_{s_2}\,d{s_1} + \int_{0}^{t}\int_{0}^{s_1}\int_{0}^{s_2}L^1L^1a(X_{s_3})\ dW_{s_3}\,dW_{s_2}\,d{s_1}\\ &+\int_{0}^{t}\int_{0}^{s_1}\int_{0}^{s_2}L^0L^0b(X_{s_3})\ d{s_3}\,d{s_2}\,dW_{s_1} + \int_{0}^{t}\int_{0}^{s_1}\int_{0}^{s_2}L^1L^0b(X_{s_3})\ dW_{s_3}\,d{s_2}\,dW_{s_1}\\ &+\int_{0}^{t}\int_{0}^{s_1}\int_{0}^{s_2}L^0L^1b(X_{s_3})\ d{s_3}\,dW_{s_2}\,dW_{s_1} + \int_{0}^{t}\int_{0}^{s_1}\int_{0}^{s_2}L^1L^1b(X_{s_3})\ dW_{s_3}\,dW_{s_2}\,dW_{s_1},\\
\end{split}
\end{equation}
where the terms other than triple integrals can be evaluated directly. Obviously, this involves permutations of multiple $t$ and $W$ integrals. For notational simplicity we define an ordered list structure to denote the permutations as \\
\begin{equation}\label{index list}
\alpha=(j_1,j_2,\dots,j_l),\quad j_i=0,1\,,\quad i=1,2,\dots,l
\end{equation}\\
and define their manipulation as\\
\begin{equation}\label{index list manipulation 1}
\alpha * \tilde{\alpha} = (j_1,j_2,\dots,j_l, \tilde{j}_1,\tilde{j}_2,\dots,\tilde{j}_{\tilde{l}}),
\end{equation}
\begin{equation}\label{index list manipulation 2}
-\alpha=(j_2,j_3,\dots,j_l) ,\quad\alpha-=(j_1,j_2,\dots,j_{l-1}), \quad()\equiv v,
\end{equation}\\
that is, we use $v$ to denote an empty list. Then we define multiple It\^o integral as\\
\begin{equation}\label{multile Ito integral 1}
I_v[f(\cdot)]_{0,t}:=f(t), 
\end{equation}
\begin{equation}\label{multile Ito integral 2}
I_{\alpha*(0)}[f(\cdot)]_{0,t}:=\int_{0}^{t}I_{\alpha}[f(\cdot)]_{0,s}ds,\quad I_{\alpha*(1)}[f(\cdot)]_{0,t}:=\int_{0}^{t}I_{\alpha}[f(\cdot)]_{0,s}dW_s.
\end{equation}\\
Thus they are defined recursively. We give the following examples for the sake of illustration.\\
\begin{equation}\label{multile Ito integral examples}
\begin{split}
I_{(0)}[f(\cdot)]_{0,t}&=\int_{0}^{t}f(s)ds,\\ I_{(1,1)}[f(\cdot)]_{0,t}&=\int_{0}^{t}\int_{0}^{s_1}f(s_2)\ dW_{s_2}\,dW_{s_1},\\
I_{(1,1,0)}[f(\cdot)]_{0,t}&=\int_{0}^{t}\int_{0}^{s_1}\int_{0}^{s_2}f(s_3)\ dW_{s_3}\,dW_{s_2}\,ds_1.
\end{split}
\end{equation}\\
Therefore, among the indices in the parenthesis $(\cdots)$ 1 stands for integrating with respect to $W$, and 0 stands for integrating with respect to $t$, and all the integrations are multiple integrals. Note that the sequence of integration should be carried out from the left to the right. From now on, when we do not write the $[f(\cdot)]_{(0,t)}$ part in the expression $I_{\alpha}[f(\cdot)]_{0,t}$, i.e. simply as $I_{\alpha}$, we mean $I_{\alpha}[1]_{0,t}$, which is a constant factor dependent only on $t$.\\

Also for simplicity, we use the list structure to denote multiple actions of $L^0$ and $L^1$:\\
\begin{equation}\label{coefficient functions}
f_v:=f,\quad f_{(j)*\alpha}:=L^j f_\alpha,\quad j=0,1\,.
\end{equation}\\
For example, we have $L^0L^0L^1f$ as $f_{(0,0,1)}$. Eq. (\ref{integral f simpler}) can be written and expanded as\\
\begin{equation}\label{integral f using indices}
\begin{split}
f(X_t)=&f(X_0)+I_{(0)}[f_{(0)}(\cdot)]_{0,t}+I_{(1)}[f_{(1)}(\cdot)]_{0,t}\\
\\
=&f(X_0)+I_{(0)}\left[f_{(0)}(X_0)+I_{(0)}[L^0f_{(0)}(\circ)]_{0,\cdot}+I_{(1)}[L^1f_{(0)}(\circ)]_{0,\cdot}\right]_{0,t}\\
&+I_{(1)}\left[f_{(1)}(X_0)+I_{(0)}[L^0f_{(1)}(\circ)]_{0,\cdot}+I_{(1)}[L^1f_{(1)}(\circ)]_{0,\cdot}\right]_{0,t}\\
\\
=&f(X_0)+I_{(0)}f_{(0)}(X_0)+I_{(0)}\left[I_{(0)}[f_{(0,0)}(\circ)]_{0,\cdot}+I_{(1)}[f_{(1,0)}(\circ)]_{0,\cdot}\right]_{0,t}\\
&+I_{(1)}f_{(1)}(X_0)+I_{(1)}\left[I_{(0)}[f_{(0,1)}(\circ)]_{0,\cdot}+I_{(1)}[f_{(1,1)}(\circ)]_{0,\cdot}\right]_{0,t}\\
\\
=&f(X_0)+I_{(0)}f_{(0)}(X_0)+I_{(0,0)}[f_{(0,0)}(\cdot)]_{0,t}+I_{(1,0)}[f_{(1,0)}(\cdot)]_{0,t}\\
&+I_{(1)}f_{(1)}(X_0)+I_{(0,1)}[f_{(0,1)}(\cdot)]_{0,t}+I_{(1,1)}[f_{(1,1)}(\cdot)]_{0,t}.
\end{split}
\end{equation}\\
This is Eq. (\ref{first stochastic expansion}). Because the multiple integrals $I_\alpha$ decrease with the length of $\alpha$ and the time scale $t$, they are supposed to decrease in value when expanded to higher orders. Note that a single $1$ index in $\alpha$ provides $\frac{1}{2}$ order while a single $0$ index provides order 1. When we expand this It\^o--Taylor expansion, the number of its terms increases exponentially with increasing the order. We now state the It\^o--Taylor expansion:\\
\begin{equation}\label{Ito-Taylor expansion}
f(X_t)=\sum_{\alpha\in\mathcal{A}}I_\alpha f_\alpha(X_0)+\sum_{\alpha\in\mathcal{B}(\mathcal{A})}I_\alpha [f_\alpha(\cdot)]_{0,t},
\end{equation}
\begin{equation}\label{expansion index lists}
\alpha\in\mathcal{A}\text{ and }\alpha\ne v \Rightarrow-\alpha\in\mathcal{A},\quad  \mathcal{B}(\mathcal{A})=\{\alpha:\alpha\notin\mathcal{A}\text{ and }-\alpha\in\mathcal{A}\},
\end{equation}\\
where $\mathcal{A}$ and $\mathcal{B}$ are respectively called a hierarchical set and a remainder set.\\

When making a numerical approximation, we throw away the remainder part and take the $\mathcal{A}$ part only, and we say the approximation is up to order $k$ if $\mathcal{B}$ only contains terms of order $k+\frac{1}{2}$ regarding time $t$. We take $t$ in the above equation as the time step $dt$ for the approximation of $X_t$ in the defining equation Eq. (\ref{defining equation of X_t}), so that we approximate $dX_t$ up to order $(dt)^k$.\\

To calculate the multiple integrals $I_\alpha$, we need an important relation:\\
\begin{equation}\label{relation between Ito multiple integrals}
\begin{split}
I_{(1)}I_{\alpha}=W_tI_{(j_1,\dots,j_l)}=&\sum_{i=0}^{l}I_{(j_1,\dots,j_i,1,j_{i+1},\dots,j_l)}+\sum_{i=1}^{l}\delta_{j_i,1}I_{(j_i,\dots,j_{i-1},0,j_{i+1},\dots,j_l)},\\
I_{(0)}I_{\alpha}&=tI_{(j_1,\dots,j_l)}=\sum_{i=0}^{l}I_{(j_1,\dots,j_i,0,j_{i+1},\dots,j_l)}.
\end{split}
\end{equation}\\
That is, when we multiply the two integrals, we first insert the index of the first integral into the index list of the second one, i.e. $\alpha$, and sum over different inserting positions, and then if the first integral is a Wiener process, it cancels each Wiener integral index 1 in the list $\alpha$ and produces a 0. We first show the $I_{(0)}$ part:\\
\begin{equation}
d(I_{(0)}I_\alpha)=d(tI_{(j_1,\dots,j_l)})=dt\,I_{(j_1,\dots,j_l)}+tI_{(j_1,\dots,j_{(l-1)})}dW^{j_l},
\end{equation}
\begin{equation}
I_{(0),t}I_{\alpha,t}=\int^{t}d(I_{(0),s}I_{\alpha,s})=I_{(j_1,\dots,j_l,0),t}+\int^{t} sI_{(j_1,\dots,j_{(l-1)}),s}dW^{j_l}_s,
\end{equation}\\
where we write $dW^0$ as $dt$. Recursively we have \\
\begin{equation}
\begin{split}
\int^{t} sI_{(j_1,\dots,j_{(l-1)}),s}dW^{j_l}_s&=\int^{t} I_{(j_1,\dots,j_{(l-1)},0),s}dW^{j_l}_s+\int^{t}\int^{s_1} sI_{(j_1,\dots,j_{(l-2)}),s}dW^{j_{l-1}}_{s_2}dW^{j_l}_{s_1}\\
&=I_{(j_1,\dots,j_{(l-1)},0,j_l),t}+\int^{t}\int^{s_1} s_2I_{(j_1,\dots,j_{(l-2)}),s_2}dW^{j_{l-1}}_{s_2}dW^{j_l}_{s_1},
\end{split}
\end{equation}\\
and therefore by induction and the fact $I_v=1$ we have \\
\begin{equation}
I_{(0)}I_{\alpha}=\sum_{i=0}^{l}I_{(j_1,\dots,j_i,0,j_{i+1},\dots,j_l)}.
\end{equation}\\
The same process applies to the case of $I_{(1)}=W_t$, with the only difference being that\\
\begin{equation}
d(WI_{(j_1,\dots,j_l)})=dW\,I_{(j_1,\dots,j_l)}+WI_{(j_1,\dots,j_{(l-1)})}dW^{j_l} +\delta_{j_l,1}dt\,I_{(j_1,\dots,j_{(l-1)})},
\end{equation}\\
due to the fact that $dW^2=dt$. The final result Eq. (\ref{relation between Ito multiple integrals}) follows straightforwardly.
\section{Order 1.5 Strong Scheme}\label{numerical update rule appendix}
In our numerical simulation, we use an order 1.5 strong approximation. The word \textit{strong} means that the simulated stochastic variable approaches a certain real trajectory with error of order 1.5 , and this directly relates to the use of It\^o--Taylor expansion. For consistency with Ref. \cite{NumericalSimulationofStochasticDE}, we rewrite $X_t$ into $Y_n$ where $n$ denotes the $n$-th time step. The update rule of $Y_n$ is essentially\\
\begin{equation}\label{1.5 Taylor}
Y_{n+1}=Y_n+aI_{(0)}+bI_{(1)}+a_{(0)}I_{(0,0)}+b_{(0)}I_{(0,1)}+a_{(1)}I_{(1,0)}+b_{(1)}I_{(1,1)}+b_{(1,1)}I_{(1,1,1)},
\end{equation}\\
where we have used the notation in Eq. (\ref{coefficient functions}). For completeness we present the terms of $a$ and $b$:\\
\begin{equation}\label{derivatives appendix}
\begin{split}
a_{(0)}=aa'+\frac{1}{2}b^2a'',&\quad b_{(0)}=ab'+\frac{1}{2}b^2b''\\
a_{(1)}=ba',\quad b_{(1)}=bb',&\quad b_{(1,1)}=b((b')^2+bb'')
\end{split}
\end{equation}\\
Then we need to evaluate the constant terms $I$. According to Eq. (\ref{relation between Ito multiple integrals}), we have\\
\begin{equation}
\begin{split}
I_{(0)}&=dt,\quad I_{(1)}={} dW,\quad I_{(0,0)}=\frac{1}{2}dt^2,\\
dW\,dW=I_{(1)}I_{(1)}&=2I_{(1,1)}+I_{(0)},\quad dW\,dt=I_{(1)}I_{(0)}=I_{(1,0)}+I_{(0,1)},\\
&dWI_{(1,1)}={}3I_{(1,1,1)}+I_{(1,0)}+I_{(0,1)},
\end{split}
\end{equation}\\
where we have used the symbol $dt$ to denote our iteration time step, which is sometimes also written as $\Delta t$ or $\Delta$. Thus both $dt$ and $dW$ are finite, and do not involve integrated values, i.e. $dW^2\ne dt$. Then we reach the result:\\
\begin{equation}\label{numerical step definition}
\begin{split}
I_{(1,1)}&=\frac{1}{2}(dW^2-dt),\quad I_{(1,0)}=:dZ,\quad I_{(0,1)}=dW\,dt-dZ,\\
I_{(1,1,1)}&=\frac{1}{3}(\frac{dW}{2}(dW^2-dt)-dW\,dt)=dW(\frac{1}{6}dW^2-\frac{1}{2}dt)=\frac{dW}{2}\left(\frac{dW^2}{3}-dt\right),
\end{split}
\end{equation}\\
where we have to define an additional random variable $dZ$. Here $dZ$ satisfies \\
\begin{equation}\label{dZ}
dZ\sim\mathcal{N}\left(0,\frac{1}{3}dt^3\right),\quad E(dZ\,dW)=\frac{1}{2}dt^2,
\end{equation}\\
which is a Gaussian variable correlated to $dW$. Its properties are discussed in Ref. \cite{NumericalSimulationofStochasticDE}. Then we can write everything explicitly:\\
\begin{equation}\label{1.5 Taylor complete}
\begin{split}
Y_{n+1}={}&Y_n+a\,dt+b\,dW+\frac{1}{2}bb'(dW^2-dt)+\frac{1}{2}\left(aa'+\frac{1}{2}b^2a''\right)dt^2\\
&+a'b\,dZ+\left(ab'+\frac{1}{2}b^2b''\right)(dW\,dt-dZ)\\
&+\frac{1}{2}b\left(bb''+(b')^2\right)dW\left(\frac{1}{3}dW^2-dt\right).
\end{split}
\end{equation}\\
This is the order 1.5 strong Taylor scheme, and we ignore arguments when they are just $Y_n$. To obtain stable and precise numerical results, the deterministic second-order term with $dt^2$ is also included, while the stochastic terms of the second order are not. Next, we need to avoid the explicit calculation of derivatives and use the numerically evaluated values. It then becomes\\
\begin{equation}\label{order 1.5 explicit}
\begin{split}
Y_{n+1}={}&Y_n+b\,dW+\frac{1}{4}\left(a(Y_+)+2a+a(Y_-)\right)dt+\frac{1}{4\sqrt{dt}}\left(b(Y_+)-b(Y_-)\right)(dW^2-dt)\\
&+\frac{1}{2\sqrt{dt}}\left(a(Y_+)-a(Y_-)\right)dZ+\frac{1}{2dt}\left(b(Y_+)-2b+b(Y_-)\right)(dW\,dt-dZ)\\
&+\frac{1}{4dt}\left(b(\Phi_+)-b(\Phi_-)-b(Y_+)+b(Y_-)\right)dW\left(\frac{1}{3}dW^2-dt\right),
\end{split}
\end{equation}
\begin{equation}
Y_\pm = Y_n+a\,dt\pm b\sqrt{dt},\quad \Phi_\pm = Y_+ \pm b(Y_+)\sqrt{dt}
\end{equation}\\
which can be confirmed by expanding $Y_\pm$ and $\Phi_\pm$ up to order 1 of $dt$. This is because the coefficients are at least of order $\frac{1}{2}$, and the $dt$ term which is supposed to approximate up to $(dt)^2$ has coefficient $dt$ that is of order 1:\\
\begin{equation}
\begin{split}
\frac{dt}{4}\left(a(Y_+)+2a+a(Y_-)\right)&=\frac{dt}{4}\left(4a+2a'a\,dt+a''b^2\,dt\right)=a\,dt+\frac{dt^2}{2}(aa'+\frac{1}{2}b^2a''),\\
\frac{1}{2\sqrt{dt}}\left(b(Y_+)-b(Y_-)\right)&=\frac{1}{2\sqrt{dt}}(b'\cdot 2b\sqrt{dt})=bb',\qquad \frac{1}{2\sqrt{dt}}a(Y_+)-a(Y_-)=a'b,\\
\frac{1}{2dt}\left(b(Y_+)-2b+b(Y_-)\right)&=\frac{1}{2dt}\left(b'\cdot 2a\,dt+b''b^2\,dt\right)=ab'+\frac{1}{2}b^2b'',\\
\frac{1}{2dt}(b(\Phi_+)-b(\Phi_-)-b(Y_+)&+b(Y_-) ) =\frac{1}{2dt}\left(2b'b(Y_+)\sqrt{dt}+2b''bb(Y_+)dt-2bb'\sqrt{dt} \right)\\
&\qquad\qquad = \frac{1}{2dt}\left(2b'(b'b\sqrt{dt})\sqrt{dt}+2b''bb(Y_+)dt \right)\\
&\qquad\qquad = b(b')^2+b^2b''.
\end{split}
\end{equation}\\
Therefore the above update rule Eq. (\ref{order 1.5 explicit}) is correct. However, this is not enough. In actual numerical approximations, the simulated vector components rotate in the complex plane, and the high-frequency components rotate faster and accumulate more error. Because the error always increases the norm of the values, it accumulates and diverges exponentially, which is especially severe for high-frequency components that are around the energy cutoff. Therefore, if possible we use an implicit method that takes the update target $Y_{n+1}$ into its arguments, which solves the high-frequency divergence problem by forcing the error not to increase the norm, and typically to shrink the values with high error so that they do not accumulate or propagate. The implicit scheme update rule is the following:\\
\begin{equation}\label{order 1.5 implicit}
\begin{split}
Y_{n+1}={}&Y_n+b\,dW+\frac{1}{2}\left(a(Y_{n+1})+a\right)dt+\frac{1}{4\sqrt{dt}}\left(b(Y_+)-b(Y_-)\right)(dW^2-dt)\\
&+\frac{1}{2\sqrt{dt}}\left(a(Y_+)-a(Y_-)\right)\left(dZ-\frac{1}{2}dW\,dt\right)+\frac{1}{2dt}\left(b(Y_+)-2b+b(Y_-)\right)(dW\,dt-dZ)\\
&+\frac{1}{4dt}\left(b(\Phi_+)-b(\Phi_-)-b(Y_+)+b(Y_-)\right)dW\left(\frac{1}{3}dW^2-dt\right),
\end{split}
\end{equation}
where $dZ,\ Y_\pm$ and $\Phi_\pm$ are the same as before. Note that there is an additional term 
$$\dfrac{1}{2\sqrt{dt}}\left(a(Y_+)-a(Y_-)\right)\left(-\frac{1}{2}dW\,dt\right),$$
which equals $-\dfrac{a'b}{2}dW\,dt$. This is to cancel the extra $b\,dW$ term inside $Y_{n+1}$, which appears if we expand $Y_{n+1}$ by It\^o--Taylor expansion (see Eq. (\ref{1.5 Taylor complete})):\\
\begin{equation}\label{Y_n+1}
Y_{n+1}=Y_n+a\,dt+b\,dW+\frac{1}{2}bb'(dW^2-dt)+\frac{1}{2}\left(aa'+\frac{1}{2}b^2a''\right)dt^2\dots,
\end{equation}
\begin{equation}
a(Y_{n+1})=a+a'\left(a\,dt+b\,dW+\frac{1}{2}bb'(dW^2-dt)\right)+\frac{1}{2}a''b^2\,dW^2,
\end{equation}
\begin{equation}
\frac{dt}{2}a(Y_{n+1})=\frac{dt}{2}a+\frac{dt^2}{2}a'a+\frac{dt\,dW}{2}a'b+\frac{dt^2}{4}a''b^2+\frac{dt(dt-dW^2)}{4}a''b^2+\dots,
\end{equation}\\
where we have ignored all stochastic terms equal to or above order 1. By taking $dW^2=dt$, we recover the order 1.5 It\^o--Taylor expansion (\ref{1.5 Taylor complete}) as the explicit update scheme case.\\

In our numerical simulation of Eq. (\ref{position measurement evolution}), we can split the function that governs deterministic evolution as $a=a_1+a_2$, where $a_1$ is the Hamiltonian part $a_1(Y)=HY$ which is linear and can be evaluated implicitly, because \\
\begin{equation}
\begin{split}
Y_{n+1}=\frac{1}{2}(HY_{n+1}+HY_n)dt\quad&\Rightarrow\quad (I-\frac{dt}{2}H)Y_{n+1}=\frac{dt}{2}HY_n\\
&\Rightarrow\quad Y_{n+1}=(I-\frac{dt}{2}H)^{-1}\,\frac{dt}{2}HY_n.
\end{split}
\end{equation}\\
However, $a_2$ is a nonlinear part. For the case of our simulated differential equation, we can only evaluate $a_1(Y_{n+1})$ implicitly but not $a_2(Y_{n+1})$, therefore we use a mix of explicit and implicit schemes. Because the terms
$$\frac{1}{2}a(Y_{n+1})dt-\dfrac{1}{2\sqrt{dt}}\left(a(Y_+)-a(Y_-)\right)\cdot\frac{1}{2}dW\,dt$$
in the implicit scheme Eq. (\ref{order 1.5 implicit}) replace the term
$$\frac{1}{4}\left(a(Y_+)+a(Y_-)\right)dt$$
in the explicit scheme Eq. (\ref{order 1.5 explicit}), we separately use them for $a_1$ and $a_2$ in our simulated equation, that is\\
\begin{equation}\label{partial implicit}
\frac{1}{2}a_1(Y_{n+1})dt-\dfrac{1}{2\sqrt{dt}}\left(a_1(Y_+)-a_1(Y_-)\right)\cdot\frac{1}{2}dW\,dt+\frac{1}{4}\left(a_2(Y_+)+a_2(Y_-)\right)dt.
\end{equation}\\
Substituting it into the corresponding part $\frac{1}{4}\left(a(Y_+)+a(Y_-)\right)dt$ of Eq. (\ref{order 1.5 explicit}), we obtain an update rule to use in our experiments. On the other hand, when the simulated state is not a vector but a density matrix, this method is hard to apply, because there seems to be no readily available method to quickly solve equations of the form $Y=-i[H,Y]dt+C$ with a banded matrix $H$.\\

Due to the simplicity of evaluating $a_1$, we add one more term concerning $a_1$ into our update rule equation. First, we have linearity $a'=a'_1+a'_2$, and we have $a''_1=0$ and $a'_1$ is a constant linear mapping. We know that the values concerning $a_1$ are typically larger than those concerning $a_2$ and $b$. Therefore, we add an additional deterministic 3-rd order term $\frac{1}{6}dt^3(a'_1)^2a$ into the update rule, which corresponds to the term $\frac{dt^3}{6}f^{(3)}$ of the Taylor expansion, and it indeed also exists in the It\^o--Taylor expansion. Because the implicit method uses $\dfrac{dt}{2}a_1(Y_{n+1})$, which already includes a term $\dfrac{dt^3}{4}a'_1a'_1a$, for implicit methods we modify it by adding a term $-\dfrac{1}{12}dt^3(a'_1)^2a$ to correct its value. This method turns out to reduce the numerical error, as it works in the direction to prevents the norm increase. This is especially important when an implicit method is not used, because Taylor expansion up to lower or higher orders around order 3 all results in update rules that increase the norm, which would make the value diverge exponentially.

		\chapter{Details of the Reinforcement Learning Algorithm}\label{experiment details appendix}
		In this appendix, we discuss the reinforcement learning strategies we have adopted as extensions of the basic Deep Q-Network reinforcement learning. We have mainly followed the Rainbow DQN in Ref. \cite{RainbowDQN} which integrates most of the recently developed DQN techniques, while there are still some differences between our implementation and theirs. In the following, we introduce the involved DQN techniques one by one in Section \ref{DQN Techniques} and present our settings in Section \ref{DQN Settings}. This appendix is intended for the background knowledge of Chapter \ref{deep reinforcement learning}.\\

\section{Extensions of DQN}\label{DQN Techniques}
\subsection{Target Network and \texorpdfstring{{\Large $\epsilon$}}{epsilon}-Greedy Strategy}\label{target network section}
As already introduced in the earliest paper of DQN \cite{DQN}, it is usually necessary to keep a target network as a reference to train the reinforcement learning network. Because the minimized loss in Eq. (\ref{TD loss}), i.e.,\\
\begin{equation}\label{TD loss (copy)}
L=Q(s_t,a_t)-\left(r(s_t,a_t)+\gamma\,\max_{a_{t+1}}Q(s_{t+1},a_{t+1})\right)\,,
\end{equation}\\
is to minimize a loss of the Q-network against the Q-network itself, a direct gradient descent method is found to be generally unstable. Therefore, to make it stable, in Eq. (\ref{TD loss (copy)}) we use a fixed target network to evaluate the $Q$-function on the right and use the currently trained network to evaluate the $Q$-function on the left, and we use the gradient descent method to modify the parameters of the currently trained network only. In order to synchronise the target network with the current learning progress, we update the target network by assigning the trained network's parameters to it after some predetermined training time, and then repeat the training. The training loss is therefore\\
\begin{equation}\label{TD loss target network}
L=Q_{\theta_1}(s_t,a_t)-\left(r(s_t,a_t)+\gamma\,\max_{a_{t+1}}Q_{\theta_2}(s_{t+1},a_{t+1})\right)\,,
\end{equation}\\
where $\theta_1$ represents the parameters of the currently trained neural network and $\theta_2$ represents the target network. Only parameters $\theta_1$ are modified by the gradient descent.\\

The $\epsilon$-greedy strategy is a strategy of taking actions for a reinforcement learning agent (or controller) during learning. The greedy strategy means to always take the action that has the highest expected reward, and the $\epsilon$-greedy strategy means to use the greedy strategy with a probability of $1-\epsilon$, and with a probability of $\epsilon$ it takes a random action. This $\epsilon$-greedy strategy uses such an $\epsilon$ random search to explore possibilities of small deviations from the greedy strategy and encourages exploration of the action space. Typically this $\epsilon$ hyperparameter is manually annealed during training. This technique is standard, and especially it is used when no other exploration strategies are adopted.\\

\subsection{Double Q-Learning}\label{DoubleDQN section}
This technique is introduced in Ref. \cite{DoubleDQN}. When we look at the loss that is optimized as in Eq. (\ref{TD loss (copy)}), we can notice that there is a $\max$ operation involved, which always takes the maximal value of the neural network output. As we know that the neural network has fluctuating parameter values, this $\max$ operation will make the neural-network estimation of the function $Q$ always higher than what it should be. In order to make this $Q$ estimation center at its correct value, we separate the decision of the best action $\max_{a_{t+1}}$ and the evaluation of the $Q$-function by using two different neural networks.\\

We keep two neural networks with different parameters $\theta_1$ and $\theta_2$. To decide the optimal action $a^*_{t+1}$, we use the network with parameters $\theta_1$:\\
\begin{equation}
a^*_{t+1}=\argmax_{a_{t+1}}Q_{\theta_1}(s_{t+1},a_{t+1}).
\end{equation}\\
Then we evaluate this optimal action $a^*_{t+1}$ on the other neural network $\theta_2$:\\
\begin{equation}
r(s_t,a_t)+\gamma\,Q_{\theta_2}(s_{t+1},a^*_{t+1}).
\end{equation}\\
Because the fluctuations in the two neural networks are typically uncorrelated, this estimated $Q$-function does not always takes the maximum of a network's fluctuation now. The above strategy does not impede training because the two networks $\theta_1$ and $\theta_2$ learn the reinforcement learning task simultaneously, and both of them output reasonable decisions throughout training. In cases where both of them work perfectly, the decision outputted from one network would be identical to that from the other, and the loss reduces to the previous case as in Eq. (\ref{TD loss (copy)}). In practice, it is found sufficient to use the target network in Section \ref{target network section} as the network $\theta_2$, and to use the trained network as $\theta_1$. Note that the choice of $a^*_{t+1}$ depends on the trained network $\theta_1$ but is not involved in the gradient computation for the gradient descent.\\

\subsection{Dueling DQN Structure}\label{Duel DQN section}
This technique is introduced in Ref. \cite{DuelDQN}. As introduced in Chapter \ref{deep reinforcement learning}, deep learning usually comes with numerical imprecision and noise. Due to its properties and its gradient-descent-based training, when the output value of the neural network is large, the gradients are large, and the numerical imprecision resulting from gradient descent is large. Therefore, if we have a generally large $Q$-function value which is the network output, it would be difficult for the network to learn small details of the $Q$-function further. However, since we use the formula $a^*_{t+1}=\argmax_{a_{t+1}}Q_{\theta_1}(s_{t+1},a_{t+1})$ to decide the best action $a^*_{t+1}$, it is necessary to reliably compare the $Q$ values for different actions $a_{t+1}$, which can possibly be very close and hard to learn. To alleviate this problem, a dueling structure of the DQN was proposed, which is used to separately predict the mean value of $Q$ for all actions and the deviations from the mean for each action choice $a$. In this way, even though the mean is large and has a large error, the deviations from the mean can be small and reliably learned, which gives a correct action choice and a better final performance.\\

\subsection{Prioritized Memory Replay}\label{Prioritized Replay}
This technique is introduced in Ref. \cite{prioritizedSampling}. Its idea is simple: if the loss on some training data is high, then there is more to learn on those data, and we should sample them more often so that they are learned for more times. This strategy is especially important when the reward is sparse, or when there are only a few moments that are crucial. As demonstrated in Ref. \cite{prioritizedSampling}, in some extreme cases this strategy even accelerates the speed of learning for $10^3$ times, which is striking. Therefore, it is almost always used together with usual reinforcement learning. \\

To implement the prioritized sampling, we first need to define the priority. As in common cases, we set the probability $p$ of a data being sampled as\\
\begin{equation}\label{sampling priority}
p\propto |L|^\alpha,
\end{equation}\\
where $L$ is the previously estimated loss on the data and $\alpha$ is a hyperparameter. In order to make the original optimization target unaffected by prioritized sampling, the loss of each data is rescaled by the probability of the data being sampled, so that the loss minimization is still done equally on the whole dataset, not merely concentrating on high loss cases.\\

In cases where the memory replay is of a limited size, when the memory replay buffer is full and new experiences should be stored, we take away the samples with the lowest loss from the replay buffer to store new experiences so that the memory is best utilized. Because this strategy would potentially change the optimization target, we do so only when the memory replay can definitely not be enlarged.\\

\subsection{Noisy DQN}\label{Noisy DQN}
The noisy DQN is introduced in Ref. \cite{NoisyDQN}, and it is an exploration strategy that helps with the $\epsilon$-greedy exploration. As explained in Section \ref{target network section}, the $\epsilon$-greedy strategy only explores small deviations from the deterministic greedy strategy and cannot explore totally different long-term strategies. Instead of completely random exploration, we may utilize the neural network itself to explore by introducing random variables into the neural network. In Ref. \cite{NoisyDQN}, noisy layers are used to replace the usual network layers. It can be written as\\
\begin{equation}
\boldsymbol{y}=(\boldsymbol{b}+\boldsymbol{b}_{\text{noisy}}\odot\epsilon_b) + (\textbf{W}+\textbf{W}_{\text{noisy}}\odot\epsilon_w)\boldsymbol{x},
\end{equation}\\
where $\epsilon_b$ and $\epsilon_w$ are sampled random variables, $\odot$ denotes the element-wise multiplication, and $\boldsymbol{b}_{\text{noisy}}$ and $\textbf{W}_{\text{noisy}}$ are learned parameters that rescale the noise. In most cases $\boldsymbol{b}_{\text{noisy}}$ and $\textbf{W}_{\text{noisy}}$ gradually shrinks during learning, showing a self-annealing behaviour. Concerning implementation, we follow Ref. \cite{NoisyDQN} to use factorized Gaussian noises to construct $\epsilon_w$ and then rescale it by the square root function while keeping its sign.\\

There is still one technical caveat when using this strategy. During training, the noise parameters $\epsilon_b$ and $\epsilon_w$ should be sampled differently for different data in a training batch; otherwise $\boldsymbol{b}_{\text{noisy}}$ and $\textbf{W}_{\text{noisy}}$ will always experience similar gradients to those of $\boldsymbol{b}$ and $\textbf{W}$ and it will not work correctly. For efficiency, one single noise can be used on multiple training data, and the number of noises used is a hyperparameter.\\

Above are the techniques that are adopted in our reinforcement learning. Two other techniques in Rainbow DQN \cite{RainbowDQN} are not used; however, for the sake of completeness, we briefly discuss and explain them in the following.\\

\subsection{Multi-Step Learning}
As introduced in Ref. \cite{ReinforcementlearningAnintroduction}, we can consider an alternative for Eq. (\ref{TD loss (copy)}) that learns multiple time steps altogether rather than from step $t$ to $t+1$. The motivation comes from the fact that, when the reinforcement learning learns step by step, the information of future rewards also propagate step by step, and if the number of steps is large, the learning process can be very slow. This results in a modified loss:\\
\begin{equation}\label{multi-step loss}
L=Q(s_t,a_t)-\sum_{k=0}^{n-1}\gamma^k r(s_{t+k},a_{t+k})+\gamma^n\,\max_{a_{t+n}}Q(s_{t+n},a_{t+n})\,,
\end{equation}\\
where $n$ is the number of steps that are considered altogether as a large step. This strategy is heuristic, and although it can result in faster learning, it no longer preserves the original definition of the $Q$-function as in Section \ref{reinforcement learning} and therefore does not theoretically guarantee the convergence to the optimal solution. Since in our experiments we compare the optimal controls, we do not adopt this strategy.\\

\subsection{Distributional Learning}
As suggested in Ref. \cite{DistributionalDQN}, it is possible to let the neural network learn the distribution of future rewards rather than the $Q$-function which is the mean of future rewards. This strategy modifies the loss function drastically, changing it from the difference between two values to the difference between two distributions, as measured by the Kullbeck-Leibler divergence. This strategy has the advantage that its loss function correctly represents how well the AI has understood the behaviour of the environment in a stochastic setting, while the loss defined with the $Q$-function may be dominated by the variance that results from environmental noises. Therefore when combined with prioritized sampling, this distributional learning strategy can have a significant increase in performance in the long term.\\

However, in order to approximately represent a probability distribution, in the original paper the authors discretized the space of future rewards, and used one output value for each discretized point per action choice \cite{DistributionalDQN}. This requires us to presume the region of rewards that can possibly be achieved, and to apply proper discretization of the rewards. These requirements contradict our original purposes of research on the quantum control problems, where we have hoped to know what rewards could be achieved and how precisely an optimal reward could be acquired. Therefore, we do not adopt this strategy.\\

\section{Hyperparameter Settings}\label{DQN Settings}
In this section, we present our hyperparameter settings and discuss the important ones. All our hyperparameter settings are determined empirically and we do not guarantee their optimality.\\
\subsection{Gradient Descent}
We use the RMSprop gradient descent optimizer provided by Pytorch \cite{RMSprop}. The learning rate is annealed from $2\times10^4$ to $1\times10^6$ by 5 steps. The 5 learning rates are: $2\times10^{-4}$, $4\times10^{-5}$, $8\times10^{-6}$, $2\times10^{-6}$, $1\times10^{-6}$. The learning rate schedules during training are different for different problems and cases, including different input cases, and we have set the learning rate schedules empirically by observing when the loss stopped to change and levelled off. The momentum parameter is set to be 0.9, and the $\epsilon$ parameter which is used to prevent numerical divergence is set to be $10^{-5}$.\\

To obtain a useful training loss for minimization, we deform the loss value $L$ as in Eq. (\ref{TD loss (copy)}) to construct the Huber loss, i.e.,\\
\begin{equation}
L'=\left\{\begin{split}
\frac{1}{2}L^2,\qquad\ \text{if } |L|<1,\\
|L|-\frac{1}{2},\quad \text{if } |L|\ge 1.
\end{split}\right.
\end{equation}\\
such that the training loss is a mean-squared-error when $|L|$ is small, and it is a L1 loss when $|L|$ is large. Unlike the usual mean-squared-error loss, this strategy guarantees that large loss values do not result in extremely large gradients that may cause instability. Also, before each update step of the parameters, we manually clip the gradient of each parameter such that the gradients have a maximal norm of 1.

\subsection{Prioritized Replay Settings}\label{Prioritized replay setting}
The memory replay buffer that supports prioritized sampling is efficiently realized using the standard Sum Tree data structure, and we have used the Sum Tree code\footnote{\url{https://github.com/jaromiru/AI-blog/blob/master/SumTree.py}} released under the MIT License on GitHub. Concerning the hyperparameter settings, our settings are different for different problems and they are summarized in Table \ref{prioritized replay settings}. Different input cases for the same problem share the same settings.\\
\begin{table}[hbt]
	\centering
	\begin{tabular}{p{8.4em}ccccc}
		\toprule
		   &  $\alpha$  &  $\beta$   &  $p_\epsilon$ &  $L_{\max}$ &  $p_{\text{replace low loss}}$  \\ \midrule
		   cooling oscillators &  0.4  &  $0.2\to 1.0$  & 0.001 &  10 & 0.9 \\ \midrule
		   stabilizing inverted oscillators &  0.8  &  $0.2\to 1.0$  &  0.001 &  10 & 0.9 \\ \midrule
		cooling quartic oscillators  & 0.4 & $0.2\to 1.0$  & 0.001 & 10 & 0.8 \\ \bottomrule
	\end{tabular}
	\caption{The prioritized replay settings used in our experiments.}
	\label{prioritized replay settings}
\end{table}\\

Among the parameters in Table \ref{prioritized replay settings}, $\beta$ denotes the extent of loss rescaling of the data with different priorities such that the optimization target is kept the same, and after each training step, it is incremented by 0.001 until it reaches 1. A small parameter $p_\epsilon$ is added to the sampling probability for every data so that all data have finite probabilities to be sampled. The parameter $L_{\max}$ is a cutoff of large losses in order to compute moderate probabilities, and for a probability $p_{\text{replace low loss}}$ we use new experience data to replace the training data that have lowest losses when the replay buffer is full; otherwise we randomly take out existing training data to store new data. Note that actually we need to sort out a portion of data that have the lowest losses at a time, not the single data. Otherwise it would be extremely computationally inefficient and almost stop the training algorithm. In our implementation we sort out the 1\% portion of the data with lowest losses each time, and then we replace them one by one.\\

\subsection{Other Reinforcement Learning Settings}\label{other settings}
The period for updating the target network discussed in Section \ref{target network section} is set to 300 training steps. However, to facilitate learning at the initial stage, we set it to 30 steps at the start, and after the simulated system has achieved a maximal time of $20\times\frac{1}{\omega_c}$ during which it does not fail, we set the target network update period to 150 steps, and after it has achieved $50\times\frac{1}{\omega_c}$, we set the update period to 300 steps. Note that if the training succeeds, it must be able to achieve the $t_{\text{max}}$ that is $100\times\frac{1}{\omega_c}$.\\

Concerning the neural networks, in order to accelerate training, we follow Ref. \cite{weightNormalization} to separate the learning of the weight matrices and the norms of the weight matrices. This strategy is applied to both usual network layers and the noisy layers as in Section \ref{Noisy DQN}. During training, we use a minibatch size of 512, and we sample 32 different noises on those data in the noisy layers. For computational efficiency, we always sample the noises in advance to use them later. The number of training steps is set to be proportional to the number of experiences that are stored into the memory buffer. For the quadratic problems, in the position and momentum input case, each experience is sampled and learned for 8 times on average, and in other input cases, each experience is sampled for 16 times on average. For the quartic problem, each experience is sampled for 8 times. \\

To obtain a trained neural network to evaluate, we track the performance of the networks during training. When the reinforcement learning agent performs better than the current performance record, we re-evaluate it for two more times, and if the average performance achieves a new performance record, we save this neural network in the hard disk for further evaluation after the training. Finally, we pick the 10 best trained networks to do more detailed evaluation as described in Section \ref{performance evaluation} to give our final performances. For the stabilizing inverted oscillator problem, it is hard to measure the performance of a reinforcement learning agent in a single trial. Therefore, whenever the reinforcement learning agent succeeds in one episode, we test it for two more episodes, and if it succeeds for all these three trials, we store the network in the hard disk when such successful networks have already appeared for 8 times. That is, we save one network per eight networks that are tested to be successful. In this way we can guarantee that the saved networks are not too close to each other and are roughly equispaced in the experienced training time. After the training, we evaluate the performances of the 10 latest neural networks that are saved.
		
	\end{appendices}

	\bibliography{references}
	\printindex
\end{document}